# Fundamentals and advances in magnetic hyperthermia

http://arxiv.org/abs/1510.06383

E. A. Périgo [a], G. Hemery [b, c], O. Sandre [b, c], D. Ortega [d, e, f], E. Garaio [g], F. Plazaola [g], F. J. Teran [d, e]

a: Physics and Material Sciences Research Unit, University of Luxembourg, L-1511, Luxembourg

b: Univ. Bordeaux, LCPO UMR 5629, ENSCBP 16 avenue Pey Berland, F-33600 Pessac, France

c: CNRS, Laboratoire de Chimie des Polymères Organiques, UMR 5629, F-33600, Pessac, France

d: iMdea-Nanociencia, Campus Universitario de Cantoblanco, 28049 Madrid, Spain

e: Nanobiotecnología (IMDEA-Nanociencia), Unidad Asociada al Centro Nacional de Biotecnología (CSIC)

f: Institute of Biomedical Engineering, University College London, Gower Street, London, WC1E 6BT, UK

g: Elektrizitatea eta Elektronika Saila, UPV/EHU, P.K. 644, Bilbao, Spain

## ABSTRACT

Nowadays, magnetic hyperthermia constitutes a complementary approach to cancer treatment. The use of magnetic particles as heating mediators, proposed in the 1950's, provides a novel strategy for improving tumor treatment and, consequently, patient quality of life. This review reports a broad overview about several aspects of magnetic hyperthermia addressing new perspectives and the progress on relevant features such as the *ad hoc* preparation of magnetic nanoparticles, physical modeling of magnetic heating, methods to determine the heat dissipation power of magnetic colloids including the development of experimental apparatus and the influence of biological matrices on the heating efficiency.

*Keywords:* Magnetic hyperthermia, iron oxide nanoparticles, superparamagnetic nanoparticles, specific absorption rate, intrinsic loss power

Corresponding author: Dr. Elio Alberto Périgo
Address: 162a, Avenue de la Faiencerie – Luxembourg – L1511 – Luxembourg
Phone: 352 46 66 44 6992 – Fax : 352 46 66 44 6330 – email : eaperigo@ieee.org

## List of symbols/acronyms

| | |
|---|---|
| $A$ | Area enclosed by the hysteresis loop |
| $AMF$ | Alternating magnetic field |
| $\boldsymbol{B}$ | Magnetic density flux |
| $B$ | Weighting coefficient |
| C | Heat capacity |
| $c$ | Nanoparticles mass concentration |
| $f$ | Frequency |
| $Fo$ | Fourier number |
| $H_0$ | Maximum magnetic field intensity |
| $ILP$ | Intrinsic loss power |
| $I_N$ | Damaged normal tissue volume above the pre-defined necrosis temperature |
| $I_T$ | Tumor volume with a temperature above a pre-defined threshold temperature |
| $Jo$ | Joule number |
| $k_B$ | Boltzmann constant |
| $K$ | Uniaxial anisotropy constant ($Jm^{-3}$) |
| $K_s$ | Surface anisotropy constant ($Jm^{-2}$) |
| L$(\xi)$ | Langevin equation |
| LRM | Linear response model |
| $m$ | Mass of nanoparticles |
| $M(t)$ | Dynamic magnetization |
| MH | Magnetic hyperthermia |
| (M)NP(s) | (Magnetic) nanoparticle(s) |
| $MRI$ | Magnetic resonance imaging |
| $Q_{met}, Q_{ext}$ | Heat flows from metabolism and spatial heating, respectively |
| *SAR* | Specific absorption rate |
| *SHP* | Specific heat power |
| *SLP* | Specific loss power |
| SPM | Superparamagnetic |
| SWM | Stoner-Wohlfarth model |
| TGA | Thermogravimetric analysis |
| $V$ | Particle volume |
| $\phi$ | Magnetic flux that crosses the pick-up coil |
| $\varphi$ | Volume fraction of the particles in the colloidal suspension |
| $\mu$ | Magnetic moment |
| $\mu_0$ | Magnetic permeability of vacuum |
| $\xi$ | Langevin's parameter |
| $\chi_0$ | Steady susceptibility |
| $\tau$ | Effective relaxation time |
| $\tau_0$ | Attempt time |
| $\tau_B$ | Brown relaxation time |
| $\tau_N$ | Néel relaxation time |
| $\Gamma$ | Ratio of the Fourier number to the Joule number |
| $\omega$ | Angular frequency of the alternating magnetic field |
| $\Omega$ | Electromagnetic-to-thermal energy efficiency parameter |
| $\rho_b, \omega_b, T_b$ | Arterial blood density, perfusion rate, and temperature |

## 1. Introduction

In medical oncology, the term *hyperthermia* refers to a therapeutic modality by which a given region of interest is subjected to a temperature (*T*) increased above 40 ℃ [1,2]. Historically, it is believed that the oldest description about the use of hyperthermia is in Edwin Smith's surgical papyrus indicating the treatment of breast cancer [3]. A more recent modality is the *magnetic hyperthermia* (MH), where the temperature increase is produced by applying an alternating magnetic field (AMF) to a magnetic material, typically iron oxide. As in many other areas—namely materials science [4], energy [5], or health [6] — the progresses made in nanotechnology have taken MH to a much higher degree of development. For example, the application of magnetic nanoparticles (MNPs) in medicine is moving towards targeting body regions otherwise difficult to reach, and chemical manipulation at the nanoscale has conferred the ability to conjugate biomolecules like antibodies for a more effective therapy or to accomplish specific targeting. In this manner, MNPs may simultaneously combine several theranostic functionalities such as drug-carriers [7], contrast agents for MRI [8] and/or magnetic heating agents [9].

Considering the extent of the treated region, hyperthermia can be classified into three types: *a-*) whole body hyperthermia (achieved by using thermal chambers or blankets), *b-*) partial hyperthermia (applied to treat locally advanced cancer by perfusion or microwaves), and *c-*) local hyperthermia (mainly for smaller volumes than organs). The temperature increase in local hyperthermia – the one most frequently evaluated – might be accomplished by distinct approaches based on the use of ultrasound, microwaves or near-infrared radiation [10,11]. Even though these modalities have been incorporated into the clinical practice to treat a relatively wide range of cancer types, nowadays MH has some fundamental advantages over these when locally dealing with solid tumors: (i) the AMF penetration depth higher than any other activation mechanism (light or acoustic waves), allowing to reach deeper tissues; (ii) administration of MNPs in a wide concentration range and sustained may stay at the tumor site for repeated therapy sessions; (iii) size-driven magnetic properties at the nanoscale determining the heating capabilities; (iv) precise control of size and morphology as well as surface modification for diverse goals including biocompatibility, providing chemical groups for attaching biomolecules, and minimizing blood proteins adsorption.

The use of MNPs as a minimally invasive agent was initially addressed by Gilchrist et al. in 1957 [12,13] giving rise to MH. This seminal work pointed out some challenges which are still under discussion by the scientific community concerning the application of MH in living beings: (i) the heat release should be the highest possible at the lowest particles dose; (ii) safety of AMF (with high voltages producing eddy currents in conducting media [14]); (iii) reliability for providing a precisely controlled intratumoral heat exposure mediated by MNPs [14].

With the proposition of MH as a potential cancer treatment, the establishment of new materials and devices has been addressed with a continuous effort. A brief analysis about MH literature illustrated in Fig. 1 shows that in the period 1973-2013, more than 3000 scientific manuscripts were published about MH, followed by an exponential growth as of the beginning of XXI[th] century. As can be seen, MH has never been so much in the spotlight as now.

Understanding MH, in a broad sense, requires the consideration of distinct perspectives. Thus, this review will address the physical, chemical, engineering, technological, and biological aspects of MH discussing fundamental features about material synthesis (basically iron oxide nanoparticles – Section 2), MH physical models (Section 3), instrumental aspects of the physical characterization of MNPs (Section 4) and the influence of biological matrices on the heating efficiency (Section 5). At last, final remarks and perspectives will raise some key aspects that deserve further clarifications.

## 2. Preparation of iron oxide-based magnetic nanoparticles, functionalization and characterization

The choice of the magnetic material to be employed in MH is virtually infinite since, in principle, every magnetic compound can be synthesized under nanoparticulate form by chemical or physical procedures. However, a set of factors has to be taken into account for accomplishing safety requirements, especially biocompatibility. In this sense, iron oxides become the most prominent candidates and are reviewed next in further details.

Iron is the fourth most common element in the Earth's crust, existing in oxidation states from -2 to +6 with common oxidation states of +2 and +3. Ochre composed of antiferromagnetic iron oxo-hydroxide MNPs (acicular haematite nano-spindles, goethite nano-laths, wüstite nanospheres) are used as natural pigment from the early ages of mankind, and thus can be seen as the forerunners of manufactured and environmental MNPs [15]. Iron oxide-based MNPs combine several physicochemical aspects leading to attractive properties. These MNPs typically have two or three dimensions under 100 nm, which brings a high surface-to-volume ratio and different properties than those from bulk iron oxide material. Human metabolism maintains the homeostasis of iron, controlling this necessary (but potentially toxic in excess) element. The human body is able to tolerate the oral administration of iron at 5 mg per kg of body mass [16], well below the limit of acute toxicity in the range of 300-600 mg per kg of body mass as determined on Wistar rats with $FeSO_4$ as the source of iron [17]. This "iron pool" of the organism consists of both molecular iron ions (hemoglobin) and in a nanoparticulate form – ferritin, which is a protein capside encapsulating an antiferromagnetic ferrihydrite core. This unique biocompatibility feature, along with its magnetic properties, make iron oxide MNPs excellent candidates for biomedical applications such as contrast agents for magnetic resonance imaging (MRI), cell labeling, magnetic separation, drug delivery assisted by DC or AC magnetic fields or magnetic heating mediators [18,19]. However, the discussion will be focused on MH in this document, and the considered materials will mostly consist in either pure iron oxides or ferrites of the general formula $MFe_2O_4$ where M stands for another transition metal, or two different metals in the case of either mixed ferrites $(M,M')Fe_2O_4$ or core-shells $MFe_2O_4@M'Fe_2O_4$.

Frenkel and Dorfman were the first to predict in 1930 that a particle of a ferromagnetic material below a critical size consists of a single magnetic domain [20]. It is accepted that a ferromagnetic particle of iron oxide with a radius under 30 nm is a single domain particle [21], meaning that under any magnetic field it will maintain a state of uniform magnetization (i.e. all the magnetic moments within the particles are pointing towards the same direction). A colloidal assembly of this type of nanoparticles suspended in a liquid is considered "ferrofluid" as long as it stays in a monophasic state (no sedimentation or aggregation). At thermal equilibrium and under no external magnetic field applied, there is no net magnetization of the ferrofluid due to thermal agitation leading to random orientation of the grains and thus of their magnetic moments when considering the whole population of MNPs. The magnetization of single domains particles in thermodynamic equilibrium is identical to that of paramagnetic atoms or ions, except that extremely large moments are involved, of several hundred to thousand Bohr magnetons [22]. Such thermal equilibrium named *superparamagnetism* follows the so-called Langevin's theory of paramagnetism when the MNPs are in a dilute state where dipolar interactions can be neglected [23]. The properties exhibited by iron oxide MNPs make them good candidates for either diagnosis or therapy as MRI contrast agents to assist diagnosis and for radiofrequency MH to remove cancerous cells by applying a thermal shock mediated by the MNPs. It is possible to engineer theranostic systems in which both of these applications are integrated in the same nanostructure for simultaneous detection and treatment of diseases [24].

### 2.1 – Synthesis of iron oxide nanocrystals

Superparamagnetic (SPM) MNPs can be obtained by various physical or chemical methods. Among others, the physical ones consist in top-down processes such as laser-induced ablation of macroscopic targets of iron or iron oxides [25] giving polycrystalline MNPs with wide size distributions, mechanical milling of bulk iron oxide [26] with subsequent mechano-chemical effect reducing the degree of crystallinity compared to the starting material (also existent in other compounds [27]). Moskowitz and Rosensweig, in collaboration with NASA, were the first in 1965 to prepare ferrofluids intended to prepare magnetically driven pumps in the Explorer-17 satellites, by a grinding procedure of iron oxide powders for several weeks in the presence of surfactants [28]. Twelve years later, the 1st International Advanced Course on the Thermomechanics of Magnetic Fluids took place in 1977 in Italy, gathering experimentalists and theorists from both sides of "the iron curtain", launching the cycle of the International Conference on Magnetic Fluids that since then was repeated worldwide on a regular basis.

Bottom-up processes consisting in the synthesis of iron oxide MNPs from iron ions or molecular precursors offer the great advantage of controlling the composition, size and shape to tune the desired properties by means of the control of the synthesis conditions. The most described synthesis routes comprise the aqueous ferrous and ferric salts alkaline co-precipitation, the thermal decomposition of organometallic complexes, the alkaline hydrolysis in a polyol solvent, and the post-synthesis hydrothermal treatment (i.e. under high pressure). These methods will be described in further details in the following part. In all synthesis methods, the so-called "LaMer model" is often evoked to interpret the size distributions of the synthesized MNPs: originally built to describe the mechanism of formation of monodisperse hydrosols [29], it was extensively used to explain the formation of any type of MNPs from (poly)atomic precursors. This model states that different processes are involved during the precipitation of MNPs: nucleation, crystal growth, and Ostwald ripening. Ideally nucleation and crystal growth steps are separated, meaning that a burst of nucleation occurs at the early synthesis stage, followed by crystal growth through diffusion of the reactants to the nuclei. Because the physical properties of the nanocrystals are strongly dependent upon their shape, size and size distributions, many publications have reported different synthetic pathways in order to produce good quality materials with narrower size distributions, leading to controlled magnetic behaviors. Some of these findings are described below.

### 2.1.1 – Co-precipitation method

The alkaline co-precipitation of ferrous and ferric salts is widely used because it is a convenient and reproducible pathway to synthesize MNPs and obtain them directly dispersed in aqueous media. It is commonly referred to as "Massart's method", as Massart reported it first in 1981 [30]. A variant was proposed quite simultaneously by Molday and Mackenzie in presence of a polysaccharide (Dextran) [31]. This aqueous route to colloidal magnetite can be scaled-up to produce even kilograms of iron oxide MNPs; thus, it is the method used by industries to produce commercial iron oxide contrast agents with well-adjusted parameters such as the mixing and addition rates of reactants to produce perfectly calibrated and reproducible batches. The precursors used are ferric ($Fe^{3+}$) and ferrous ($Fe^{2+}$) chlorides, sulfates or nitrates, first dissolved in an acidic aqueous solution to prevent the individual precipitation of hydroxides whose solubility products are very high, respectively pKs=34 for $Fe(OH)_2$ and pKs=44 for $Fe(OH)_3$, respectively [32]. Then they are "co-precipitated" (meaning the two valences of iron ions together) under the addition of a strong base (commonly $NH_4OH$ or $NaOH$), according to the reaction

$$2\ Fe^{3+} + Fe^{2+} + 8\ OH^- \rightarrow Fe_3O_4 + 4\ H_2O. \qquad (2.1)$$

Controlling the salt metathesis of iron precursors into iron hydroxides followed by sol-gel reaction is not a straightforward task because it occurs instantaneously upon mixing; thus, the conditions have to be adequately set. The $Fe^{3+}/\ Fe^{2+}$ ratio, the nature of the anions in the salts, along with the final pH of the solution (dictated by the molar ratio $R$ of $OH^-$ ions to total iron ions compared to the stoichiometric value $R = 8/3$), temperature, mixing rates, ionic strength and optional presence of ligands (citrate, tartrate, etc.) greatly affect the nature of the nanocrystals obtained, including their size and shape. Magnetite ($Fe_3O_4$) is commonly synthesized this way, but an inadequate procedure can also lead to other non-magnetic iron oxo-hydroxide phase (goethite $\alpha$-FeOOH or akaganeite $Fe_8O_8OH_8Cl_{1.35}$) or oxide (hematite $\alpha$-Fe$_2$O$_3$) phase [32]. Maghemite ($\gamma$-Fe$_2$O$_3$) can be obtained from magnetite by simple oxidation in an acidic medium with $Fe^{3+}$ nitrate salts, or by leaving magnetite nanocrystals in contact with oxygen from ambient air, accelerating the formation of the thermodynamically favored maghemite compound.

The use of stabilizers during the co-precipitation process has been reported as a way to produce good quality materials with a narrower size distribution. Efficient stabilizers include polyvinyl alcohol (PVA), tri-sodium salt citrate, tartrate, or other multivalent carboxylate ions. An important criterion toward the selection of these organic additives is their hydrophilic or lipophilic affinity, which determines the final solubility of the MNPs in organic or aqueous solvents. For biomedical applications, hydrophilic ligands are used to ease the dispersion of the resulting particles in aqueous systems. Regarding shape and size control, co-precipitated samples under electron microscope usually exhibit "rock-like" MNPs with broad size-dispersity, corresponding to diameters ranging from 5 to 15 nm. As a post-synthesis process, a size-grading procedure based on the addition of an electrolyte allows to obtain narrow size-dispersities. The addition of an electrolyte in excess screens out the electrostatic repulsions between the iron oxide

MNPs and leads to a liquid-liquid phase separation in two fractions respectively named S ("supernatant") and C ("culot"), which are enriched with the lower sizes (respectively larger sizes) fractions of the initial distribution as shown in Fig. 2. This was the method employed in one of the first articles describing the size effects on magnetic heating efficiency [33], in agreement with Rosensweig's linear relaxation model [34].

### 2.1.2 – Nano-template methods

Several template synthesis methods have been described in the literature in order to orient the particles' geometry and to narrow the size distribution, as compared to co-precipitation in batch leading to a broad range of diameters. The use of pre-existing nanostructures as nano-molds was recently reviewed for organic templates, i.e. surfactants and polymers [35]. Inorganic templates such as mesoporous silica matrixes synthesized by the sol-gel route have also been tried out, since they enable to perform combustion synthesis at temperatures as high as 400°C while avoiding the issue of aggregation and sintering of the nanocrystals [36]. Regarding micro-emulsions, these are composed of a nano-sized, thermodynamically stable dispersion of two immiscible solvents (water / oil) stabilized with a surfactant (dodecyl sulfate, Triton X-100, etc.) that can be used as medium to produce MNPs. In water-in-oil micro-emulsions, the iron precursor is solubilized in the water droplets forming confined reactors. The molar ratio of water-to-surfactant determines the size of the reverse micelles. Upon addition of a strong base, co-precipitation is initiated. The precipitating agent can be introduced directly in the emulsion of precursors or as a stabilized emulsion. In the latter case, the droplets containing the iron salts and the droplets containing the base collide and coalesce together, allowing the formation of MNPs. Since the size of the micelles is in the nanometer range, MNPs as small as 7.4 nm can be produced [37]. However, this method requires large amounts of solvents for small yields with a relatively low control of shape, crystallinity and monodispersity.

### 2.1.3 – Hydrothermal method

Single crystals can be synthesized from aqueous solutions at mild temperature and then transferred into a Teflon-lined stainless autoclave to perform a hydrothermal treatment. Combining a high temperature (usually around 200°C) and a high vapor pressure, such treatment favors the Ostwald ripening, by which the smallest crystallites are dissolved into the largest ones, increasing the average size and the crystallinity. This cost-effective and environmentally friendly route initially developed by von Schafhäutl in 1845 to grow microscopic quartz crystals has been used to produce various ferrites [38,39], magnetite [40-41] and maghemite [42] nanocrystals with good water solubility and high crystallinity. Typically ferric and ferrous salts are mixed with a controlled molar ratio in an aqueous solution and precipitated upon addition of a base, similarly as in the co-precipitation method. The MNPs are then transferred into an autoclave for further aging at temperatures ranging from 150°C to 200°C under pressure. Starting with nanocrystals of 12 nm, the hydrothermal treatment leads to particles of 39 nm with ferrimagnetic behavior at room temperature [40]. The hydrothermal method leads to MNPs with very large sizes, good crystallinity and high saturation magnetization. However, although the nucleation and growth steps are well separated (different conditions of temperature and pressure), the particles after hydrothermal treatment exhibit rounded (as opposed to faceted) yet rather ill-defined shapes as depicted in Fig. 3. Therefore, this method still needs adjustments to limit the growth at the step of quasi-spherical particles of intermediate sizes (e.g. 20 nm).

### 2.1.4 – Thermal decomposition method

Originally introduced by Heyon [43] and concomitantly by Sun [44], the thermal decomposition of various organometallic complexes (iron pentacarbonyl [40], acetylacetonate [44], oleate [45], or stearate [46]) in apolar organic solvents in the presence of ligands (oleic acid and/or oleyl amine) was reported as a synthesis path leading to the best geometrically-defined nanocrystals. These syntheses are performed at reflux of high boiling point solvents, commonly 1- octadecene ($T_{eb}$ = 318°C), octyl ether ($T_{eb}$ = 288°C), or diphenyl ether ($T_{eb}$ = 268°C). Several morphologies can be obtained from perfectly spherical [43-46] to slightly polyhedral [46] or prismatic [47] and cubic [48, 49, 50] by controlling the synthesis parameters.

As seen on Fig. 4, the nanocube morphology decreases the surface disorder and/or spin canting effect, which was interpreted as the origin of the observed decrease of the magnetic anisotropy constant as compared to a spherical particle of same volume [51], and possibly of the overall magnetic heating properties of the dispersions [48, 52]. For all these syntheses, the reactants can be introduced following a "hot injection" protocol at the high reaction temperature, leading to a rapid formation of nuclei called "burst", caused by the sudden supersaturation of the solution with precursors. Alternatively, reactants can be introduced following a "heating-up" protocol: the solvent, the precursors and the ligands are mixed at a lower temperature before being heated at a controlled rate up to the reaction temperature, leading to the formation of nanocrystals. Compared to the other routes, the thermal decomposition method has the superior advantage to dissociate the nucleation step and crystal-growth step which occur at different temperatures (ca. 200-240°C for the decomposition of the organometallic complex leading to precursors and up to 300°C for the growth, respectively). Whatever the source of iron (FeCO$_5$, Fe(acac)$_3$, FeCl$_3$, etc.), it has been hypothesized that at the temperature at which pyrolysis occurred, iron carboxylate salts of the ligand used (e.g. iron oleate) were the real precursors [53]. Iron oleate complex can be produced before reaction from the inexpensive FeCl$_3$·6H$_2$O chemical [49] but it has to be purified beforehand in order to remove chlorine anions from the medium. Fatty acids such as oleic, decanoic or lauric acid, possibly mixed with fatty amines like oleylamine or hexadecylamine, are used as surfactants chemisorbed on the surface of the MNPs: first they can orient towards a specific morphology of the MNPs by blocking the growth of certain crystallographic facets compared to others; then, at the end of reaction, they assist the MNPs' dispersion in organic solvents, and prevent aggregation by pointing outside into the solvent their non-polar chains. It should be noted however that the seed-and-growth technique consisting in adding more precursors to initially synthesized seeds enables to increase the sizes, but an imperfect epithaxial growth (i.e. internal defects existing within the crystals) can result in poor magnetic response [54].

In principle, the ratio of organometallic reagents, surfactant(s) and solvent drives the morphology and sizes obtained, but also other parameters like increasing the reaction time in absence of stirring as indicated in Fig. 5 [49]. In some cases, further oxidation improves the crystallinity of the nanograins [43], as it is also the case by applying a magnetic hyperthermia treatment that acts as an annealing process [54]. MNPs obtained this way are dispersible in polar solvents and are not water miscible, which is a limitation for biological applications. Following a ligand-exchange strategy, fatty acids can be exchanged with polar molecules, such as tetramethylammonium hydroxide (TMAOH) followed by adsorption of a synthetic polypeptide with a poly(aspartic acid) block [55], or chemical grafting of charged organosilanes [56]. Amphiphilic polymers can also be used as phase transfer agents while keeping a good size-dispersity and colloidal dispersion, such as poly-(maleic anhydride *alt*-1-octadecene) [57]. The use of multidendate ligands such as meso-2,3-dimercaptosuccinic acid has also been reported [49,58].

Lee et al. described the synthesis of ferrite@ferrite core-shell MNPs by thermal decomposition of $MnCl_2$ and $Fe(acac)_3$ in the presence of oleic acid, oleylamine and trioctylamine and $CoFe_2O_4$ seeds suspended in hexane [59]. The magnetically hard core coupled with the magnetically soft shell leads to an exchange interaction resulting in thirty-four times enhancement of the specific absorption rate (*SAR*, physical magnitude related to MNPs heat dissipation which is discussed in details in the next sections) with respect to the commercial FDA approved iron oxide Feridex™ MNPs at same field intensity and frequency. This approach was further extended by Gavrilov-Isaac et al. who recently described the synthesis of trimagnetic multishell $MnFe_2O_4@CoFe_2O_4@NiFe_2O_4$ in order to tune the coercive field and, ultimately, *SAR* [60].

### 2.1.5 – Polyol method

In this method, the metal precursors (acetates or chlorides) are added to a polyol solvent (diol, triol), usually diethylene glycol (DEG), 1,2-propylene glycol (PG) or ethylene glycol (EG), exhibiting good chelating properties and stabilizing the precursors. Iron hydroxide precursors can be produced *in situ* by addition of NaOH to the ferrous and ferric iron chlorides (one equivalent relatively to chloride anions). Heating and mixing the solution helps to solubilize the precursors before reaching temperatures

higher than 150°C, at which the polyol molecules chelated to the metal cations undergo a nucleophilic

substitution reaction by water molecules, with the formation of a hydroxide according to the following

equation (in the case of DEG) [61]:

complexes and further made hydrophilic or lipophilic with suitable phosphonic and hydroxamic acids [64]. But the polyol route can also lead to much larger MNPs, in particular to flower-like assemblies of smaller iron oxide crystallites. These so-called nanoflowers, as depicted on Fig. 6, are multi-cores but behaving as magnetic single domains of size equivalent to the outer diameter, thus presenting very high *SAR* values. They can be synthesized from iron chlorides precursors in a mixture of DEG with an organic base such as N-methyldiethanolamine (NMDEA) [65,66] or tetraethylene tetramine (TETA) [67]. Another study shows that these structures result from well oriented attachment of the cores into flower-like clusters when the reaction is carried in EG and poor oriented attachment of the individual grains when the reaction is carried in PG [68]. This is explained by the formation rate of nanocrystals: in EG the formation and growth is slower, allowing the MNPs to assemble and organize by rotation resulting in crystal alignment and oriented aggregation, while in PG the mesostructure is less coherent and stable, with possible Ostwald ripening process meaning re-dissolution of the non-aligned crystals and growth of the organized crystals, the end-product being large MNPs exhibiting pores and magnetic multi-domains.

### 2.1.6 – Combustion methods

Iron oxides can be produced in ultrafine (nanometric) powder from a gas reactant ($FeCO_5$) by laser-assisted pyrolysis [69-70], however post-treatments are necessary to collect the MNPs in a non-aggregated state [71]. An exothermic, fast and self-sustaining combustion reaction between +III manganese and a mixture of lanthanide nitrates and glycine also called Glycine Nitrate Process (GNP) can produce $La_{0.82}Sr_{0.18}MnO_{3+\delta}$ (lanthanum strontium manganite) MNPs, with a perovskite structure and a mean crystallite size of 22 nm [72]. This method leads to aggregated MNPs as the final product is composed of dry MNPs; milling steps can be performed to favor their disaggregation [73]. These ferromagnetic particles are of particular interest for MH as their Curie temperature can be well adjusted by the lanthanide composition, thereby leading to "self-limited" nano-heaters. However, the high toxicity of lanthanides and manganese cations leads to the necessary coating of these MNPs by an inert shell such as silica in order to be used in biomedical applications.

To complete this short brief overview of synthesis routes towards MNPs and, more particularly, iron oxide MNPs, some of them in this panel have been optimized to achieve larger scale. With this goal of scale-up in mind, microwave heating has been tried out instead of conventional heating in order to perform a homogeneous heating even with a large batch, as for the thermal decomposition of $Fe(acac)_3$ [74] and the alkaline co-precipitation [75]. Overall, all these syntheses present various advantages and limitations, and a compromise must be chosen between the amount of MNPs produced and the degree of control of sizes and/or shapes and, of utmost importance for MH, of their magnetic properties. One more aspect needing to be highlighted in this part deals with the coating methods of MNPs. This is not only important to disperse them in hydrophilic media for biological applications, but also on the physical point of view, to tune the average distance between the MNPs in the dispersion: the ratio of mean particle diameter to their average center-to-center distance indeed controls the amplitude of their magnetic dipolar interaction, scaling like the $3^{rd}$ power of this ratio for ferri- or ferromagnetic MNPs, and the $6^{th}$ power for SPM MNPs [76]. As will be developed in Section 3.3, these dipolar interactions that are weak in dilute individually dispersed MNPs, and on the contrary very high in densely clustered MNPs, have a non-negligible impact on the MH efficiency by slowing down the relaxation dynamics of the moments and/or increasing the shape component of the magnetic anisotropy and the hysteresis losses.

*2.2 – Surface modification of iron oxide MNPs*

Several natural and synthetic polymers have been tested to modify the surface of MNPs to improve their colloidal stability first *in vitro* (in pH buffers, then in cell culture media supplemented with serum proteins) and eventually *in vivo* (in blood circulation), while being biocompatible: polysaccharides (e.g. Dextran [31]), poly(vinylpyrrolidone) (PVP) or poly(ethylene glycol)s (PEG), poly(ethers) (e.g. poly(ethylene oxide), poly(oxazonline) (POxa), or poly(glycidol)) have been investigated for giving "stealth" properties to MNPs, i.e. repulsion towards certain proteins of blood serum (opsonins), whose role is to mark foreign bodies by strongly adsorbing onto their surface so that they are recognized by white blood cells and go through elimination pathways. Once the circulation half-life is improved, it is possible to graft ligands with a specific affinity toward target sites, or to direct them near the tumor site with an external static or AMF gradient. The classical methods to accomplish the hydrophilic coating steps will be presented thereafter, while the topic of bio-conjugations will be addressed further in Section 5.

Iron oxide-based MNPs must be stabilized in aqueous media in order to be used in MH applications, i.e. they must exhibit and keep a proper state of dispersion under given physicochemical conditions (pH, ionic strength, adsorbing proteins, etc.) and under an applied static field or AMF, whatever the field strength. One way is the coating with polyelectrolytes such as polyacrylic acid (PAA) in order to provide strong adsorption (resisting to dilution) and electro-steric repulsion between the MNPs, overcoming van der Waals and magnetic dipole-dipole attractive forces. Compared to smaller multivalent ligands like citric acid, the PAA coating offers superior stability, especially in high salinity buffers [77] and impedes the adsorption of blood plasma proteins [78]. In practical terms, PAA chains are adsorbed at acidic pH on the positively charged MNPs before modifying the pH to neutral or basic values, a process referred to as "precipitation-redispersion". Strong polyelectrolytes such as poly(4-vinylbenzenesulfonate sodium (PSS) can also be used [79]. In that case, the resulting negatively charged MNPs are stabilized by an electro-steric repulsion provided by the grafted chains whatever the pH value.

Polysaccharides such as heparin, starch, hyaluronan, dextran, carboxydextran and chitosan can also be adsorbed on MNPs to provide steric or electro-steric stabilization. Instead of adsorbing polymer chains at multiple sites but through weak bonds, other authors used molecules presenting a catechol function at one chain-end. This function derived from mussel adhesive protein can bind to the surface of iron oxide MNPs through direct chelating bond of the surface ferric irons. Dopamine is a natural molecule that presents a catechol function and, through chemical modification of its primary amine, other molecules can be chemically grafted onto the MNPs [80]. Another option for efficient grafting on the iron oxide MNPs is to combine strong iron-complexing ligands and multivalence, as with a copolymer of PEG, poly(ethyleneimine) and poly(L-dopamine) [81],a graft copolymer with PEG pendants groups and phosphonate moieties that enable to make PEG coating very resistant to blood protein adsorption with outstanding stability in cell culture media supplemented with serum [82], or PEG oligomers with dendritic architecture strongly anchored to the iron oxide surface through a phosphonate ligand thereby bringing outstanding stealth properties *in vivo* [83].

Several chemical moieties can be added through surface modification of the MNPs with surface-complexing agents. Condensation of alkoxysilanes directly on the MNPs or after silica coating with tetraethylorthosilicate (TEOS) can produce a dense outer layer on the MNPs, following the Stöber process in acidic or alkali media [84]. The alkoxysilanes undergo hydrolysis forming silanols, which can polymerize and condense on the surface of the MNPs presenting hydroxyl groups (either the original Fe-OH moieties of the raw iron oxide surface or the Si-OH brought by an intermediate silica shell). Although thick, the grown silica layer can be made mesoporous and its permeability to water can be adjusted using a sacrificial organic template, leading to functional nanoparticles for sustained drug-delivery [85] or as MRI contrast agents where the accessibility of water molecules is important [86]. The MNPs resulting from sol-gel coating are water-dispersible and exhibit different moieties such as amino, cyano, isocyanate, aldehyde or carboxylic depending on the silane ligand chosen, as a first step toward further functionalization [56]. Once these chemical moieties are introduced on the surface of MNPs, further modification can take place following "orthogonal" coupling reactions. One possibility is amide coupling chemistry via formation of N-hydroxysuccinimide (NHS) ester after carboxylic acid activation by a carbodiimide, e.g. 1-ethyl-3-(3-dimethylaminopropyl)carbodiimide (EDC). As an example, amine functions are introduced this way e.g. from 3-amino-propyltriethoxysilane (APTES) and then covalently attached to molecules of interest such as carboxy-dextran through classical amide bond formation coupling [87]. Another possibility to covalently graft dextran chains is reductive amination, using the aldehyde end-function presented by any polysaccharide [88]. A maleimide function can also be introduced in MNPs for further reaction with thiol groups of biomolecules (protein, antibody, etc.), either preexisting (such as a cysteine aminoacid) [89] or introduced by the so-called Traut reaction [90].

To conclude on this part, the various syntheses and functionalizations of iron oxide MNPs developed in the last decades have provided nanocrystals with optimal physical and chemical properties of interest for MH. Crystallinity that determines the magnetic order of the magnetic moments within the particles and hence their magnetization is a parameter to consider, along with size-distribution. To avoid the clustering of MNPs and dipolar interactions between the magnetic moments leading to demagnetization effect, molecules of interest can be grafted on the surface of the MNPs to keep them individually dispersed by providing steric, electrostatic or electro-steric repulsions. Finally, a parameter that needs to be taken into account when going for biological tests is the availability of a sufficient amount of MNPs (meaning a few grams) in order to perform all the control experiments, especially when experimenting on animals. More precisely, one can estimate that 1 mg nanoparticles are sufficient and safe (ten-fold lower than the toxic dose) for an injection to a mouse (20 g average body weight), but 25 mg is required for a rat (500 g) and 100 mg for a rabbit (2 kg). Considering that dozens of animals are needed to obtain a significant statistics of tumor decrease results, one comes to the conclusion that between several tens and hundred grams of well-defined iron oxide cores must be synthesized, in a reproducible way. In the case of a clinical assay on humans, the need is even at the kg level. This is crucial to obtain the authorization of a magnetic nanoparticle system on market, as it has been already delivered in the past for iron oxide based contrast agents (e.g. Resovist™, Endorem™).

## 3. Progress in magnetic hyperthermia physical models

Physical models do not only constitute a key element in revealing the most relevant parameters for developing better MNPs for MH, but also help to draw general behaviors from the acquired *in vitro* and *in vivo* experimental data. Once validated, these models may result in more reliable particle post-production benchmarking or treatment planning and monitoring.

From a physical point of view, the phenomenology related to MH consists of the electromagnetic energy conversion into heat when MNPs are subjected to AMF, the subsequent heat transmission to the surrounding medium and the heat flow through it. The last decade has witnessed a significant increase in theoretical research related to magnetic hyperthermia. Of particular note is the effort made in adapting the existing models or even proposing new ones for heat distribution in tumors. This situation has been possible thanks to the research made in understanding the physics related to magnetic heating mediated by MNPs from both theoretical and experimental points of view, as well as the instrumental development.

A number of review articles and book chapters have already dealt with the many facets of the physical aspects of MH from its early days with special emphasis, as appropriate in each case, on the magnetic properties of the MNPs used in this technique, the associated heating mechanisms or the existing modalities [91,92,93,94,95,96]. In this section we will concentrate just on those aspects standing in the way of more reliable MH models for treatment planning, mainly those published within the last five-year period.

### 3.1 – Modeling heat dissipation in MNPs

Before going through any biological testing, the heating power of MNPs intended to be used in MH needs to be quantified as accurately as possible. This is usually accomplished by measuring *SAR*, or their equivalents/complementary parameters—*specific loss power* (*SLP*) and *specific heating power* (*SHP*) – under controlled conditions. In view of the evident uncertainty around the nomenclature, it has also been recommended the use of the *intrinsic loss power* (*ILP*) when reporting MH measurements, as its value is virtually independent from the AMF frequency and intensity used during the experiments [97,98] (the reader is referred to consult Section 4 for a more complete description about measurement setups). Power losses in MNPs under an AMF are roughly given by the area enclosed by the corresponding hysteresis loop. Taking into account the nanoparticle size range typically considered in MH—tens of nanometers, even though in distinct cases the size has reached a few hundred—two models are usually employed to describe the hysteresis loop of a system of MNPs from a theoretical perspective: the so-called *linear response model* (LRM) and the *Stoner-Wohlfarth model* (SWM). The LRM is better suited for single-domain MNPs in the SPM regime, whereas those models derived from the SWM are valid for single-domain MNPs with blocked magnetic moments (either ferro- or ferrimagnetic). A more detailed treatment of both can be found elsewhere [92, 99]. Even though LRM and SWM may explain the behavior observed at low fields, they do not cover the whole range and further approaches are needed for relatively high fields [100].

As already noted by Carrey et al. [99] there has been some confusion regarding the nomenclature and the origin of the power losses in MNPs. Following is a brief description of key concepts that may help to clarify this matter. The total energy $E$ of a magnetic particle is given by the sum of the Zeeman and the anisotropy energy terms, which in a simplified form, reads [101]

$$E = KV\sin^2\theta - \mu H_0 \sin(2\pi f \cdot t)\cos\varphi \qquad (3.1)$$

where $K$ is the anisotropy constant (with several contributions: magneto-crystalline energy, surface disorder or shape anisotropy), $V$ is the particle volume, $\mu$ is the magnetic moment, $\theta$ is the angle between $\mu$ and the anisotropy axis, $\varphi$ the angle between $\mu$ and the external field, $t$ is the time, and $H_0$ is the intensity of the AMF . At zero field, SPM MNPs undergo a *magnetic relaxation* process, also known as *Néel relaxation* (with a characteristic time given by $\tau_N = \tau_o \exp(KV/k_BT)$, where $\tau_0$ is the attempt time $\sim 10^{-9} - 10^{-13}$ seconds and $k_B$ is the Boltzmann constant), due to thermally-driven continuous fluctuations of their magnetic moments. Thermal energy ($k_BT$) can surpass the anisotropy energy barrier that separates orientation states (local minima) of the magnetic moments. As opposed to SPM MNPs, both multi-domain and single-domain MNPs that have their moment(s) blocked do not show Néel relaxation, as their anisotropy energy cannot be overcome by the thermal energy. Nevertheless, both blocked and SPM MNPs present a *Brownian relaxation*, whose characteristic time $\tau_B$ is strictly associated to their random hydrodynamic interaction with molecules of the liquid medium and their corresponding rotational diffusion time [92]

$$\tau_B = \frac{3\eta V_H}{k_B T} \qquad (3.2)$$

where $\eta$ is the dynamic viscosity of the medium and $V_H$ is the hydrodynamic diameter of the suspended MNPs. It can be seen from Eq. (3.2) that $\tau_B$ is not directly related to the magnetism of the SPM MNPs, but it can increasingly modulate it as the particle diameter increases within the SPM range. A number of models exist for Brownian relaxation in MNPs, their corresponding comparison with experimental data and measuring protocols having been published during the past years [102,103,104]. Both relaxation processes are related by Shliomis' equation to give the effective relaxation time $\tau$ [105]

$$\tau = \frac{\tau_B \tau_N}{\tau_B + \tau_N} \tag{3.3}$$

Eq. (3.3) may lead to think that Brown and Néel relaxations are decoupled, but they are not [106]. At non-zero fields, for blocked and SPM MNPs the balance between both terms in Eq. (3.1) will dictate the magnetization reversal of the MNPs and hence the shape of the corresponding hysteresis loop. At a constant temperature, the anisotropy barrier will remain the same and any variations in the applied field will introduce an imbalance in the total energy (Eq. (3.1)). If this imbalance is large enough, it will overcome the anisotropy barrier/energy of the system and the magnetization will reverse. Upon cycling this process for different field values with opposite signs, one obtains the hysteresis loop of the system. This means that the Néel relaxation, *per se*, is not a power loss mechanism. In other words, relaxation drives the magnetization dynamics towards equilibrium and, at the same time, somehow modulates the processes that shape the hysteresis loop of the system in the presence of an external field, but it is not the only driving force behind the power losses.

New approaches to work out the possible relevance of Brown and Néel relaxations for a particular nanoparticle system are still been devised. One of the latest approaches is the construction of *SAR* equipotential diagrams that would help in finding out the reigning relaxation mechanism and the *SAR* values for a given field frequency and amplitude [107]. The main idea is to plot the magnetic energy term against a viscous energy term (both proportional to the respective relaxation mechanism) and then plot also a set of distinct curves, some acting as regime limits. The result is a diagram depicting the prevalence domain of Néel and Brown processes, as well as the coexistence region as shown on Fig. 7. From here, the equipotential lines are calculated using LRM and therefore inheriting the limitations of the latter. A very recent experimental *in vitro* study aimed at demonstrating the prevalence of Brown/Néel processes have been carried out by Soukup et al. in MG-63 osteoblasts loaded with iron oxide MNPs [108]. Using AC magnetometry to detect any variations of relaxational processes, blocked MNPs showed no Brownian relaxation upon cell internalization due to either aggregation or immobilization. Nonetheless, when MNPs were released to the medium again by cell lysis, Brownian relaxation reappeared. In the case of SPM MNPs, susceptibility curves remained invariable despite the changes in their environment, reflecting the invariability of Néel relaxation in SPM MNPs with respect to their environment inside cells. These results are along the same research line that those previously reported for *in vitro* tests of MNPs inside human adenocarcinoma cells [109], where a systematic decay of the heating capabilities of MNPs were observed upon interaction with the cellular membrane and/or cell internalization.

The different methods for calculating AC hysteresis loops and their areas have been thoroughly described [99,110] and analytical expressions have been proposed. Mamiya reviewed the theoretical criteria to choose the proper field intensity and frequency that maximize the power losses and hence increase the heat release by MNPs [110]. In blocked MNPs, $H_0$ has to be adjusted above the anisotropy field ($H_K$) value of the system, given by $H_K = 2K / (\mu_0 M_S)$ and $f$, the field frequency, has to be maximized because the heating power is given by the product of the hysteresis loop surface area by $\pi \cdot f$. For MH with SPM MNPs, $H_0$ has to be maximized, while $f$ has to be set to $\tau^{-1}$.

A more recent model has delved deeper into the non-linear behavior in the dynamical susceptibility in magnetic colloids by considering the field dependence of relaxation times. Raikher and Stepanov obtained an exact expression for the dynamic susceptibility of MNPs forming a colloid under the action of an AMF field considering the LRM and the low-frequency approximation, the latter implying that $f <<$ $\tau_D^{-1}$, where $\tau_D$ is the characteristic time of the rotational diffusion of the particle magnetic moments [111].

The obtained exact expression for the dynamic susceptibility (Eq. 3.4), which takes into account both Brown and Néel dissipation mechanisms, constitutes a relatively better approach towards optimization of the nanoparticle-mediated heating than many of the other previously published models

$$\chi = \chi_0 \left[ \frac{B}{1+i\omega\tau} + (1-B) \right], \text{ where } B \text{ is a weighting coefficient} \qquad (3.4)$$

Considering a log-normal particle size distribution $g(d)$, Eq. (3.4) reads

$$\overline{\chi} = \overline{\chi_0} \left[ 1 - \frac{i\omega}{d^6} \int \frac{\tau B d^6}{1+i\omega\tau} g(d) \mathrm{d}d \right], \text{ with } \overline{\chi_0} = \frac{\pi\varphi M_S^2}{18 k_B T} \overline{\frac{d^6}{d^3}} \qquad (3.5)$$

where $M_S$ is the saturation magnetization, $d$ is the diameter distributed along a probability function $g(d)$, $\omega$ is the angular frequency of AMF ($\omega=2\pi f$), $\chi_0$ is the steady susceptibility and $\varphi$ is the volume fraction of the particles in the colloid. It is shown that the heat dissipation caused by the particle rotation in the solvent distinctly contributes to the *SAR* in the considered sample

$$SAR = \left( \frac{\omega H_0^2}{2\rho} \right) \overline{\chi} \qquad (3.6)$$

*SAR* values obtained with (Eq. 3.4), the exact expression, are compared to those calculated following other two approaches, designated by the authors as the *plain heuristic model* (PHM, Eq. (3.7)) and the *modified heuristic model* (MHM, Eq. 3.8)

$$\chi = \chi_0 \left[ 1 + i\omega\tau \right]^{-1} \qquad (3.7)$$

$$\chi = \left[ \chi_0 \cdot 3L(\xi)/\xi \right] \left[ 1 + i\omega t \right]^{-1}, \text{ with } \xi = \mu_0 \mu H / k_B T \text{ defined as the Langevin's parameter} \qquad (3.8)$$

where L($\xi$) is the Langevin equation and $\mu$=$MV$ is the magnetic moment [112]. They discuss the influence of the type of anisotropy (magnetocrystalline, $K$ and surface, $K_s$) and polydispersity ($\sigma$) on *SAR* values as depicted in Fig. 8. There is a relatively noticeable discrepancy between exact and PHM approaches, and an even larger departure of these two from the MHM approach towards higher particle sizes, especially in the cases with bulk anisotropy, i.e. discarding surface effects, as depicted in Figs. 8a and c. Unlike the case of MNPs with just bulk anisotropy, *SAR* maxima are better defined when considering surface anisotropy (see Figs. 8b and d). In all cases, $\sigma$ tends to broaden the *SAR* maxima, as expected. The model comparison seems to capture a distinct feature frequently overlooked by many of the heat dissipation models published so far: Néel and Brown relaxations become comparable for a given average particle size, and from that value onwards, *SAR* becomes almost independent from particle size. Several years before, Carrey et al. discussed in further detail the field dependency of *SAR* for different particle sizes based on the comparison of physical models with experimental data [92]. After carrying out numerical simulations of hysteresis loops for MNPs with either random orientation of anisotropy axes or parallel to the external field, they fitted the corresponding areas to a power law of *SAR* versus $H_0$ with different exponents depending on the range of particle sizes, as shown on Fig. 9a. The results reveal a radically different trend wherever a substantial change in the relevant magnetic behavior arises from increasing particle size. Starting from a square power for the lowest size, the exponent decreases within the 3.5 - 9 nm particle size range, where MNPs are superparamagnetic and the LRM is valid. Then at larger sizes, the power law exponent values abruptly increase for a threshold diameter, above which MNPs enter the ferromagnetic state and exit the LRM validity range.

In addition to particle size, particle concentration also affects *SAR* (see also Section 3.3), even though the positive or negative character of the effect depends on the intensity of the applied field; less concentrated samples show lower *SAR* values at the lower field end, whereas the trend reverses for increasing fields [113], showing a linear relationship for the most concentrated samples. The most interesting finding is that regardless the concentration, all the *SAR* curves tend to saturate to a given value as depicted in Fig. 9b.

Transforming electromagnetic energy (applied AC field) into thermal energy (heat dissipated by the MNPs) lies at the heart of MH, but not so many models consider the efficiency of a system in doing so. Within the framework of a model based on the Landau-Lifshitz equation, Landi and Bakuzis introduced an electromagnetic-to-thermal energy efficiency parameter $\Omega$ of the form [114]

$$\Omega = \frac{A}{H_0^2} \qquad (3.9)$$

where $A$ is the area enclosed by the hysteresis loop. In the LRM, $\Omega$ is given by the equation

$$\Omega_0 = \pi \chi_0 \frac{\omega\tau}{1 + (\omega\tau)^2} \qquad (3.10)$$

where $\chi_0$ is the already defined static susceptibility. For a particular nanoparticle system under given field conditions, any deviation from linearity of a $\Omega$ versus $H_0$ plot could be taken as an indication of departure from the LRM, since $\Omega_0$ is independent of $H_0$. In addition, $\Omega$ provides some glimpses on the anisotropy energy of the MNPs. This efficiency parameter could be then taken as a quick check of the suitability of the LRM to calculate the heating capabilities of MNPs in view of their intrinsic magnetic properties and the field conditions chosen, or to "tune" the latter to obtain the maximum heat out of the system.

Knowing how the heat transport takes place in real magnetic colloids is needed to validate all the models proposed so far by different research groups. In this sense, some progress has been made by designing different ways of measuring temperature differences occurring at the nanoscale. Riedinger et al. compared calculated and experimental temperature gradients at the surface of iron oxide MNPs by tracking the thermal decomposition of a thermo-cleavable molecule attached to their surface (Fig. 10a) [115]. There is a fast temperature decay from the nanoparticle surface outwards (Fig. 10b), and the discrepancies between experimental (Fig. 10c) and calculated values (Fig. 10d) are explained in terms of the unsuitability of Fourier's law to work out temperature gradients in the ballistic regime, which is the dominant one at the nanoscale. Using "nanothermometers", Dong and Zink measured the inner temperature of silica NPs containing smaller iron oxide MNPs by taking advantage of the temperature dependence of the up-conversion emission of rare earth-doped MNPs also hosted by the silica particles [116]. It remains a challenge to demonstrate theoretical findings such as the large power dissipation (exceeding the macroscopic *SAR*) and the temperature gap between the core and the surface of MNPs.

### 3.2 – Modeling in vivo heat transfer

The difference between applied field (cause) and heat generated (effect) experimentally seen in *in vitro* and *in vivo* experiments is overwhelming, let alone the heat distribution. Bringing together theory and experiment in these cases requires new efforts in terms of heat measurement techniques and simulation approaches. Once the match between the latter two is achieved, then there will be real chances of a routine use of MH as a therapy adapted to the particular requirements of each cancer case. One of the main shortcomings of many *in vitro* models is the omission of both tumor and body physiological conditions. In this subsection we will give an overview on how researchers have dealt with this situation.

For the sake of simplicity, biological media are usually considered as a blood-flooded matrix composed of cells and interstitial space in physical models of hyperthermia. Tumors with a well-defined geometry, typically spherical (with radius $r$, Fig. 11) or cylindrical, are then added to this simple picture so that they remain surrounded by a finite (with radius $R$, Fig. 11) or infinite layer of healthy tissue (with radius $R = \infty$, Fig. 11). Specific heat, thermal and electrical conductivities, mass density as well as the dielectric constant of the involved biological media—tumors, viscera, muscle, fat, skin, etc…—must be also included in models and simulations. In the case of deep-seated tumors, the perturbing effects of bones (with low dielectric constant and thermal conductivity) are usually neglected due to the added complexity and its relatively limited influence on the heat transfer. Another important element is the vascularization. The heat transfer process greatly depends on blood perfusion, which in turn is different for tumors and normal tissue. Moreover, bifurcations in vessels have an impact on the cooling effect of blood [117]. Finally, the heat sources, i.e. MNPs, must be included to complete the basic tumor model of MH. These may exhibit size distribution, which influence their magnetic properties and hence the heat generation process, and spatial distribution, which determines the formation of "hot spots" and the uniformity of heat deposition in the tissue where the MNPs are infused [91, 118].

Leaving aside for now the role of MNPs as heat sources in our basic model above, the heat exchange processes involved in any hyperthermia treatment can be initially modeled using the *bioheat transfer equation*—sometimes referred as *parabolic bioheat equation* or simply *Pennes' equation*—described by Pennes in his seminal paper on the tissue and blood temperature of the human forearm [119]. The general modern form of this equation reads

$$\delta_{ts}\rho C \frac{\partial T}{\partial t} + \nabla \cdot \left( -k \nabla T \right) = \rho_b C_b \omega_b \left( T_b - T \right) + Q_{met} + Q_{ext} \tag{3.11}$$

where $\delta_{ts}$ is a time-scaling coefficient (typically equals to 1), $\rho$ is the tissue mass density, $C$ is the tissue's specific heat and $k$ is the thermal conductivity, $\rho_b$ is the blood's density, $C_b$ is the blood's specific heat, $\omega_b$ is the perfusion rate, $T_b$ the arterial blood temperature, $Q_{met}$ and $Q_{ext}$ are the heat sources from metabolism and spatial heating, respectively. The terms on the left of Eq. (3.11) represents the thermal energy storage and the thermal energy diffusion, respectively. At the right, there are the terms referred to blood perfusion, metabolic heat and external heat, respectively. The original Pennes' equation makes some assumptions that limit its applicability to biological tissues. For example, thermal equilibrium is only attained at capillaries, neglecting any heat transfer between skin and larger blood vessels. Eq. (3.11) assumes an infinite heat propagation rate, because it is based on the Fourier's law of thermal conduction. Additionally, it presupposes a homogeneous and isotropic sample volume (and blood flow). These conditions are rarely met in real tissues, and heterogeneities make them exhibit a non-Fourier behavior that results in a thermal response lag upon a temperature change. In other words, there is a heat relaxation time of the tissue, which may reach values of ~100 seconds for certain biological materials [120], due to the difference between the occurrence of temperature gradient (cause) and heat propagation processes (effect). This is taken into account in the *dual-phase-lag* equation of the namesake model, or *hyperbolic bioheat equation*

$$\rho C \left( \tau \frac{\partial^2 T}{\partial t^2} + \frac{\partial T}{\partial t} \right) + \nabla \cdot \left( -k \nabla T \right) = \rho_b c_b \left( T_b - T \right) + Q_{ext} + Q_{met} + \tau \left( \omega_b C_b \frac{\partial T}{\partial t} + \frac{\partial Q_{met}}{\partial t} + \frac{\partial Q_{ext}}{\partial t} \right) \quad (3.12)$$

Note that when $\tau \to 0$, Eq. (3.12) reduces to Eq. (3.11) but for the time-scaling coefficient $\delta_s$. Despite its limitations, equation (3.11) has been validated in the ensuing years for different types of living tissue using experimental data [121], and has also been through subsequent corrections for the isotropic blood perfusion term [122], and small-scale microvascular contributions to the overall temperature [123], among others. Some comprehensive compendia of classical and modern heat transfer models in vascularized tissues have been reviewed by Charny, Arkin and Bhowmik [120,124,125].

Considering again the role of MNPs as heat sources, a great deal of the specific models for MH derives from the aforementioned bioheat models or improved versions. Frequently, the corresponding equations cannot be analytically solved for the imposed boundary conditions, and a range of numerical methods already employed in fluid dynamics may be used instead (Monte Carlo, finite element method [126], finite difference time domain method [127]). In some cases, simulated data are then compared with tissue phantom models made of materials that mimic the actual physical properties of biological tissues (thermal conductivity, heat capacity, etc.) as close as possible [128]. Further, it has been stated that the most sought-after properties in a phantom for evaluating MH are [129]: (i) homogeneous structure, (ii) long-term stability, (iii) thermal stability up to 100 °C, and (iv) immobilization of magnetic particles in the material. A word of caution regarding the last point: MNPs in tissues are not immobilized to the same extent as in a polymer phantom model. Nanoparticle distribution inside tissues tends to be inhomogeneous due to the presence of many different biological media (cells, interstitial space, blood vessels, etc.); these tissue changes are hardly reproducible inside a phantom even in the case of tissue equivalent materials [130]. A phantom model made of magnetite MNPs dispersed in a polyurethane matrix is shown in Figs. 12a and 12b. Temperature is measured with thermocouples inserted through the phantom (Fig. 12c).

Yamada et al. compared simulations and experimental data to find out suitable heat doses to treat cell pellets—as tumor models—from three different pancreatic cancer cell lines, namely SUIT-2, BxPC-3 and AsPC-1 [131]. Simulations are performed by solving the bioheat equation using a finite element method. Since the hepatic blood flow can be partially interrupted by the so-called Pringle's maneuver, heat losses should be minimized and it follows that only the static thermal diffusion is regarded in this model for pancreatic tumors. As expected, simulations show that the induced temperature difference between tumor center and periphery is larger for larger tumors (~18%), three times that found for smaller tumors (~6%).

The *Lattice Boltzmann method* (LBM), which is a computational method to simulate Newtonian fluids, has been lately applied to solve the bioheat equation for the first time [132] and subsequently study heat transfer in MH [133-134]. The representation of the system is very similar to that in Fig. 11. In essence, LBM computes the temperature and heat flux through the internal energy evolution considering the probabilities of finding a particle at a certain position and at a given time along the directions of a predefined lattice. Using the LBM, Lahonian and Golneshan [133] have matched the results obtained by Lin and Liu [135] for 9 nm FePt MNPs and 19 nm magnetite MNPs employing a hybrid numerical method. They also show how the type of nanoparticle volume distribution (see Table 1) inside tumors affects their temperature profile. Whereas the temperature of the surrounding healthy tissue remains unchanged in all the cases, a nanoparticle concentration profile decreasing linearly with radius from the center allows the highest temperature at the tumor center under a field of $H_0$=10 kAm$^{-1}$ (~ 125 Oe) at 100 kHz as compared to other spatial distributions of the MNPs in the tumor. Under the same conditions, a homogeneous distribution reaches a lower maximal temperature at the tumor center. Despite the advantages of this approach to study temperature distributions in tissues considering nanoparticle volume distributions, tissue discontinuities are not considered and, like for many other models, the ability to accommodate them in future improvements will determine its usability in preclinical and/or clinical MH.

Liangruksa et al. proposed a model of a spherical tumor similar to the generic one depicted in Fig. 11, with radius $r$ inside a portion of isotropic tissue ($R = \infty$) and taking into account the blood perfusion rate [136]. Three parameters are deemed crucial for optimal MH conditions in this model based on the solutions of the bioheat equation for both steady and transient states: the Péclet number ($Pe$), which is the rate between convective and conductive transport in fluids, the Fourier number ($Fo$, here taken as the ratio between "thermal" time (thermal conduction rate) to magnetic relaxation time, and the Joule number ($Jo$), which represents the ratio between Joule heating to the magnetic energy. In practical terms, those two parameters containing the magnetic contribution from the MNPs are combined into a new single parameter, namely "capital gamma" ($\Gamma = Fo/Jo$), in such a way that most of the estimations are actually made on the basis of $Pe$ and $\Gamma$. Inside the tumor, $\Gamma$ is given by

$$\Gamma = \frac{\pi \mu_0 f R^2 \omega \tau \chi_0 H_0^2}{k_t T_b [1 + (\omega \tau)^2]} \qquad (3.13)$$

where $\mu_0$ is the magnetic permeability of vacuum, $\omega$ the angular frequency of the AMF, $\chi_0$ the static magnetic susceptibility of a nanoparticle-containing tumor, $k_t$ the thermal conductivity of the tumor and $T_b$ the basal temperature. Predictions from the model correlate well with the experimental tumor temperature values measured for different nanoparticle concentrations by Moroz et al. (see Fig. 13a) in a previous *in vivo* embolization hyperthermia study [137]. Perhaps one of the most interesting points of this model compared to others is that the authors propose two quality parameters for MH treatments, namely the tumor volume with a temperature above a pre-defined threshold temperature ($I_T$), and the damaged normal tissue volume above the pre-defined necrosis temperature ($I_N$). According to Fig. 13b, a tumor is treated when $\Gamma = 1$ and $Pe = 1$, since $\Gamma < 1$ implies longer treatment times and $\Gamma > 1$ is associated to $I_N \neq 0$ values, i.e. damaging the healthy tissue. Since the aim of the therapy is to achieve the therapy temperature as quick as possible ($I_T \rightarrow 1$) while inducing the least possible damage ($I_N \rightarrow 0$), $I_T$ should be around 1 for at least 30 minutes. Regarding the influence of the particle size distribution on the therapy efficacy at steady state, the model predicts that a homogeneous distribution would reach a higher maximum temperature at the tumor center than a Gaussian one, whereas an exponential distribution would induce more damage to the surrounding healthy tissue. Even though this model suffers from the aforementioned drawbacks of assuming an infinite isotropic tissue surrounding the tumor and do not contemplate a multi-layer tumor-tissue interface (fat, muscle, etc.), it succeeds in proposing interesting design parameters for planning MH treatments different than those encountered in the majority of the models, even allowing an estimation of the exposure time to attain a full tumor treatment for a given set of conditions.

Most of the models described in this section and many others found in the literature usually consider well defined boundaries between tumor and healthy tissue—very likely to simplify the computational complexity and to reduce the computation time. Since the real tumor/healthy tissue interface is diffuse, with variable extension, and often causes tumor relapse if not properly removed, improving the existing biophysical methods to model that intricate interface is a priority for advancing the applicability of MH in cancer therapy.

### 3.3 – Current trends: modeling interparticle interactions

Until very recently, the influence of dipole interactions between particles has been almost systematically ignored. This situation may have been motivated by the lack of experimental evidences supporting their prominent role in earlier studies or their difficult implementation depending on the physical model of choice. During the last few years significant progresses have been made both theoretically and experimentally [135,138,139,140,141], but a clear consensus on this issue is yet to be reached.

The reported efficacy on some *in vivo* MH experiments have been ascribed to the interplay of several factors that leads to a collective behavior of MNPs [142], where the contribution from dipole interactions has been considered especially relevant. Although other dedicated works share a similar vision on the need of having a collective behavior to achieve a therapeutic effect, many others find dissimilar evidences on the precise role of interactions. The primary effect of interactions is the decrease of the external AMF strength "seen" by the particles, also called "demagnetization effect", giving place to an effective field with lower intensity. In addition, strong interactions occurring in a concentrated nanoparticles colloidal suspension may result in a collective behavior closer to the system as a whole rather than the sum of the properties of the individual MNPs. Gudoshnikov et al. [143] found that strongly interacting systems can be represented by a single demagnetizing factor of the whole sample. More specifically, *SAR* decreases more than four times when the sample aspect ratio goes from 11.4 to 1. The aggregation degree and the local spatial arrangement of MNPs have been also a focus of debate. Nanoparticle concentration in magnetic colloids is known to profoundly affect *SAR* values [144], often in a detrimental way; however, the mean inter-particle distance may not only decrease from just an increase in the average number-concentration of MNPs in the colloidal suspension, but also from an increase of the local volume fraction $\varphi$, induced by a partial aggregation degree. The so-called *multicore* structures—nanoparticle clusters of variable size—have been presented as an alternative to single MNPs in biomedical applications due to stability reasons, as the latter are more likely to show unspecific aggregation and clumping [145]. This controlled clustering also keeps remanence at a minimum without increasing the average particle size too much. Multicores show a noticeable magnetoviscous effect [146], which influences the way magnetic colloids flow through and interact with biological media. However, are multicores necessarily better heaters than single MNPs? Monte Carlo simulations by Serantes et al. have shown the heat dissipation enhancement in magnetic colloids for MH introduced by a controlled assembly of MNPs through dipole interactions [147]. Different geometries were explored and the formation of linear chains of individual MNPs were found to be the most efficient heaters due to the gain in anisotropy energy, which supports the high *SAR* values presented by biogenic magnetite MNPs in magnetostatic bacteria [148]. Reported Monte Carlo simulations support experimental data linking chain formation and chain length with a generalized *SAR* decrease [131]. Going beyond this general trend, Mehdaoui et al. proposed in a more categorical fashion that the only way one can get the highest possible

*SAR* from a particular material is to have chains of MNPs with uniaxial anisotropy [149].

The type of magnetic interactions (i.e. dipolar and exchange) between MNPs is also an important aspect in determining the heat dissipation in multicore structures. Recently, *SAR* measurements of citric acid coated iron oxide MNPs have shown better heating capabilities in smaller multicore structures composed of bigger MNPs than in larger multicore structures with smaller MNPs [75]. Data from Henkel plots reveal that the latter multicores present a higher proportion of demagnetizing interactions between constituent MNPs ——also regarded as *cores* when referring to nanoparticle aggregates—than the former ones. Furthermore, "interactions engineering" has been suggested as a means to obtain better colloids for MH in light of experimental and theoretical results [75, 138].

Given the intricate fate of MNPs inside the body and their inherent short circulation time—unless avoided through a proper coating—due to capture by the reticuloendothelial system (RES) or macrophage endocytosis, MH has been mainly performed locally by intratumoral injection. Experimental studies with iron oxide MNPs reveal that approximately 89% of the injected MNPs get immobilized in the tumor tissue forming homogeneous spots, remaining as such even after a MH session using a field intensity of 25 kAm$^{-1}$ (~ 314 Oe) and a frequency of 400 kHz [150]. Moreover, MNPs are not only retained in the tumor interstitia, but also tightly packed in tumor cells and macrophages associated to the tumor [151] inside endosomes as depicted in Fig. 14. These findings imply that (i) Brownian relaxation is of little relevance in the therapeutic practice, and (ii) the role of interparticle magnetic interactions in heat generation cannot be neglected. Consequently, assuming that MNPs are magnetically isolated from each other upon intratumoral injection [110] may lead to deviations in simulated models.

Kinetic Monte Carlo methods are becoming more popular for computational models of magnetic heating considering interparticle interactions, as they incorporate the time dimension of the considered system. Special emphasis is being put in developing models suitable for that size range comprised between the validity limits of LRM and SWM; an example is the unified model of magnetic heating proposed by Ruta et al. [152] where optimum heating is found to be located in the non-linear region between SPM and fully hysteretic regimes. In an effort to model the real *in vitro* scenario depicted in Fig. 15, Tan et al. have used an improved version of a previously published kinetic Monte Carlo model, now taking into account inter-particle interactions [153]. The study is mainly focused on how *SAR* is influenced by the concentration of the nearest neighbors around each particle inside a lysosome. An interesting point of this work usually absent in other models is the influence of the spatial distribution of the MNPs inside lysosomes, where the particles are known to form largely distorted agglomerates of variable size. Compared to the ideal situation of an ordered 3D NP array, significant differences are observed in the heat dissipation of the agglomerate by positioning MNPs in a cubic lattice and then introducing a disorder parameter. Simulations illustrate the existing difference in heating power between those MNPs situated at the surface of lysosomes and those in the inner core (see Fig. 15). Such behavior is attributed to the reduced number of nearest neighbors of the particles at the ensemble's surface and therefore to frustrated interactions. The authors use these results to support the role of the induced damage by MNPs in the lysosomal membrane, a cell death mechanism proposed to cause the "cold" MH, i.e. with no global temperature rise, reported by several authors [154,155].

### 4. Physical characterization of MNPs and its technical approach

As important as the preparation of MNPs with physical properties suitable for MH and the development and understanding of better models to describe the heat dissipation of magnetic colloids, the proper physical characterization of MNPs and its respective instrumentation play an important role. The following subsection addresses aspects of utmost importance for MH experiments, as more conventional methods in magnetometry of molecular and inorganic magnets can be read in textbooks.

The accuracy of experimental *SAR* values to better understand the heating mechanisms and thus optimize their performances is fundamental in MH. Electromagnetic applicators able to generate AMF well-characterized in a certain volume are mandatory in order to perform experimental research in the field of magnetic hyperthermia. The *usable volume* or the *usable gap* of such an electromagnetic applicator determines the dimensions of the spatial region where the sample would be placed. By requirement, the intensity of the applied AMF must be homogeneous inside the usable volume.

The common frequency range employed in MH experiments extends from 50 kHz to 1 MHz [18, 19]. Due to the spatial dimensions of the problem and the very much larger wavelengths ($\lambda > 300$ m) associated to these AMF frequencies, a suitable way to produce AMF with large intensity values is to circulate an AC current of the same frequency across a conductor. To concentrate the magnetic flux **B** and to increase the AMF intensity, the conductor can be wound forming a coil or inductor. Two different strategies can be followed to generate an intense and homogeneous AMF in the usable volume. The first one involves air-coils with the usable volume and the sample located in the center of the coil, where the field is more homogeneous and intense [33,156,157,158,162] (see Fig. 16a). The second one involves using soft ferromagnetic cores to concentrate the magnetic flux **B** inside the sample and to produce a more homogeneous field [18,159] (see Fig. 16b). A miniaturized version of this ferrite core design with a gap as small as 370 µm was recently described by Connord et al. [160], allowing to observe the MH effects on cells by using confocal microscopy. This opens new possibilities to follow the metabolic effect on cells in real time (like the permeabilization of the lysosomes). When large usable gaps or usable volumes are required (i.e. whole human body applicators), electromagnetic applicators based on soft ferromagnetic cores are preferred [159]. However, using these cores implies additional power consumption, especially at high frequencies.

Due to the strong AC currents and AMF involved, special care has to be taken to the conductors otherwise additional parasitic heating of the samples or malfunctioning of the electromagnetic applicator may happen. In principle, using water cooled wires ensures a safe operation with a stable temperature [161]. For short operation times ($< 1$ s), water cooling is not necessary. Other authors reported the usage of so-called Litz-wires [162,163] to avoid the requirement of water cooling and enhance the efficiency of the apparatus. Litz-wires consist of isolated wire strands twisted specifically in order to minimize heat losses at AC frequencies up to 1 MHz. The rationale of this approach is that, according to Maxwell equations, the current density is confined in a thin layer at the surface of the conducting medium (skin effect), therefore having multiple strands rather than a single wire offer a higher cross sectional area and thus a lower current density, minimizing heating by Joule effect.

An AC electrical current ($I_{AC}$) has to flow across the coil in order to generate the AMF. When $I_{AC}$ flows across the coil, it dissipates a power $W$ expressed by

$$W = \frac{1}{2} R I_{AC}^2 \tag{4.1}$$

where $R$ is the equivalent series resistance of the inductor. In any case, the power $W$ must be provided by a power amplifier. Given the inductive character of the coil, LC resonant circuits connected to power amplifiers are a plausible approach. Basically, two different circuit topologies can be used: series and parallel resonant circuits (as shown in Fig. 17). These circuits must be fed by radiofrequency power amplifiers. The output characteristics of the amplifiers depend on the resonant circuit used: the output impedance must match the circuit impedance in order to transmit the maximum power to the circuit [161]. The power matching can also be achieved using transformers [162], although they consume an additional power and hence, may reduce the efficiency of the electromagnetic applicator.

In recent years, an increasing number of works have been published about *SAR* measurements of different MNPs, as well as magnetic hyperthermia experiments performed *in vitro* and *in vivo*. However, few of them give detail about electromagnetic applicators used to generate the required magnetic field. Chieng-Chi Tai et al. [164] reported an electromagnetic applicator based on a series resonant circuit fed by two MOSFET switchers in half-bridge inverter configuration. The inductor consisted of a Litz-wire coil and a ferrite core. Cano et al. [165] constructed an induction heater device for studies of magnetic hyperthermia and specific absorption rate measurements based on a series resonant circuit fed by four MOSFET switchers in full-bridge inverter configuration. The device worked at a single frequency of 206 kHz and it was able to generate $H_0$=11.9 kAm$^{-1}$ (~ 150 Oe) within the usable volume of 0.6 × 0.5 cm, centered inside the air-coil.

The amplifiers based in MOSFET switchers apply a pulsed output signal to the resonant LRC circuit (see Fig. 18a), which filters part of high-order harmonics and the resultant AC current across the main inductor is nearly sinusoidal. Other types of electromagnetic applicators fed by linear amplifiers have been also reported [161, 162, 163,166] (see Fig. 18b). In this case, the input signal that feeds the resonant circuit is almost sinusoidal and hence, the resulting AC current has a smaller component of high-order harmonics. In general, power linear amplifiers have more restricted power limitations, chiefly because the transistors have to dissipate more energy (unlike MOSFET switchers).

Although MH experiments are strongly frequency dependent, few applicators with adjustable magnetic field frequency have been reported in the literature. Indeed, due to the frequency dependent skin effects and the resistance of magnetic applicators, the building of systems with the availability of frequency scan is technically challenging in this broad frequency range. Lacroix *el al.* [162] described a cost-effective device based on a series LC circuit able to generate AMF with intensities up to 3.82 kAm$^{-1}$ (48 Oe) and with frequencies ranging from 100 to 500 kHz in a useful gap within the ferrite core of 1.1 cm. In this case, a Litz-wire was used as conductor. Garaio et al. [161] built an electromagnetic applicator working in the 149 – 1030 kHz frequency range with AMF amplitude up to 35 kAm$^{-1}$ (~ 440 Oe) at the lowest frequencies. The air core coil was able to apply the AMF to a cylindral usable volume of 31 mm in height and 18 mm in diameter. Bekovic et al. [166] reported an electromagnetic applicator able to apply a rotational AMF of $H_0$= 4.1 kAm$^{-1}$ (~ 52 Oe) intensity in the 20 – 160 kHz frequency range. In all the cases, the multi-frequency feature was achieved using variable capacitors that modify the resonant frequency of the LCR circuits.

Electromagnetic applicators for experiments with large laboratory animals have been also presented. Dürr et al. [167] designed an AMF generator based on a parallel LC circuit fed by a 50 Ω output impedance power amplifier. The animal is placed between two flat pancake coils with variable distance (40 -100 mm) and a maximum AMF intensity of 6.76 kAm$^{-1}$ (~ 85 Oe). This applicator also worked at a single-frequency (200 kHz). The first human-sized prototype (MFHW® 300 F) was made by A. Jordan, U. Gneveckow and collaborators [159]. The device works at an AMF frequency of 100 kHz and a variable intensity from 2.5 to 18 kAm⁻¹ (32 to 226 Oe), using a soft ferromagnetic core to concentrate the field lines and to produce a homogeneous AMF intensity in the gap [159].

The AMF intensity has to be controlled during magnetic hyperthermia experiments. Garaio et al. [161] proposed the use of an external control coil electromagnetically coupled with the inductor that generates AMF. This coil is placed out of the inductor, leaving free space for the sample inside it. Bekovic et al. [166] and Connord et al. [163] proposed a control coil with the same purpose but located inside the inductor that generates the AMF. The electromotive force induced in these coils is proportional to the product of AMF intensity and frequency, it is thus possible to obtain the magnetic field intensity with a data acquisition system. To insure non-invasive measurement, the conducting wire of a scout coil must be narrow (usually a diameter less than 250 µm) in order to avoid heating by eddy currents induced by the AMF. In addition, the AMF phase can be obtained from the phase of the induced signal. A summary of the cited electromagnetic applicators is listed in Table 2.

*4.1 – SAR measurement*

*SAR* is one of the most important parameters in the design of MNPs for MH. This parameter is defined as the absorbed power, normalized by the mass of MNPs, under an applied AMF of certain frequency and intensity $H_0$ (see eq. (4.2)). In other words, *SAR* quantifies the efficiency of MNPs colloids to transform magnetic energy into heat.

$$SAR = \frac{Absorbed\ power}{Mass\ of\ MNPs} \qquad (4.2)$$

*SAR* units are sometimes referred to watts per gram of iron because it is the parameter related to MNPs mass which can be most directly calculated from the molar concentration [Fe] determined by spectroscopic assays titrating elemental iron (UV-visible, ICP-MS, ICP-AES…) or relaxometric methods using the molar mass of Fe (55.845 gmol⁻¹). However a better suited way (although leading to lower numerical values than the previous) is to express *SAR* in watts per gram of MNPs (including oxygen) as from thermo-gravimetric analysis (TGA). *SAR* can be related to the volume power density ($P_{vd}$) by

$$P_{vd} = SAR \cdot c \qquad (4.3)$$

where $c$ is the nanoparticle mass concentration: mass per unit volume. In any case, *SAR* is a crucial parameter to determine $P_{vd}$ and hence the tissue temperature during hyperthermia treatments. Overheating the tumor may result in serious damage to the surrounding healthy cells or in uncontrolled necrosis. On the contrary, the desired therapeutic effect cannot be achieved if the temperature rise is not high enough. *SAR* in MNPs strongly depends on the frequency and intensity of AMF as well as on the chemical, physical and magnetic properties of the material [33, 156,168]. Moreover, it also depends on the dispersion media and the agglomeration degree [34,169]. *SAR* for the same nanoparticle batch can be different in a colloidal dispersion, powder sample or inserted into a biological tissue; therefore, more reliable protocols for measuring *SAR* in MNPs need to be established before MH becomes a feasible cancer therapy. There are currently two main groups under which *SAR* measurements methods can be classified: calorimetric and magnetometric methods.

### 4.1.1 – Calorimetric methods for SAR measurement

MNPs placed into an AMF absorb energy from the field which is subsequently transformed into heat. If the field is strong enough, and also thermal losses are small enough, the generated heat rises the sample temperature. *SAR* is calculated from the temperature derivative over time at instant $t = 0$ as

$$SAR = \frac{C_{p,s}}{m_{MNPs}} \left| \frac{dT}{dt} \right|_{t=0} \qquad (4.4)$$

where $C_{p,s}$ is the heat capacity of the sample and $m_{NP}$ is the mass of the MNPs present in the sample. When MNPs are dispersed in a medium, $C_{p,s}$ in Eq. (4.4) is related to the specific heat capacity of the dispersion medium, $C_{p,d}$, and the specific heat capacity of the MNPs themselves, $C_{p,NP}$, by means of

$$C_{p,s} = C_{p,d} m_d + C_{p,NP} m_{NP} \qquad (4.5)$$

If the time evolution of sample temperature is recorded, the time derivative of temperature at $t = 0$ can be obtained. Thereafter, the *SAR* value of the sample can finally be determined by means of Eq. (4.4). This concept is the basis of calorimetric methods [157]. In order to obtain the time derivative of temperature in Eq. (4.4), the dynamic equation of temperature evolution over time, $T(t)$, must be known (see Fig. 19). As a first approximation, the sample holder can be considered completely adiabatic. Then, the temperature evolution of the sample is

$$T(t) = t \cdot \left| \frac{dT}{dt} \right|_{t=0} + T_0 \qquad (4.6)$$

where $t$ is the time and $T_0$ the initial temperature (at $t = 0$). The time derivative of temperature can be obtained simply by a linear regression of the measured temperature curve. However, in order to achieve such adiabatic conditions, complex isolating systems are required. Natividad et al. [170,171,172] developed an adiabatic magneto-thermal setup for samples that featured such adiabatic conditions. Mendo et al. achieved adiabatic conditions with natural cork as insulating material [173]. However, a sufficiently good isolated sample holder requires a larger inductor and hence, more power to generate the same $H_0$, which reduces the performance of the electromagnetic applicator resulting in a smaller maximum $H_0$. In addition, the design of such adiabatic systems is rather complex and difficult to implement. Therefore, most works found in literature about calorimetric *SAR* measurements are carried out using electromagnetic applicators with non-adiabatic sample holders. In this way, it is interesting to cite the work by Iacob et al. who proposed to correct the *SAR* from thermal losses by alternating ON and OFF periods of the AMF, getting a saw-tooth profile for the curve of temperature versus time, and to subtract, at any temperature, the negative segment (AMF OFF) from the positive slope (AMF ON) [174].

The effects of thermal losses in non-adiabatic conditions cause the decrease of the time derivative of temperature until steady state is reached as shown in Fig. 19. As a consequence, non-adiabatic effects such as radiation losses favored by large colloidal suspension volumes [158], bad thermal isolation, or non-thermal equilibrium conditions [175] lead to an underestimation of *SAR* values when Eq. (4.6) is used [170]. Identifying and quantifying heat losses in the measurement setup lead to higher accuracy for determining the *SAR* value of MNPs colloids. These system losses have to be considered when calculating temperature evolution equations. Assuming Newton's law of cooling, sample temperature evolution over time $T(t)$ is given by [175,176]

$$T(t) = \lambda_Q \cdot \left| \frac{dT}{dt} \right|_{t=0} \cdot \left( 1 - e\exp\left( -\frac{t}{\lambda_Q} \right) \right) \qquad (4.7)$$

where $\lambda_Q$ is a relaxation constant which depends on the heat capacity, the surface of the sample and the heat transfer coefficient between the sample and the medium [177]. In this way, the slope of the initial temperature can be obtained in non-adiabatic situations by performing a non-linear curve fitting of the temperature evolution data to the exponential in Eq. (4.7), resulting in $dT/dt|_{t=0} = \Delta T/\lambda_Q$, where $\Delta T$ is the temperature increment after switching AMF on [175]. Another way of calculating the initial temperature derivative over time in non-adiabatic conditions is to fit the temperature evolution curve to a second-order polynomial equation, such as

$$T(t) = T_0 + |dT/dt|_{(t=0)} \cdot t - a \cdot t^2 \qquad (4.8)$$

where the second-order coefficient $a$ is related to the heat losses of the sample [178]. An alternative method for obtaining the initial time slope to calculate the *SAR* by means of Eq. (4.4) consists in measuring the temperature derivative over time since the initial temperature slope corresponds to the maximum measured derivative.

Many parameters affecting the accuracy of calorimetric methods such as thermal losses, sample volume, magnetic field gradient, or measurement methodology are found in the literature [157, 158]. When dealing with large sample volumes with high heat dissipation, a fast heating causes large temperature gradients throughout the sample and thus, the recorded *SAR* can be very dependent on the temperature sensor positioning [157]. For samples with high magnetization values, demagnetizing fields may reduce the applied $H_0$ intensity inside the sample [158], resulting in the underestimation of the measured *SAR* values. The method and time interval used to determine the initial temperature slope in Eq. (4.4) add uncertainty as well [157,178]. *SAR* measurements by calorimetry are based, indeed, on the measurement of the initial temperature slope. However, the extension of the linear regime is always limited (except for ideal adiabatic conditions) by a plateau value $\Delta T_{max}$ as indicated in Fig. 19, which occurs after an elapsed time that increases with the sample volume. This maximum temperature increment that can be reached by MH directly depends on the surface area of the sample, as demonstrated by the diameter square dependence on magnetic droplets in a micro-fluidic circuit [179].

A "*thermal inertia*" on the measuring system is observed in the initial temperature increment as depicted in Fig. 20, where the temperature initially rises up slowly and it starts to increase rapidly after a few seconds [158,178]. Due to this uncertainty about the initial time, *SAR* values obtained by calorimetric methods can be underestimated. In addition, *SAR* variation with temperature may also increase the error of calorimetric methods, for example if, due to the slowness of calorimetric experiments, several SAR measurements are averaged although not having the same starting temperature. When deducing Eqs. (4.6 – 4.8) for the temperature evolution, *SAR* is assumed temperature independent. However, *SAR* values of self-regulated MNPs have strong temperature dependence in the 300 – 330 K range due to their low Curie temperature (~27 ºC). Temperature variations of *SAR* up to 40% were also found in iron oxide MNPs in the 10 – 50ºC range [166,180,181]. Clearly, calorimetry-based methods are not suitable to measure the temperature dependence of *SAR* values. The thermal dependence can be measured only when the setup is perfectly adiabatic [171, 172]. In this case, the slope corresponds to the initial time slope in Eqs. (4.4) and (4.6) at any time and temperature. In addition, precise measurements of the heat capacity of the sample are necessary to achieve an accurate *SAR* by means of Eq. (4.4).

*4.2.1 – Magnetometric methods for SAR measurement*

The second approach to *SAR* measurement consists in measuring the dynamic magnetization $M(t)$. *SAR* values are subsequently obtained by integrating the dynamic magnetization with respect to the magnetic field [181]

$$SAR = \frac{f}{c}\mu_0 \oint M(t) \cdot dH_0 \qquad (4.9)$$

The integral in Eq. (4.9) is performed over one period ($2\pi/f$). Note that in order to solve the integral in Eq. (4.9), the applied AMF intensity $H_0$ has to be measured during the corresponding period. It follows that *SAR* is proportional to the area of the AC hysteresis loop as indicated in Fig. 21.

There are different methods to measure $M(t)$ in the frequency range of MH. The electromagnetic method based on Faraday's law of induction is the most direct and used one. Experimentally, $M(t)$ (as well as the applied AMF) is measured by different pick-up coil systems. According to Faraday's law of induction, the following voltage $e$ is induced in the pick-up coil

$$e = -\frac{\partial \phi}{\partial t} \qquad (4.10)$$

where $\phi$ is the magnetic flux that crosses the pick-up coils. Hence, using Eq. (4.10) $\phi$ across the sample can be obtained by measuring the voltage induced in the pick-up coils surrounding the MNPs sample, as shown in Fig. 22. However, the magnetic flux density (*B*) is the sum of the AMF intensity and the magnetization of the sample

$$B = \mu_0\big(H_0 + M(t)\big) \qquad (4.11)$$

Therefore, the voltage signal $e$ is caused by the combination of the applied AMF and the magnetization of the sample. In order to measure $M(t)$, the component related with the $H_0$ intensity must be removed from the induced voltage. The direct subtraction involves serious technical difficulties because usually the applied field in MH is much larger than the magnetization $M(t)$ of the sample. Thus, the signal produced by the applied AMF must be removed from the induced voltage signal, by means of a compensation coil. Some authors [161, 163,182,183] proposed the use of two coils wound oppositely in such a way that the upper coil surrounds the sample as indicated in Fig. 22a. Becovik et al. [166,176] used a system of *J*-compensated concentric pick-up coils as indicated in Fig. 22b. If the pick-up coils are properly compensated, there is no induced voltage in the absence of magnetic sample ($M = 0$) and the induced voltage is related to the magnetization in the sample by

$$e = -\xi \frac{\partial M}{\partial t} \tag{4.12}$$

where $\xi$ is a parameter that depends on the geometry of the sample coil.

In principle, the electromagnetic method is more accurate, reproducible and usual than others to measure *SAR* values [176,178], even allowing to measure the temperature dependence of *SAR* more conveniently than using calorimetric methods [181]. Despite its virtues, the electromagnetic method also has some technical difficulties. The voltage signal induced by nanoparticle samples has to be transmitted to a data acquisition system, usually an oscilloscope. The transmission lines or coaxial cables that perform this task can resonate with the inductances and capacitances of the pick-up coils, leading to a linear distortion of the voltage signal induced by the sample [161]. The effects of the distortion are more appreciable when measuring at high frequencies (above 500 kHz). In addition, a capacitive coupling can arise between the carrier liquid where MNPs are dispersed and the windings of the pick-up coils [161,184]. This coupling is important when the carrier liquid is water, a highly polar solvent. At this point, it is also worth mentioning that it is fundamental to correctly measure and control the field intensity generated by AMF generators. As in general, *SAR* values of MNPs strongly vary with AMF intensity, only adequately calibrated MH setups can provide results that can be compared to the models and between different laboratories. A good and accurate way to measure this intensity is using mobile pick-up coils where a voltage signal (electromotive force) is induced in the presence of the AMF. This signal is proportional to the angular frequency and to the magnetic flux enclosed by the coil circuit, and the AMF intensity can be obtained with prior calibration. In addition, the output impedance and available power of the amplifiers that feed electromagnetic applicators, as well as load impedances, usually present a thermal shift. This causes the generated field intensity to vary with operating time. Therefore, a continuous monitoring of the intensity is necessary to set its value to a constant value during the experiment by a closed-loop control (i.e. PID controller).

Also using magnetometry, Ahrentorp et al. [184] obtained *SAR* values by combining AC magnetic susceptibility and static magnetization curves. Another approach based on the magneto-optical Faraday effect was also used to determine the AC magnetic susceptibility of MNPs [185,186]. In this case, the measurement method is based on the Faraday rotation of the polarization plane of light crossing the magnetic fluid. Hence it is possible to measure $M(t)$ by measuring the polarization of a laser beam upon crossing the magnetic fluid. However, magneto-optical measurements are limited to low MNP concentration conditions in order to keep the dispersion transparent.

## 5.-Influence of biological matrices on the magnetic heating efficiency

As mentioned in Section 2, many efforts have been paid to synthesize MNPs with outstanding magnetic properties [48, 49, 75] in order to release the highest heat power at the lowest MNP dose into tumoral tissues. However, a significant amount of experimental results evidences a substantial reduction of heating efficiency related to alterations of the MNPs magnetic properties when either biomolecules are adsorbed onto the surface of MNPs or MNPs are located inside cells or tissues. These changes of their magnetic response are strongly reflected by distinct characterization techniques such as magnetization loops [109], zero-field cooling and field cooling (ZFC-FC) magnetic susceptibilities [187,188], magnetic relaxation processes [108], heat dissipation power [109,139], and relaxivity, that is the physical efficiency parameter related to the MRI contrast signal [139,189, 190,191,192]. Hence, all the efforts made for improving the magnetic response of MNPs may be in vain, for instance, in view of their often poor intracellular magnetic response. The underlying reasons of the alteration of MNP magnetic properties, and consequently the reduction of their magnetic heating, are yet to be clarified. For that purpose, it is crucial to understand the physical reasons, which influences the MNPs magnetic response inside any biological matrices (i.e. cells, tissues, subcellular vesicles) and/or fluids (i.e. blood, urine, cell media) in order to preserve the heating efficiency from MNPs regardless of the host medium. Only in this manner it would be possible to engineer MNPs with well-defined heating efficiency or contrast signal acting as reliable theragnostic agents for clinical applications. In the following, we review different works analyzing the influence of the biological matrix on the heating efficiency of MNPs.

### 5.1 – Protein adsorption onto MNP surface

As mentioned in Section 2.2, it is well-recognized that the surface of MNPs is covered by biomolecules (proteins, sugars and lipids) upon coming into contact with biological fluids such as blood. The formation of a protein "corona" results from the strong interaction between MNPs' surface and biomolecules present in blood [193,194]. Such protein corona can be considered as the living organism response to MNPs. During the opsonization process, plasma proteins circulating in the blood stream are mainly adsorbed onto the surface of MNPs by electrostatic interactions with the invasive entities onto their surface, acting as markers to warn macrophages of the reticuloendothelial system for removing unwanted entities from blood stream and possibly to metabolize them. The composition of the protein corona plays a crucial role in the biological fate of the MNPs [195,196]. One of the first changes related to protein corona is reflected on the hydrodynamic size of MNPs. Dynamic light scattering measurements of MNPs colloids dispersed in biological fluids show an enhancement of the hydrodynamic size after few minutes/hours, resulting in changes of the colloidal stability that may favor the formation of micrometer size MNP aggregates [192,197].

The type and amount of protein absorbed onto MNPs coating surface have been shown to affect the colloidal stability of the MNPs in biological media and in consequence, leading to changes in their magnetic response [190] that can be even used as sensor signal [198,199]. Several works report on the variation of transverse relaxivity of MNP colloids when dispersed in cell media containing serum protein compared to the same particles suspended in water. These changes are indirectly related to the changes introduced in the magnetic properties of MNPs by protein absorption [190], which eventually influences the proton spin relaxation of the surrounding water molecules [190,191, 197]. In addition, the presence of the protein corona leads to a 30% reduction of saturation magnetization. Hence, variations of magnetic properties induced by protein adsorption should consequently influence the dynamical magnetic response [200]. Krishnan's group reports how the heat dissipation power of MNPs dispersed in biological media is reduced up to 30%. Authors suggest that this is due to MNPs agglomeration induced by protein absorption, which significantly alter their colloidal stability. These results are in agreement with those observed by Aires et al. revealing a 50% decrease of *SAR* values of dimercaptosuccinic acid coated $\gamma-Fe_2O_3$ MNP colloids when dispersed in a cell medium such as fetal bovine serum (FBS) in comparison to data in aqueous dispersion (see inset of Fig. 23) [201]. Due to protein adsorption, the MNP dispersion in FBS media leads to strong changes on their MNP colloidal stability as reflected in the hydrodynamic size, which increases from 60 nm (in water dispersion) to 193 nm (in FBS dispersion). TGA reveals that such increase of hydrodynamic size is mainly related to protein adsorption onto MNP surface and in a lower extent due to MNP agglomeration. However, the precipitation of MNP colloids dispersed in biological fluids is a matter of time due to their loss of colloidal stability soon followed by chemical degradation as first shown by De Cuyper et al. in the case of a citrate coating [202]. As expected, the aforementioned *SAR* changes are significantly reflected on the corresponding AC hysteresis loops. As shown in Fig. 23, AC hysteresis loops are more elliptical in the case of MNPs dispersed in biological media than in water dispersions. The main reason behind those changes of the AC magnetic response can be understood in terms of MNP agglomeration favored when MNPs colloidal stability is altered after protein corona formation. In the light of these observations, further studies are needed to understand the physicochemical mechanisms responsible of such variation of the AC magnetic response induced by protein adsorption.

### 5.2- MNP cellular processing

The interaction of magnetic MNPs with cells or tissues have shown to significantly alter their magnetic response compared to the initial aqueous colloidal suspensions of MNPs. Recent results have shown that *SAR* values shrink between 50% and 90% when MNPs are internalized into cells, depending on their size, shape and coating [109,139], as shown in Fig. 24. Different origins are attributed to this remarkable effect. On the one hand, Di Corato et al. argue that the heating reduction is due to the switching of magnetic relaxation processes from Brown to Néel due to the increase of viscosity into the intracellular environment [109]. The role of aggregation is also considered, but without establishing if it is more or less dominant than the MNP physical blocking in the cell environment. Contrarily, Etheridge et al. firmly demonstrate that MNPs aggregation plays a highly relevant role on the reduction of their magnetic heating capabilities [139], as shown in Fig. 25. These authors show that aggregation-induced *SAR* reduction becomes more pronounced when viscosity increases, but the viscosity effect is responsible of the ~25% of the total *SAR* reduction. The changes of MNP properties induced by cell environment are clearly exemplified by magnetization loops showing distinct hysteretic behaviors depending on the MNP environment (aqueous dispersion, cell membrane, and into cell). Thus, irreversible magnetization reversal processes show up when MNPs are internalized into cells as shown in Fig. 26. In this case, blocking temperatures follow a general upward trend observed also when the MNPs are in tissues and becomes more pronounced with MNPs concentration and size [188] as shown in Fig. 27. Different works have analyzed MNPs uptake and their trafficking into the cytoplasm [203,204,205,206]. In both cases, MNPs clustering is favored by cellular processing, resulting in an increase of magnetic dipolar interactions [187]. This effect of cellular uptake can be observed on the temperature dependence of ZFC-FC magnetic susceptibility, which is known to be sensitive to magnetic interactions [207,208]. Such sensitivity of MNP magnetic properties to cell environment risks the reliability of MNPs as heating mediators, contrast agents or magnetic labels in biomedical applications. The methodology employed by Etheridge et al. allows mimicking to a certain extent the intracellular transit of iron oxide MNPs from early endosomes to lysosomes [206]. In this context, magnetic dipolar interactions are expected to be significantly favored inside MNPs aggregates, resembling the situation of large MNPs concentrations. Martinez-Boubeta et al. showed a non-monotonic dependence of *SAR* values with MNPs concentration [141]. This dependence can be visualized as a universal curve when normalizing *SAR* values to certain intrinsic magnetic parameters *viz.* magnetic anisotropy and/or saturation magnetization. This behavior is due to dipolar interactions, which profoundly influences *SAR* values upon increasing MNP concentration [209,210]. Despite the advances made so far, further theoretical models would be needed to encompass the broad variety of experimental results reported on the related literature.

The viscosity of the medium is another important parameter also linked to magnetic dipolar interactions, but it has not received the attention it deserves. In a first approach, it is widely accepted that medium viscosity may strongly influence the magnetic relaxation processes governing the heat generation by MNPs [33]. In the case of SPM MNPs, magnetic moment relaxation processes follow Brown and Néel mechanisms [211]. Fortin et al. reported on *SAR* values of iron oxide MNPs dispersed in solvents with different viscosity to discern the contributions from Néel and Brownian mechanisms to heat generation [33]. Recent AC susceptometry results show the *in situ* magnetic response of model systems of blocked and SPM NP, following their cellular internalization and subsequent release by freeze-thaw lysis [108]. The AC susceptibility signal from internalized MNP in live cells showed only Néel relaxation, consistent with measurements of a suspension of immobilized nanoparticles. However, Brownian relaxation was restored after cell lysis, indicating that the immobilization effect was reversible and that MNP integrity was maintained inside cells. In addition, the role of the AMF intensity on modifying those relaxation processes is not totally clear, especially in the case of the Brownian one. Besides its influence on determining the effective magnetic relaxation times, medium viscosity also affects magnetic dipolar interparticle interactions in ferromagnetic and SPM MNPs [147,212]. This is usually ascribed to the alignment degree between the particle magnetic moment direction and its anisotropy easy axis. The degree of this alignment strongly influences the shape of the MNPs hysteresis loops [140] and hence the associated *SAR* values [140, 147]. Indeed, decreasing *SAR* values in viscous MNPs colloids needs further explanation. Considering that $SAR = \pi \cdot A \cdot f$ [99] – where $A$ is the area of the hysteresis loop - one may expect strong variations in the shape, coercive field and/or remanence of the with MNPs concentration and viscosity of the medium. Preliminary studies indicate that the shape of the hysteresis loops becomes more elliptical when viscosity increases. This viscosity-induced shape variation can be understood in terms of the alignment between the particle magnetization and its anisotropy easy axis, as proposed by Landi and Serantes et al. [140,147]. Also, the variation of the coercive fields and magnetization values at the highest field amplitude plays a crucial role for defining the hysteretic area and consequently the heat dissipation under AMF. Further experimental studies are required in order to verify how the shape of AC hysteresis loops (i.e. coercive field and remanence) of MNPs colloids varies with the particle concentration and viscosity media. Explaining this variation would place us in position to clarify the magnetic phenomena occurring inside cells or tissues, which is mandatory to engineer nanostructures based on MNPs for heating with controlled energy release independently of the biological matrix where MNPs are hosted.

### 5.3- MNP biotransformation

In the previous section, we have discussed on the phenomenology related to NP cell uptake, i.e. the enhancement of MNPs clustering and media viscosity, which seems to be responsible of the drastic reduction of the heat dissipated from MNPs inside biological matrices. Other important effect altering the MNP magnetic response is related to the exposure to acid environments and/or biochemical reactions into subcellular vesicles (endosomes and/or lysosomes) or at the interstitial extracellular location. Such conditions in biological environments lead to the MNP transformation, which can be totally metabolized inside living organisms [213,214,215]. In general, the latter takes place when different types of proteins (such as iron storage proteins) or enzymes (such as lysosomal digestive enzymes) act to decide MNP fate, which is tightly dependent on their surface properties/coating, charge, size and morphology. MNPs biodegradation processes in living organisms [216] have been proven to modify the MNPs atomic structure resulting in a size reduction and therefore, resulting in unavoidable changes of their magnetic properties [187,188,214,217], as shown in Fig. 27. The structural degradation of individual MNPs is driven by a stochastic corrosion process that proceeds depending on the nature and the distribution of particle coating, which controls surface reactivity with chelating agents. Indeed, the availability of chelating agents and their accessibility to the nanoparticle core are the key factors regulating degradation kinetics. Lartigue et al. hypothesize that cells could timely orchestrate the redistribution of a nanoparticle from a dense assembly in early endosomes to a more dispersed and exposed state into lysosomal compartments [216]. This intracellular trafficking mechanism results in reducing MNP size and increasing polydispersity. In consequence, the magnetic properties of biodegraded MNPs are significantly altered with time under acidic conditions as shown in Table 3 [217]. On the one hand, biodegradation leads to large presence of defects onto the MNPs surface, which significantly contribute to increase spin disorder [218] at the particle surface and values of surface magnetic anisotropy constant $K_s$ [212,214]. On the other hand, the magnetic MNPs size polydispersity implies unsuitable magnetic properties, and in consequence tends to lower their heating efficiency [49,219].

Concerning these biotransformation aspects of MNPs determining the biodegradation, dynamics is crucial to define the time window where the intrinsic MH characteristics are preserved before undergoing chemical changes that may affect their heating capabilities. Recently, Javed et al. have shown that the exposure of iron oxide MNPs to an acidic medium (citrate buffer) of pH similar to that found inside lysosomes has a dramatic effect on *SAR* [217] (see Table 3). After six days, the heating power decreases by 70%, and 99% after 23 days (Table 3). This effect testifies not only the early alteration of magnetic properties due to the degradation of MNPs in a few days but also the potential deterioration of their heating efficiency. Recent works have developed a suitable methodology to follow *in vivo* transformations of both magnetic and structural properties of iron oxide MNPs. On one hand, structural degradation of individual MNPs was monitored at the atomic scale with aberration-corrected high-resolution transmission electron microscopy [216] and also it has been corroborated by the observation of a decrease of relaxometric properties [217]. This successful *in situ* monitoring of nanoparticle transformation might be easily applicable to other nanomaterials to provide critical insight about their fate in the living organism. In addition, their approach reduces the gap between *in situ* nanoscale observations in mimicking biological environments and *in vivo* real tracking of nanoparticle fate also called "life cycle analysis". On the other hand, AC magnetic susceptibility studies provide a powerful tool to monitor time scale of magnetic MNPs biodegradation by directly looking at their magnetic signal [220]. This particular research niche should move forward in the coming years to clarify the biodegradation dynamics for the sake of i) establishing the suitable time period where heating exposure mediated by MNP can be controlled, ii) engineering new nanostructures based on MNP resistant to transformation processes during a given treatment period.

## 6. Final remarks and outlook

In order to render this review as a useful tutorial on the MH topic, we summarize here the main advances reported in the last few years. Furthermore, we have selected the next challenges to be faced in near future by the scientific community working on MH.

First of all, a standardization is required for different purposes, from the scaled-up production of MNPs with optimal magnetic characteristics for MH to the fabrication and use of adequate AMF applicators. There is a strong need to define which are the suitable experimental conditions with minimal error sources (calorimetric or magnetic measurements, adiabatic or non-adiabatic conditions, MNPs dispersion media, thermal probe, AMF generator, etc...), the methodology (in order to obtain an accurate value of $dT/dt|_{t=0}$ for calorimetric methods, how to calibrate AC magnetometry measurements) and the calibration standard. Only in this manner it would be possible to compare measurements independently of user's set-up, methodology and experimental conditions.

Regarding the production of optimized MNPs, criteria can be summarized as quantity (gram-scale synthesis) without losing on quality, meaning: reproducibility and scalability of the chemical synthesis, bio-safety of the MNPs (assessment of their long-term *in vivo* fate and bio-elimination), accurate control of dimensions and low size-dispersity and, most importantly, high heating efficiency in magnetic field conditions (field strength and frequency) recommended for *in vivo* experiments. It is not the goal of this review to make a definitive choice between the different classes of synthesis routes that have been listed in Section 2 (alkaline co-precipitation, hydrothermal treatment, polyol route, thermal decomposition of metal-complexes, laser-assisted pyrolysis, etc…) since they all have their advantages and drawbacks.

Chemists need to adapt these methods to make them fit the quantity and quality requirements for MH applications. A thumb-rule to reach high *SAR* values could be to use magnetic nanocrystals near the superparamagnetic / ferrimagnetic transition, which in case of pure iron oxide cores corresponds to sizes around 18 nm [212]. This is somehow just above the 12-15 nm range of values already chosen by MagForce™ Company in their 2011 clinical trial on MH treatment combined with radiotherapy of high-grade glioblastoma tumors. Although it is certainly better to speak of volume power (Wm$^{-3}$) which is the physical parameter in the bio-heat equation, the *SAR* value (Wg$^{-1}$) is still the most practical one to compare the heating efficiency of different nanoparticles, under well specified magnetic field conditions (field strength and frequency), the *ILP* concept being not agreed on among users since *SAR* rarely appears as quadratic in field intensity. The threshold size range of 12-18 nm lies just between two alternative options offered: either relying on pure Néel relaxations of SPM MNPs, meaning higher frequencies in the 400-700 kHz range, but limiting the field to 10 kAm$^{-1}$ (~ 125 Oe) to minimize physiological side effects; otherwise playing with the larger hysteresis loops of ferri- or ferromagnetic cores (described by Stoner-Wolhfarth model), thereby decreasing the frequency in the 100-200 kHz range, but at the cost of a higher field strength (at least 20 kAm$^{-1}$ (~ 250 Oe)) to overcome the coercive field $H_c$ of the hysteretic magnetization curve. Please note that the "15-18 nm rule" is valid only for quasi-spherical iron oxide MNPs, since other shapes (nanocubes, nanoflowers…), or other magnetic materials of different values of the magnetic anisotropy have a shifted magnetic transition of different MNPs sizes.

Many physical parameters are known to affect the heating capabilities of MNPs, but from some recent results it is clear that magnetic anisotropy is key-one: MNPs with high anisotropy are less sensitive to dipolar interaction effects and larger *SAR* values are obtained when increasing $H_0$ above a threshold field. Equally, most of the predicted behaviors for selected MNPs have been derived for AMF parameters producing major hysteresis loops, indicating that field conditions lead to different hysteretic behaviors. In any case, more effort should be put in further exploring the non-linear region between SPM and hysteretic regimes, where the most popular models may behave unpredictably and, at the same time, most of the experimental data have been obtained.

Physical models should be seen as complementary to real-time imaging and other ancillary techniques for treatment planning and monitoring, since they could hardly model—at least for now—biochemical reactions like the immunological call effect of heat in tissues or the activation of heat-shock proteins under changing conditions. Moreover, there is an unpredictable and complex relationship between the different physical properties of biological entities, which further complicates the task of bridging the gap between theory and experiments. Depending on the body region or tissue to be modeled, a proper multi-layer interface has to be set in order to choose the appropriate boundary conditions. Neglecting abrupt variations of physical properties across biological tissues (bones, blood vessels, etc.) lead to results with little applicability; this also applies to the patient condition: for instance, it is known that diabetes cause a different heat transfer depending on the body region. Other important point to follow up during the ensuing years is tuning the magnetic interparticle interactions, an aspect that is still being neglected in a good deal of the new physical MH models. Despite the advances made so far in computing interparticle interactions towards modeling heat dissipation in MNPs systems, more realistic *in vivo* simulations are yet to come through their progressive integration into the existing (more) physiological models. Finally, there is still a lack of systematic comparison between simulated and *in vivo* data to validate the proposed physical models instead of also being checked for other datasets, which seems to confirm that a "one-fits-all" approach is not a realistic option for treating different cancer types by MH.

Special attention has also to be paid to the spatial distribution of the $H_0$ intensity. Ideally, the experiments need a homogeneous field. A suitable way to calculate the spatial distribution of AMF is using finite element methods (FEM). There are several software packages that implement this method for electromagnetic field simulations. Axial symmetry can be assumed when calculating the fields produced by coils, which reduces the computational cost dramatically. It is recommended to compare the theoretically obtained $H_0$ values with the experimental ones.

Regarding the *SAR* measurements of MNPs, special attention has to be paid to the thermal losses when using calorimetric methods. In this case, adiabatic sample holders are preferred. Furthermore, the thermal dependence of *SAR* has to be taken into account when using these magneto-thermal approaches.

In general, the electromagnetic methods like AC susceptibility and AC magnetometry are more accurate methods than calorimetry. Although not discussed in the present article, thermometry is one of the MH aspects requiring more attention in the near future. Besides the typical *invasive methods* based on the use of inserted thermal probes, there are promising *non-invasive* methods, among which magnetic resonance (MRI) and computerized tomography are the most popular choices. Taking advantage of the temperature dependence of some physical parameters (longitudinal relaxation time, proton resonance frequency shift, diffusion constant of water) [221], MRI thermometry methods currently offer a sub-millimeter spatial resolution but less precision on temperature and scan rate than when using a matrix (yet invasive) of inserted thermal probes. Another promising bio-imaging method that can potentially enable non-invasive thermometry is magnetic particle imaging (MPI) which is specific to the presence of the MNPs in tissues: the thermal variation of the high-order harmonics of AC magnetization was recently proposed to monitor temperature [222]. The calibration process is very important for the development of better temperature measurement techniques, as different tissues/organs will show different calibration curves for the same patient. New thermometry methods are focused on taking advantage of the nanoscale, more specifically on the use of "nano-thermometers" that consist in quantum dots or lanthanide complexes whose fluorescent emission depends on the temperature [223]. When coupled to the MNPs, such luminescent probes can improve spatial resolution and offer a reliable means to monitor temperature during MH treatments at a subcellular scale, potentially solving the paradox of "cold hyperthermia" (cellular death in the absence of a perceptible rise of the macroscopic temperature of the medium).

We have also gathered experimental evidences on the influence of the biological matrices and fluids on the magnetic heating capabilities of MNPs. On the one hand, the formation of a protein corona onto the MNP surface leads to changes in the magnetic heating efficiency explained by the deterioration of the MNP colloidal stability. In this respect, future challenges will require to engineer MNP surface to avoid or minimize protein adsorption into blood stream resulting in changes of their magnetic response. Simultaneously, MNP surface will be mandatory kept hydrophilic and neutral to provide biocompatibility and provided with chemical groups favoring biomolecule conjugation. On the other hand, the magnetic properties and response of MNPs inside cells and tissues may be strongly altered due to MNP cellular processing and biotransformation. Understanding of the influence of the protein corona, intracellular nanoparticle transit and transformation on the MNP magnetic properties is a key issue for the efficient use of MNP in MH, as well as many other biomedical applications. For that reason, it is recommended to check the actual magnetic response of MNP colloids in true biological environments and physiological conditions in order to engineer MNPs with robust and reliable magnetic heating capabilities.

Finally, a word of caution must be raised regarding the AMF generation technologies, specially, concerning the tolerance and/or safety limits of AMF exposure. Since both $H_0$ and $f$ play a relevant role in the magnetic heating capabilities of MNP, the range of tolerable or adequate field amplitudes/frequencies for humans needs to be established. In the last 20 years there has been a remarkable technological progress in the use of high performance AMF generators delivering AC magnetic fields of high frequency (up to tens of MHz) and intensity (up to 310 kAm$^{-1}$) as shown in Fig. 28a. Comparing the experimental $H_0$ and $f$ values for different magnetic heating studies—namely *SAR* measurement as well as *in vitro* and *in vivo* MH—reported in the analyzed set of references, there is around three times more variation in $f$ than in $H_0$. On the one hand most of the reported $H_0$ values are below 50 kAm$^{-1}$, with a median of ~16 kAm$^{-1}$, and the highest intensities achieved have increased almost monotonically from the second half of the 2000s towards the present. On the other hand, the most popular $f$ values are within the 100-400 kHz range, with a median of ~233 kHz. Considering the type of experiment carried out, the number of *in vitro* MH experiments has remained approximately constant throughout the last 20 years. However, there is a noticeable concentration of articles concerning *SAR* measurements during the second half of 2000s, whereas those dealing with *in vivo* experiments became ubiquitous since 2010 (Fig. 28b).

The only reliable tolerance limit for AMF exposure in MH treatments that is available to date was suggested by Brezovich and coworkers, who experimentally estimated three decades ago that the $H_0 \cdot f$ product should not go beyond $4.85 \times 10^8$ A(ms)$^{-1}$ for a safe AMF use in humans [14]. However, this experimental finding sometimes quoted "Brezovich limit" must be revised considering the underlying premises and also the current technologies for AMF generation for clinical applications. Firstly, and most importantly, it relates to the treatment of an entire human thorax using a coil wrapped around it. In many practical cases, the tumors to be treated are localized in a particular region that by no means requires such an extensive application. In addition, this particular configuration is known to favor the occurrence of eddy currents and therefore non-specific heating, unlike other systems that project the field into the body [159]. Although these limitations make the Brezovich criterion to be taken as a mere upper exposure limit, yet 45% of the experimental $H_0 \cdot f$ values exceed it (Fig. 28a). Remarkably, many of them were reported well after the Berzovich criterion was first published. Whatever the cause may be—impetus for getting well positioned in the race for the MNPs with the highest *SAR* value, inadequate experimental setup or simply ignorance of the limit—this observation evidences the lack of tolerance limits specific to MH, which in turn hinders the design of clinical trials. Therefore, a consensus must be urgently reached by the MH research community to make real progresses towards its acceptance as a recognized treatment method in the clinical practice.

## Acknowledgements

The authors thank the European COST action TD1402 (RADIOMAG). This work has been partially supported by European Commission (MULTIFUN, nº 262943), Spanish Ministry of Economy and Competitiveness (MAT2013-47395-C4-3-R) and Madrid Regional Government (NANOFRONTMAG-CM S2013/MIT-2850). F. J. Teran acknowledges financial support from Ramon y Cajal subprogram (RYC-2011-09617). Dr. Antonio Aires and David Cabrera are deeply acknowledged for kindly supplying data shown in Fig. 23. D. Ortega is grateful for the support received through the AMAROUT-II Marie Curie Action under the European Commission's FP7 PEOPLE-COFUND program and a Research Fund grant from the Royal Society of Chemistry. E. Garaio and F. Plazaola acknowledge the Basque Government for financial support under grant IT - 443 – 10. O. Sandre thanks the department of Science & Technology of Univ. Bordeaux for supporting G. Hemery's PhD thesis grant. E.A. Périgo

This document is the Accepted Manuscript version of a Published Work that appeared in final form in *Appl. Phys. Rev.* 2, 041302 (2015) © AIP Publishing LLC (American Institute of Physics, USA) after peer review and technical editing by the publisher. To access the final edited and published work see http://dx.doi.org/10.1063/1.4935688

acknowledges the financial support from the National Research Fund of Luxembourg (ATTRACT project

no. FNR/A09/01) and from the University of Luxembourg (INTERFACE 2012 project).

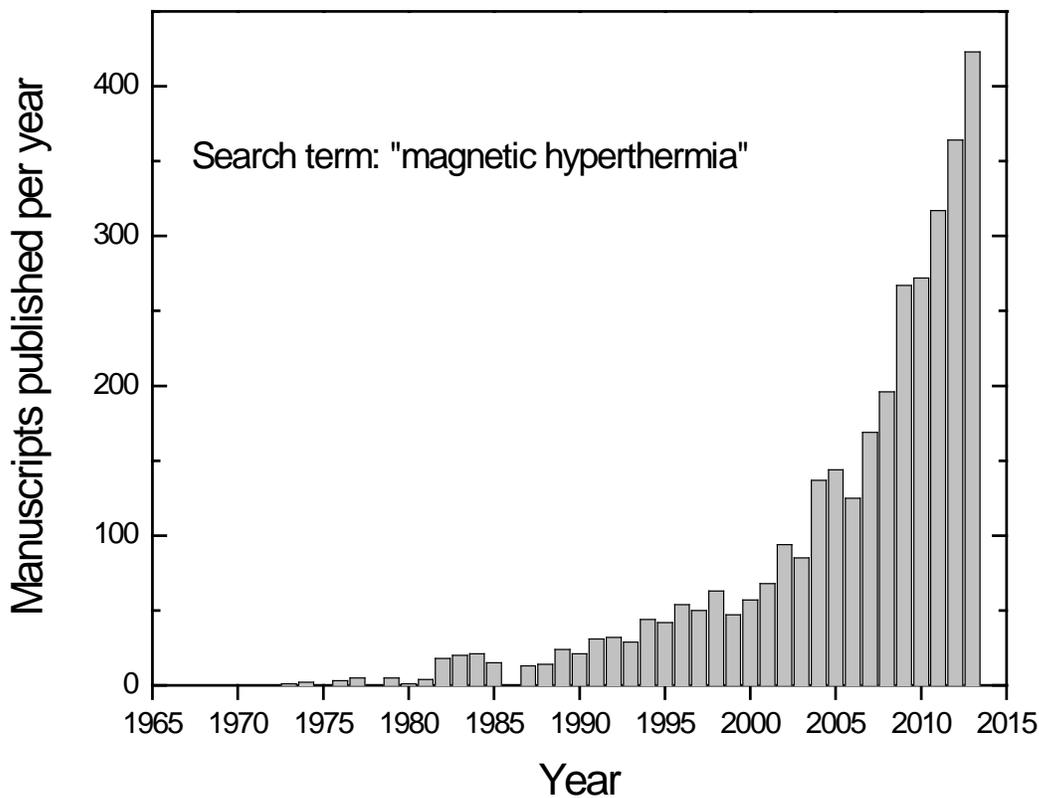

**Figure 1** – Number of published scientific manuscripts during the period 1973-2013 using the search term "magnetic hyperthermia". © ISI Web of Knowledge.

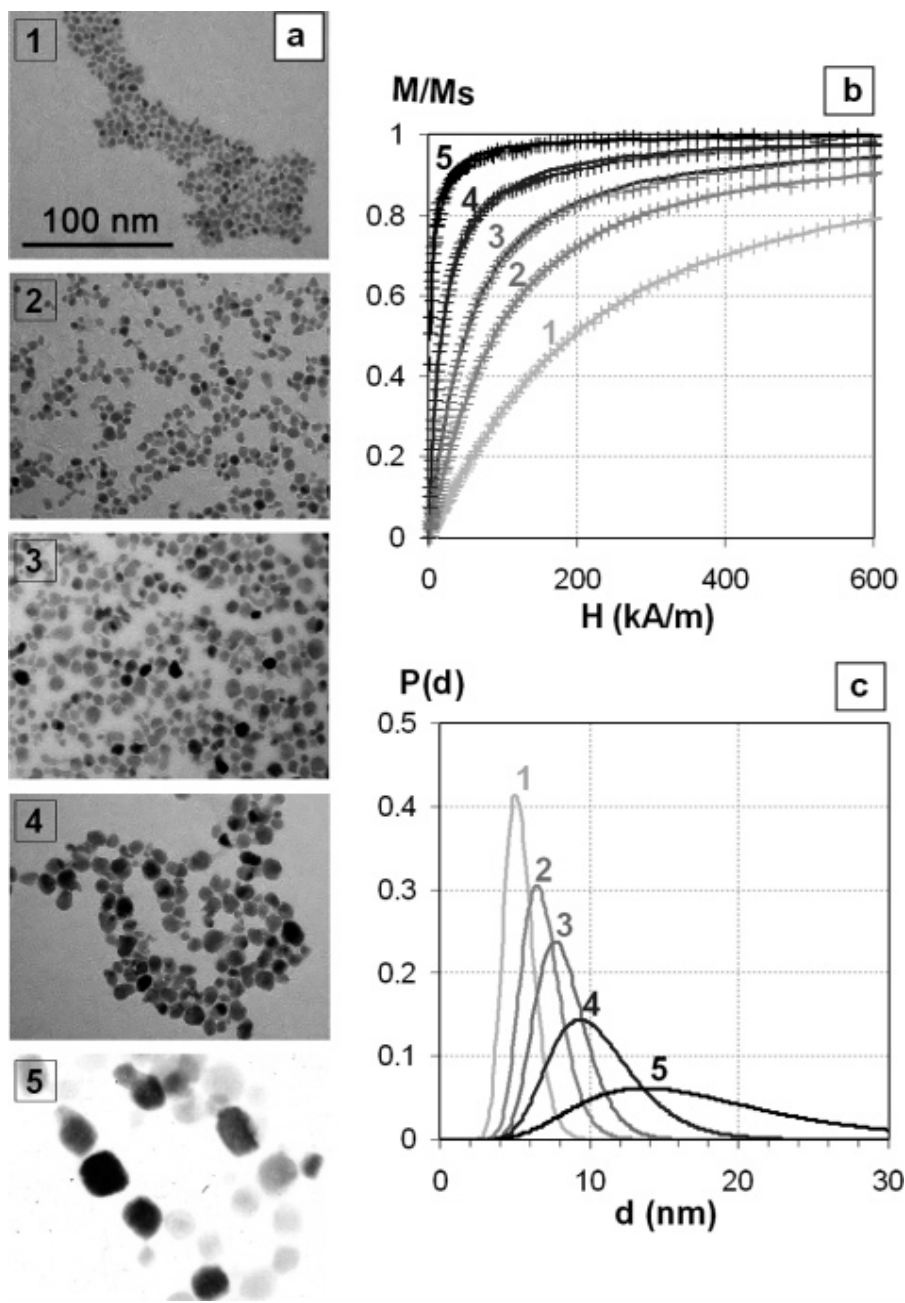

**Figure 2 –** Transmission electron microscopy images of iron oxide nanoparticles synthesized by aqueous alkaline co-precipitation followed by a size-sorting method based on ionic force induced phase-separation of (1) $5.3 \pm 1.0$ nm, (2) $6.7 \pm 1.4$ nm, (3) $8 \pm 1.7$ nm, (4) $10.2 \pm 2.9$ nm, and (5) $16.5 \pm 7.5$ nm [33]. Reproduced with permission from J. Amer. Chem. Soc. 129 (9), 2628 (2007). Copyright 2007 American Chemical Society.

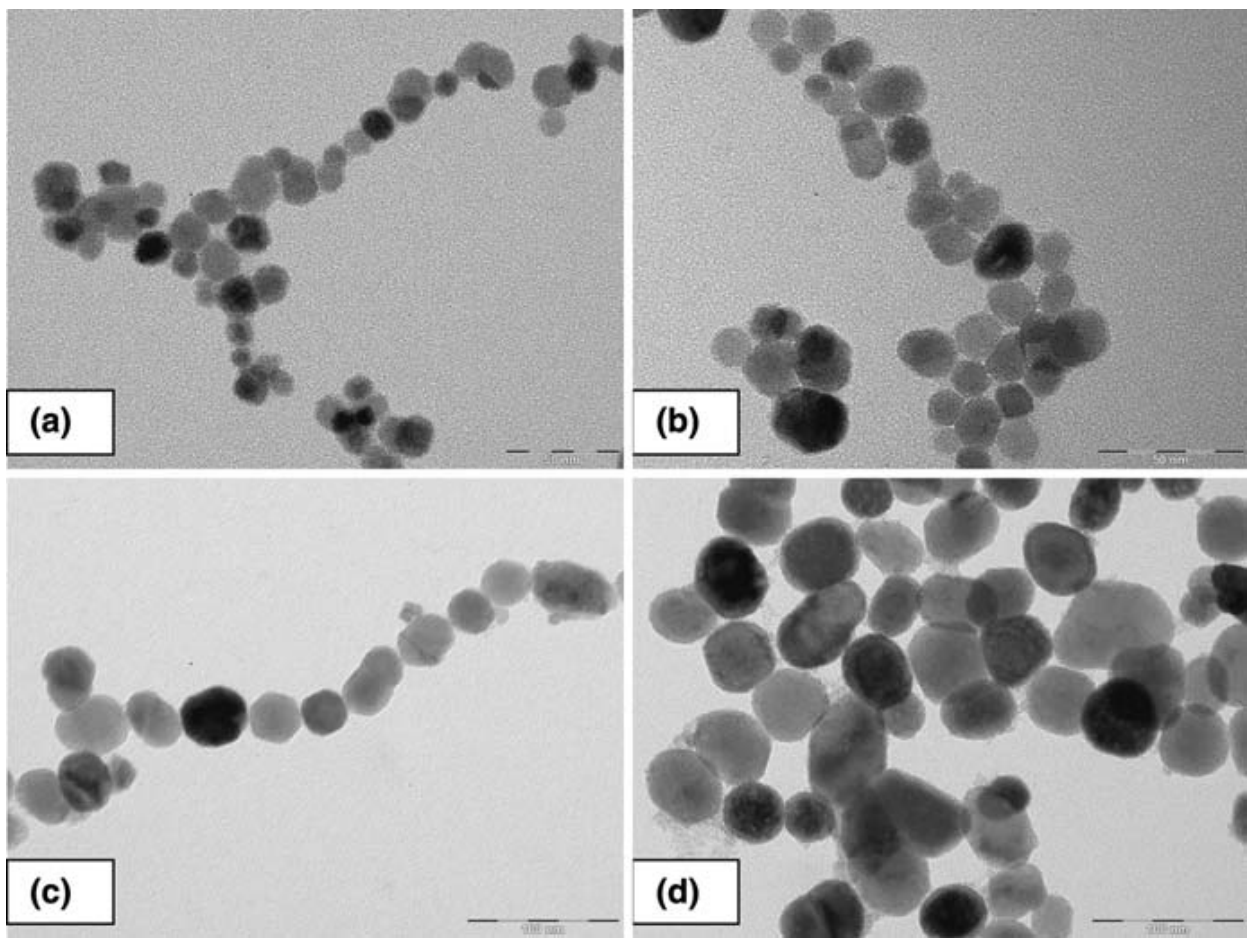

**Figure 3** – Transmission electron microscopy images of maghemite ($\gamma$-Fe$_2$O$_3$) nanoparticles obtained by hydrothermal treatment (T = 200 °C, ratio of Fe$^{2+}$/( Fe$^{2+}$+Fe$^{3+}$) = 0.7) for different duration and pH: a) 2 h, pH = 12 (particles that are re-dissolved at pH = 3); b) 2 h, pH = 12 (insoluble fraction); c) 24 h, pH = 12; d) 2 h, pH = 14 [42]. Average diameters by TEM are: a) 13 nm; b) 22 nm; c) 55 nm; d) 52 nm. Reproduced with permission from J. Nanopart. Res. 11 (5), 1247 (2009). Copyright 2009 Springer.

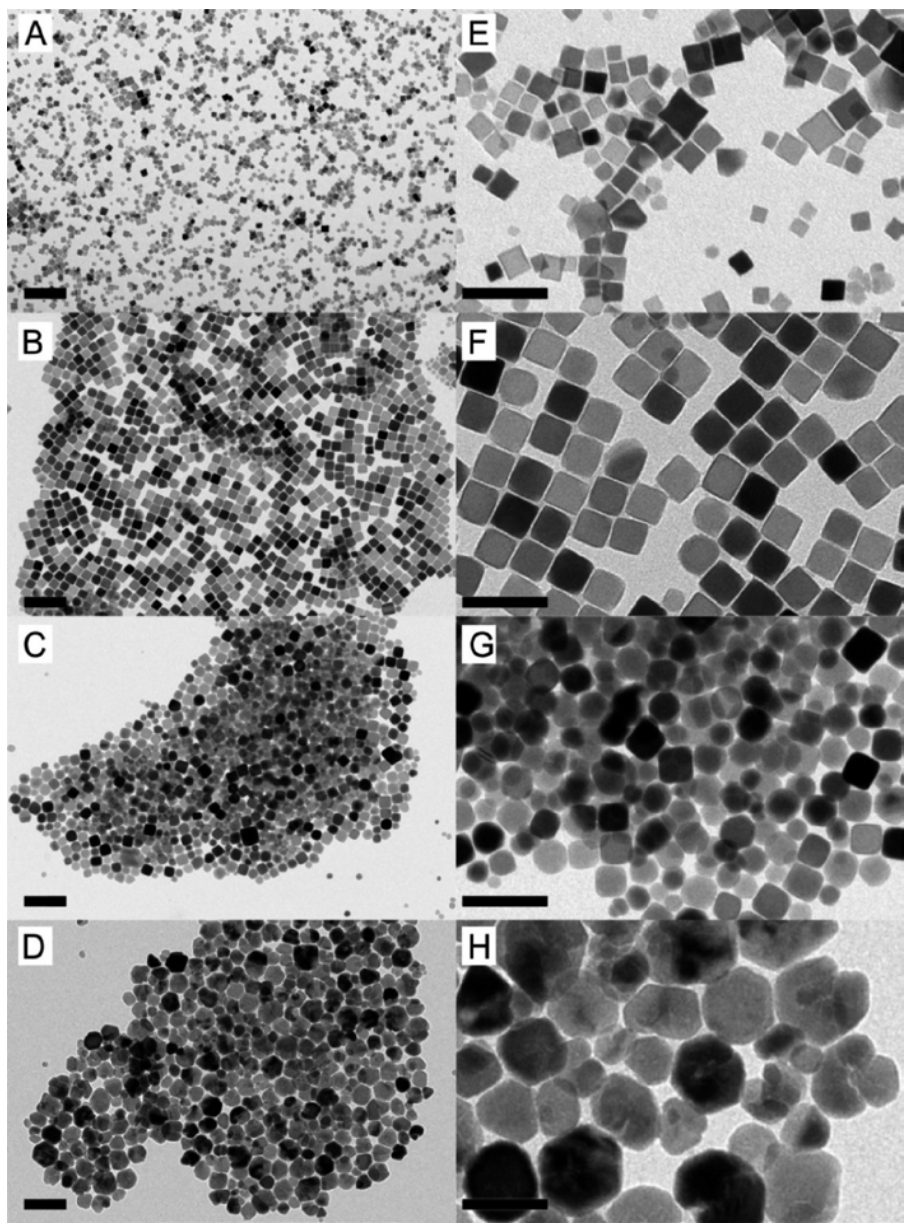

**Figure 4** – Transmission electron microscopy images of iron oxide nanocubes synthesized by thermal decomposition of iron acetylacetonate, for cube edge lengths of (A) 12 ± 1 nm, (B) 19 ± 3 nm, (C) 25 ± 4 nm, and (D) 38 ± 9 nm. Panels from E to H (scale bars of 50 nm) are higher magnifications of samples shown in panels from A to D (scale bar of 100 nm) [48]. Reproduced with permission from ACS Nano 6 (4), 3080 (2012). Copyright 2012 American Chemical Society.

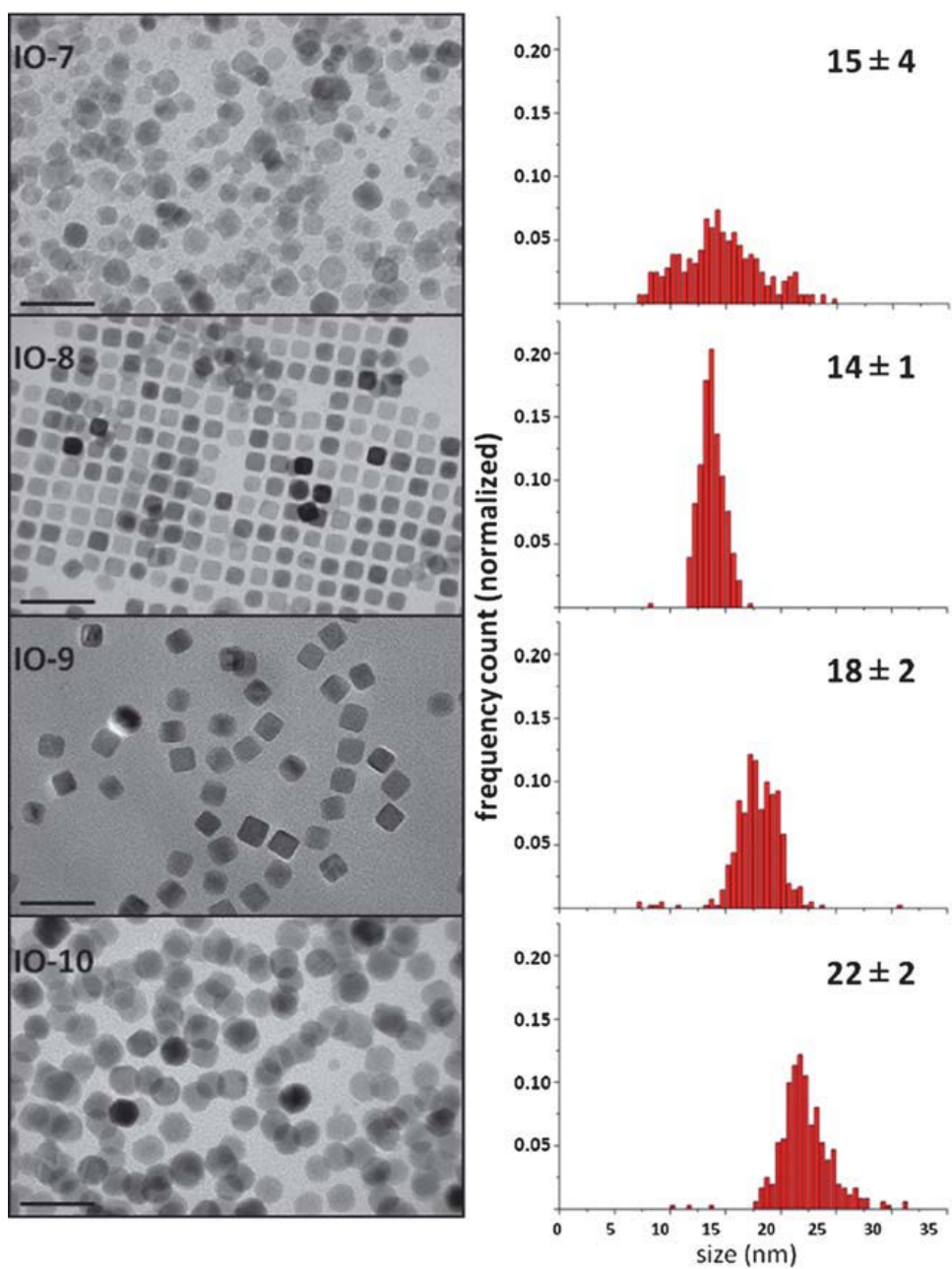

**Figure 5 –** Transmission electron microscopy micrographs and normalized size distribution histograms of iron oxide nanoparticles synthesized by thermal decomposition of iron-oleate in 1-octadecene without stirring. Scale bars of 50 nm [49]. ©2012 The Royal Society of Chemistry, with permission. Reproduced with permission from J. Mater. Chem. 22 (39), 21065 (2012). Copyright 2012 Royal Society of Chemistry.

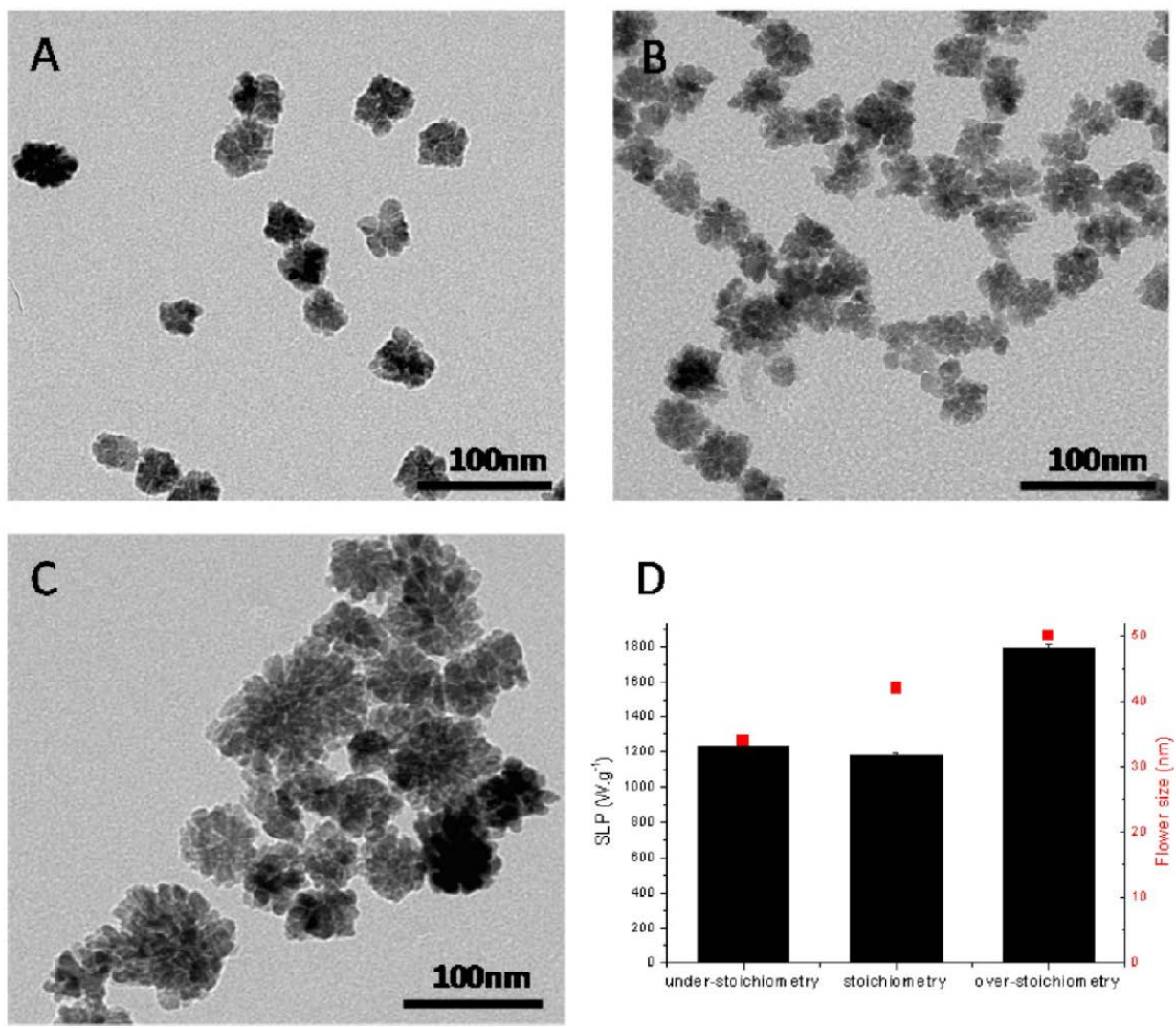

**Figure 6 –** Transmission electron microscopy micrographs of nanoflowers with sizes tuned by adjusting NaOH equivalents relative to iron chloride precursors (A), (B), (C). (D) *SAR* (black bars) and mean diameter (red squares) of the samples [66]. Reproduced with permission from J. Phys. Chem. C 116 (29), 15702 (2012). Copyright 2012 American Chemical Society.

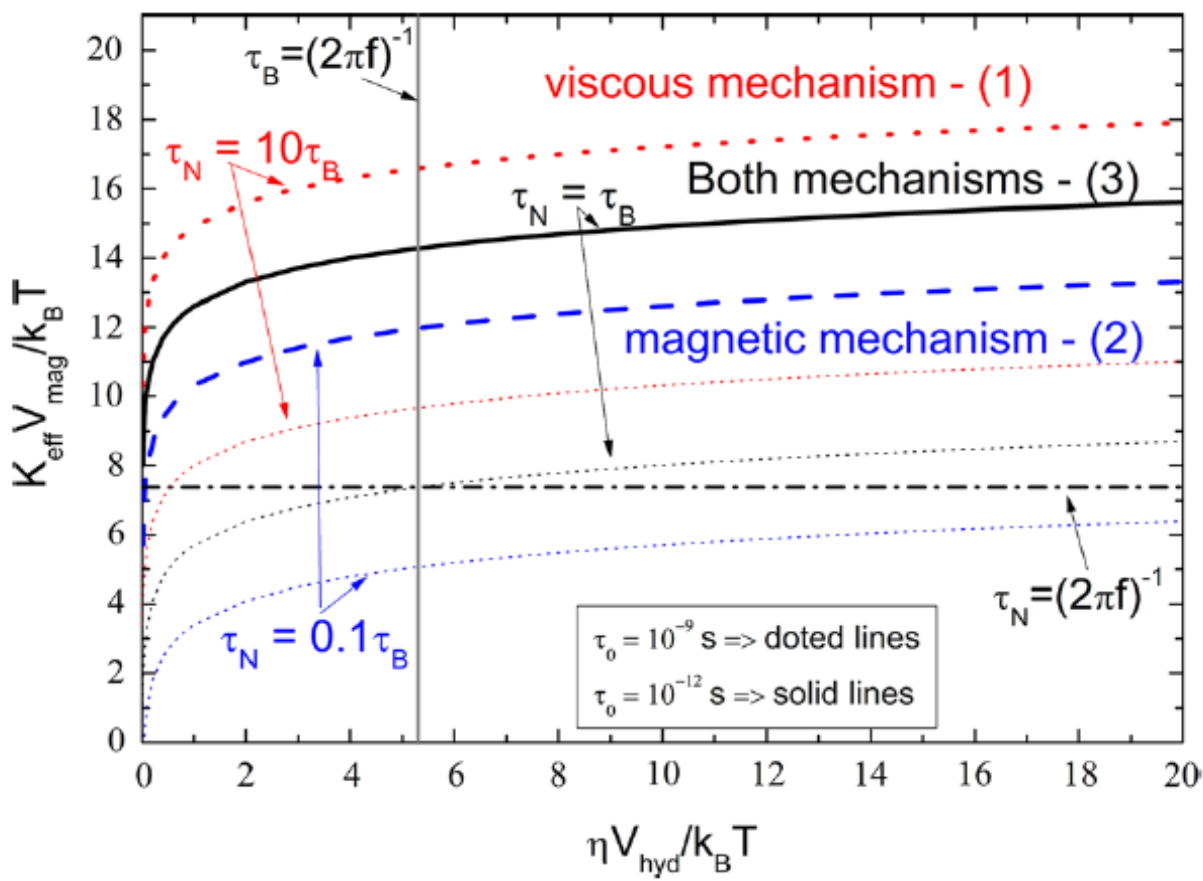

**Figure 7** – Proposed diagram where $y$ and $x$ axes correspond to the dimensionless quantities $K_{eff}V_{mag}/k_BT$ (proportional to $\ln(\tau_N/\tau_0)$) and $\eta V_{hyd}/k_BT$ (proportional to $\tau_B$), respectively. Black, blue, and red lines correspond to $\tau_N = \tau_B$, $\tau_N < 0.1\ \tau_B$ and $\tau_N > 10\ \tau_B$, respectively, when considering $\tau_0 = 10^{-9}$ and $10^{-12}$ s (dashed and solid lines, respectively) [107]. Reproduced with permission from J. Nanopart. Res. 16, 2791 (2014). Copyright 2014 Springer.

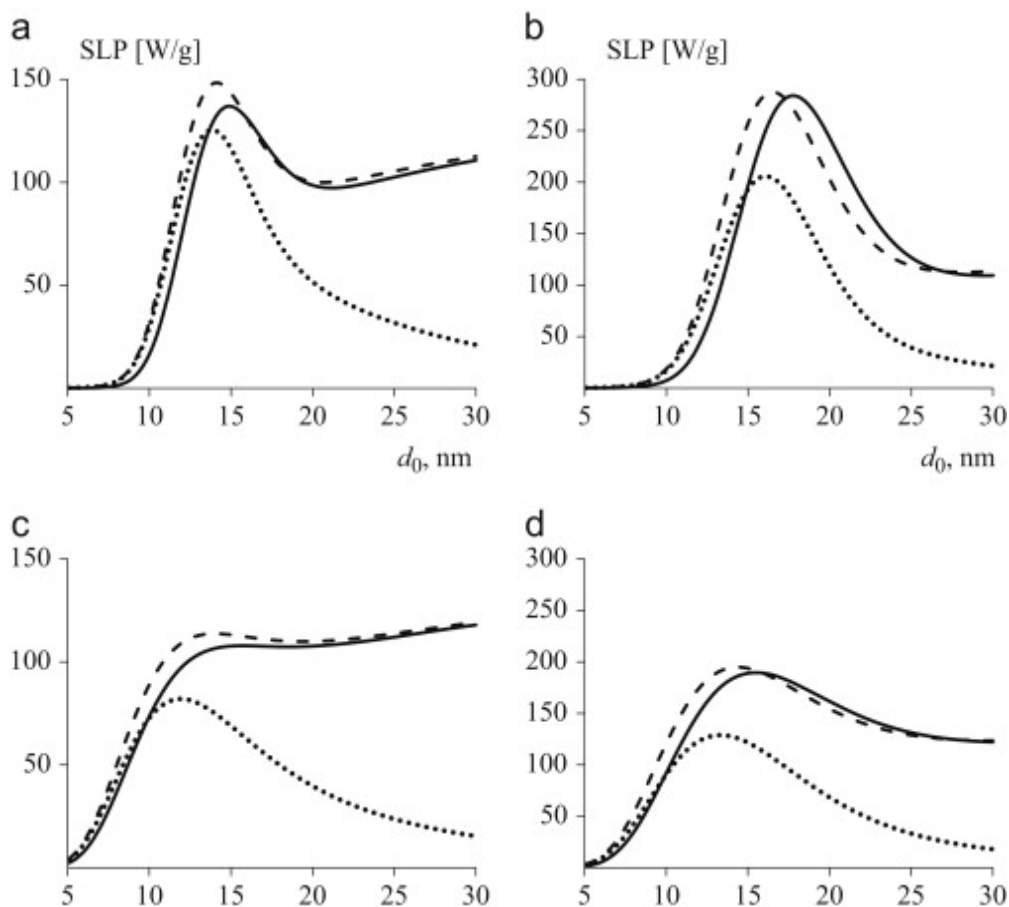

**Figure 8 –** Particle size dependence of *SLP* (or *SAR*) at $f$ = 500 kHz (maghemite colloid) obtained with exact (solid lines) and PHM (dashed lines) and MHM (dotted lines) approaches; AC field amplitude $H_0$ = 8 kAm$^{-1}$ (~100 Oe). Left column (a and c): particles with bulk anisotropy $K$ = $1.6 \times 10^4$ Jm$^{-3}$; right column (b and d): particles with surface anisotropy $K_S$ = $2.7 \times 10^{-5}$ Jm$^{-2}$; size polydispersity coefficients are $s$ = 0.15 (a and b) and 0.3 (c and d); maghemite mass density is 5000 kg/m$^3$; fluid viscosity is 0.001 Pa·s [111]. Reproduced with permission from J. Magn. Magn. Mater. 368, 421 (2014). Copyright 2014 Elsevier.

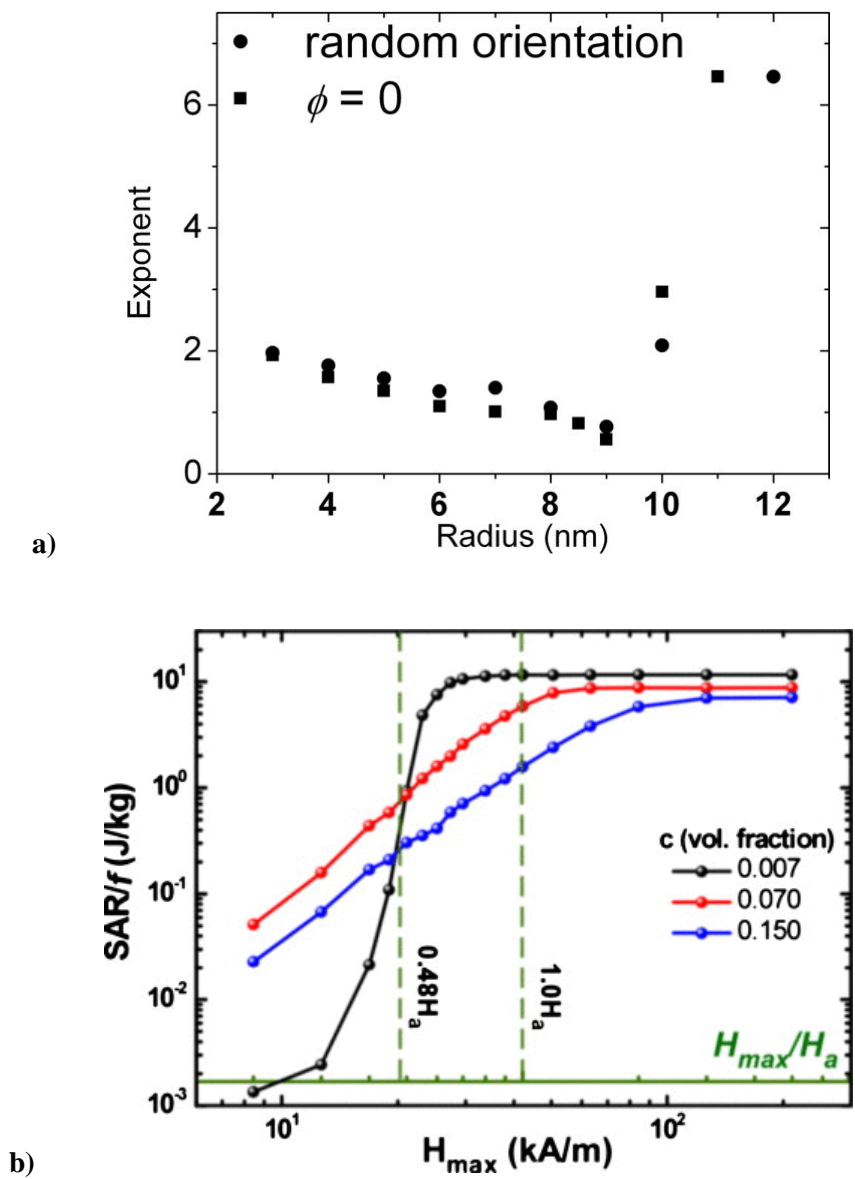

**Figure 9 –** a) Exponent of the best power law fit to numerical simulations of hysteresis loop areas for nanoparticles with random orientation of anisotropy axes (full circles) and parallel to the external field (full squares) [99]. When there was an inflection point in the curve, the fit was performed only up to this point. Reproduced with permission from J. Appl. Phys. 109, 083921 (2011). Copyright 2011 AIP Publishing LLC. b) Field-dependence of the *SAR* for different nanoparticle concentrations (*c* = 0.007, 0.070, 0.150) [113]. Reproduced with permission from J. Appl. Phys.108, 073918 (2010). Copyright 2010 AIP Publishing LLC.

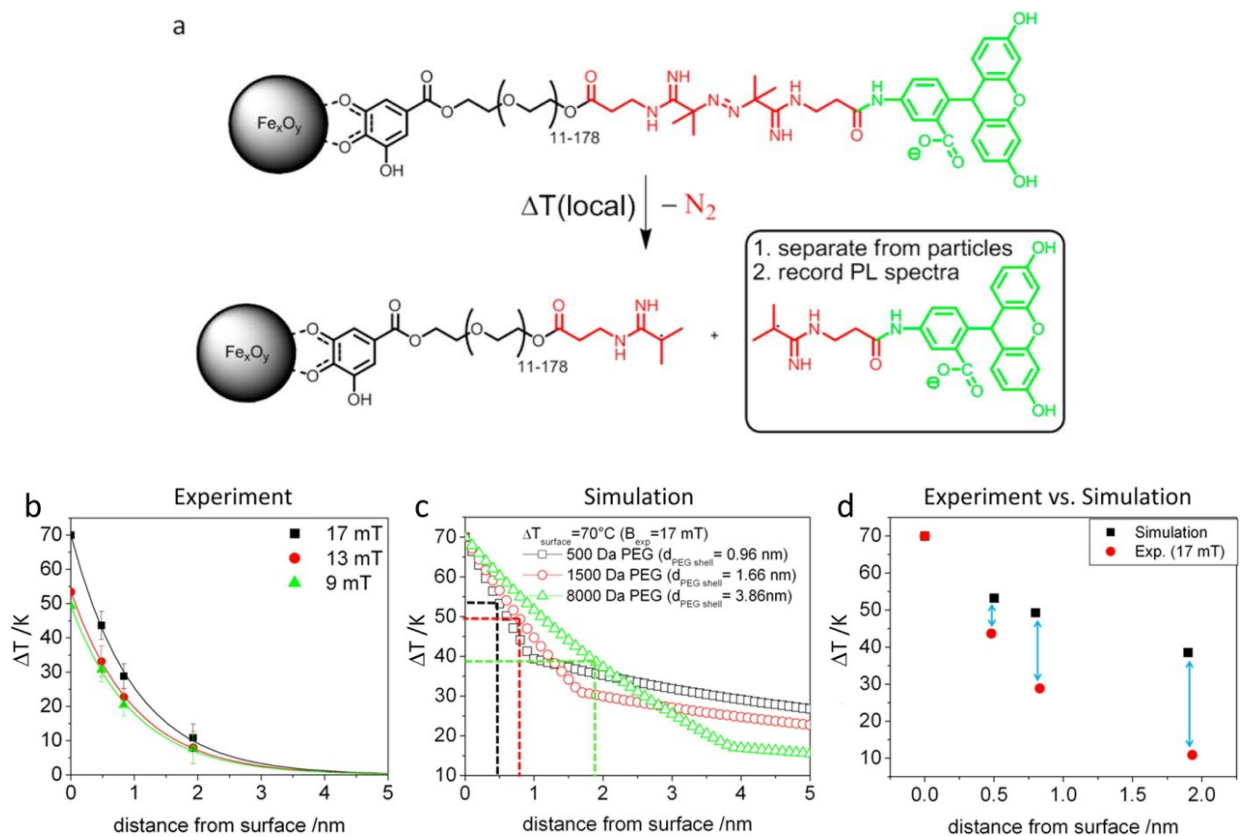

**Figure 10 –** (a) Sketch of the readily functionalized iron oxide nanoparticles bearing FA (fluoresceineamine) connected through VA057 azo molecule to the tails of PEG spacers of different molecular weights. An increase in temperature results in increased cleavage of the azo group and release of the dye from the particles. The released FA is separated from the nanoparticles by centrifugation, and the PL spectra of the downstreams are recorded. (b) Experimental temperature gradients for all field amplitudes: significant local-to-global temperature differences can be found at distances shorter than 3 nm. (c) Temperature gradients calculated by applying the Fourier law for three different shell thicknesses (PEG500, PEG1500, and PEG8000). (d) Comparison of $\Delta T$ values at 0, 0.47, 0.83, and 1.9 nm. The discrepancy between experimental data and the diffusive heat transport model at this nanoscale regime can be observed and is increasing with distance [115]. Reproduced with permission from Nano Lett. 13 (6), 2399 (2013). Copyright 2013 American Chemical Society.

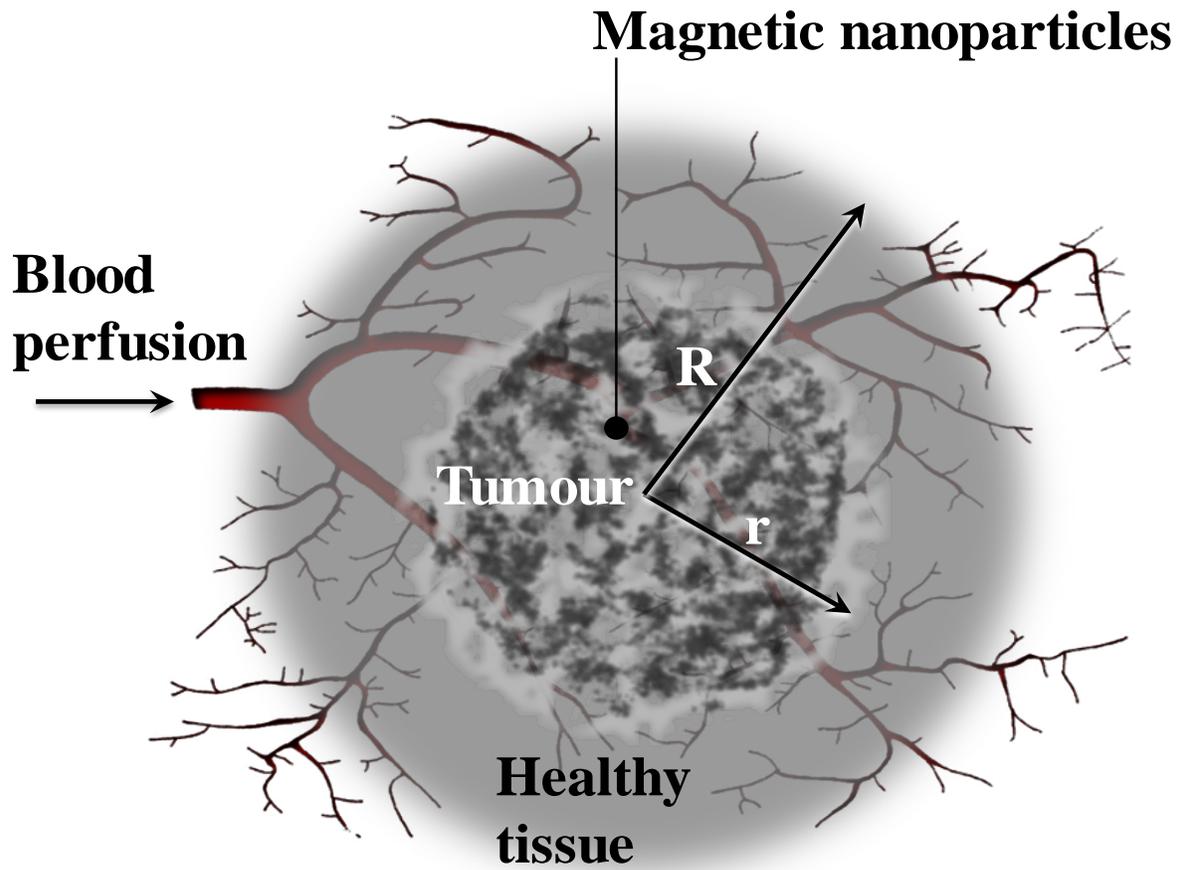

**Figure 11** – Typical concentric tumor model, where *r* is the tumor radius and *R* the radius of the sample

body region. Blood perfusion is considered homogeneous throughout both the tumor and healthy tissue. Magnetic nanoparticles are homogeneously distributed only inside the tumor.

Tumor own microvasculature is not depicted.

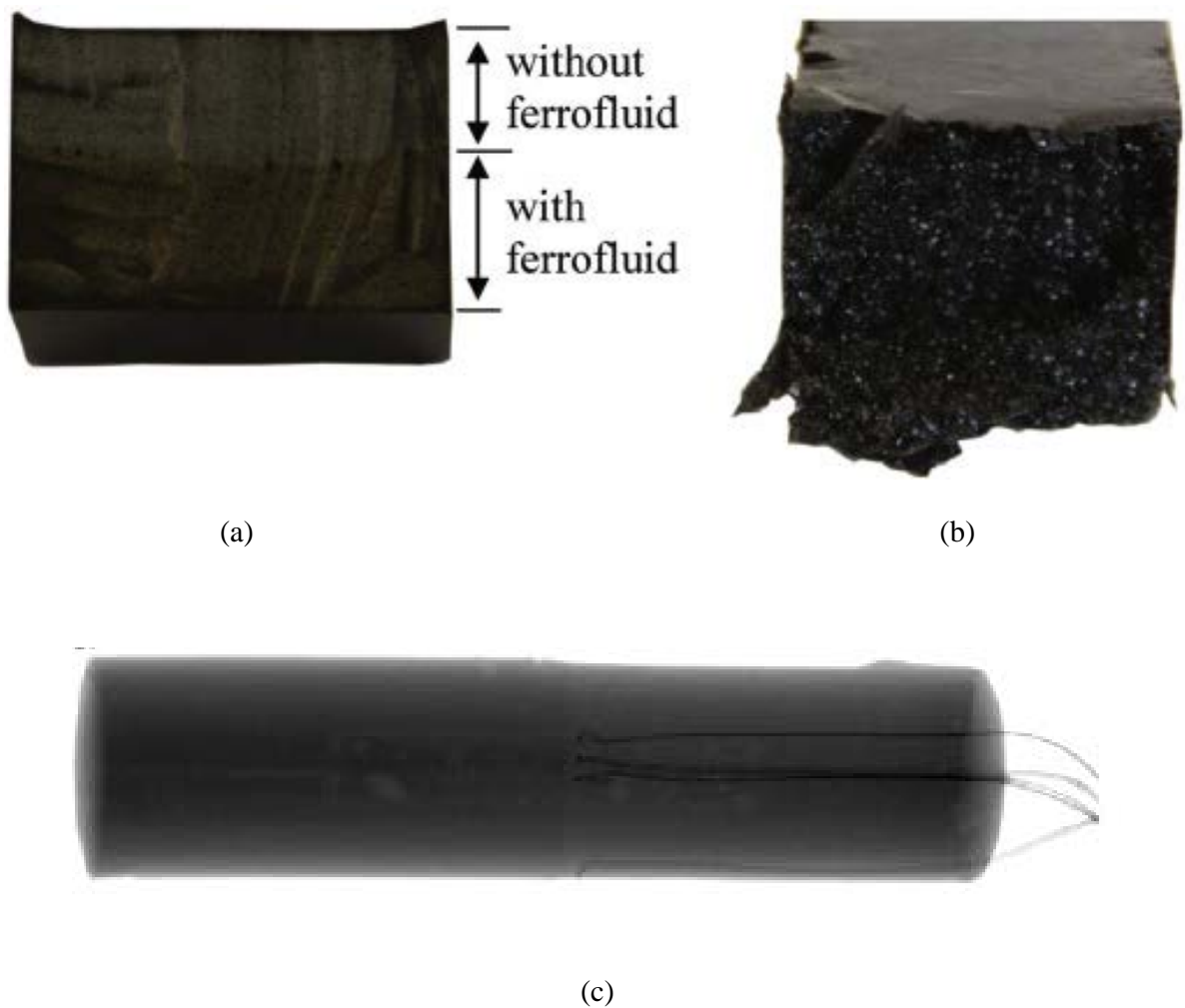

(a)  (b)

(c)

**Figure 12 –** Pictures of polyurethane with suspended magnetic particles: (a) an un-foamed solid gel consisting of a section enriched with MNPs at the bottom, and pure polyurethane at the top and (b) a polyurethane foam created using a biocompatible MNPs. c) Cross section of a tomogram reconstructed from 720 radiographs taken with an angular resolution of 0.5°. The spatial resolution of the tomogram allows a precise determination of the thermocouple positions [129]. Reproduced with permission from J. Magn. Magn. Mater. 351, 1 (2014). Copyright 2014 Elsevier.

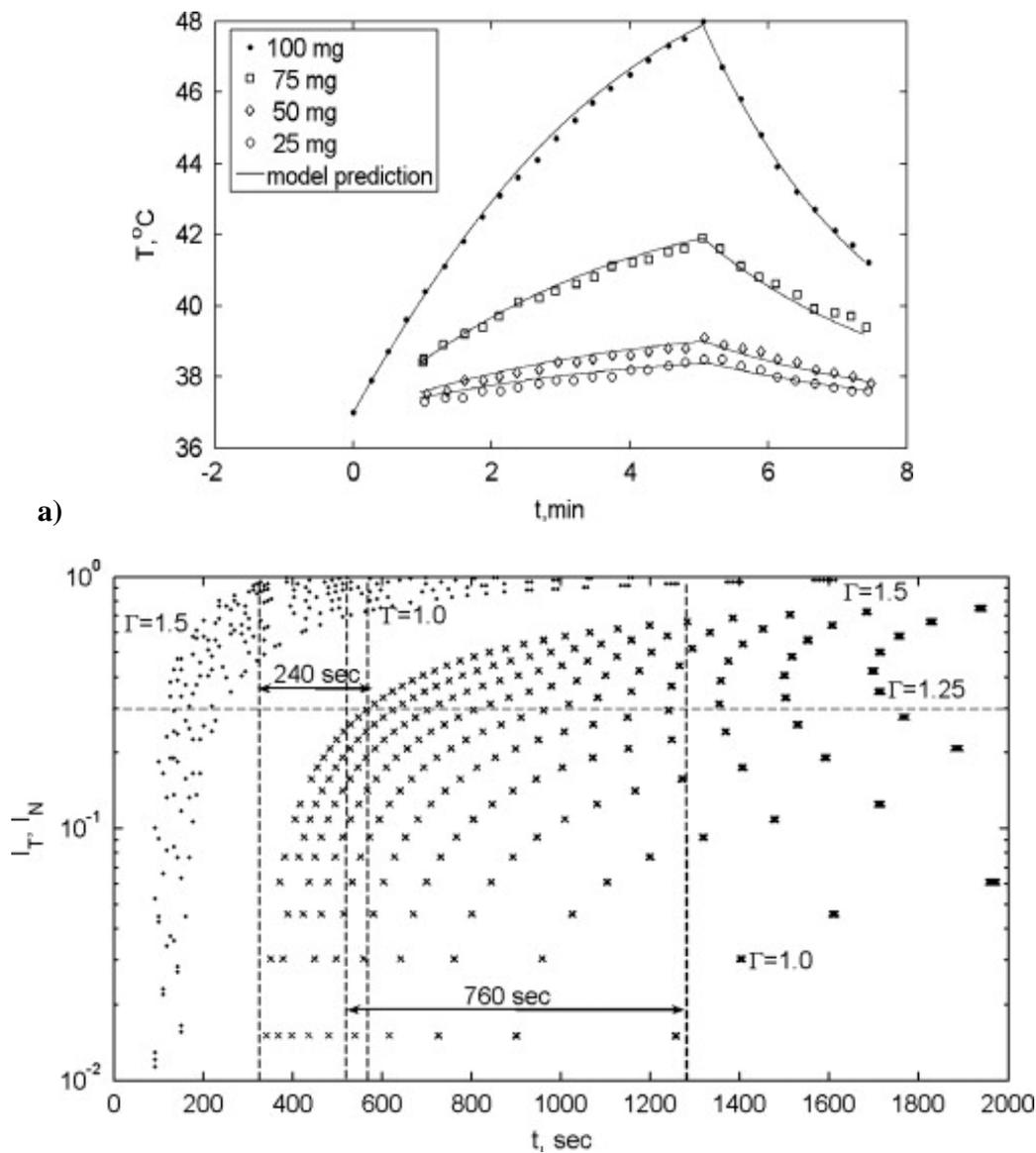

a)

b)

**Figure 13 –** Comparison between model predictions (solid lines) and a previous experimental investigation of ferromagnetic embolization hyperthermia [136]. The scattered symbols represent the mean tumor temperature monitored over the five subjects in each treatment group receiving different doses of ferrous particles, while the lines represent the corresponding results from the model. b) Temporal evolution of $I_T$ and $I_N$ for varying $\Gamma$ at $Pe$=1. The dots represent $I_T$ and the crosses $I_N$. See ref. [136] for further information. Reproduced with permission from J. Magn. Magn. Mater. 323, 708 (2011). Copyright 2011 Elsevier.

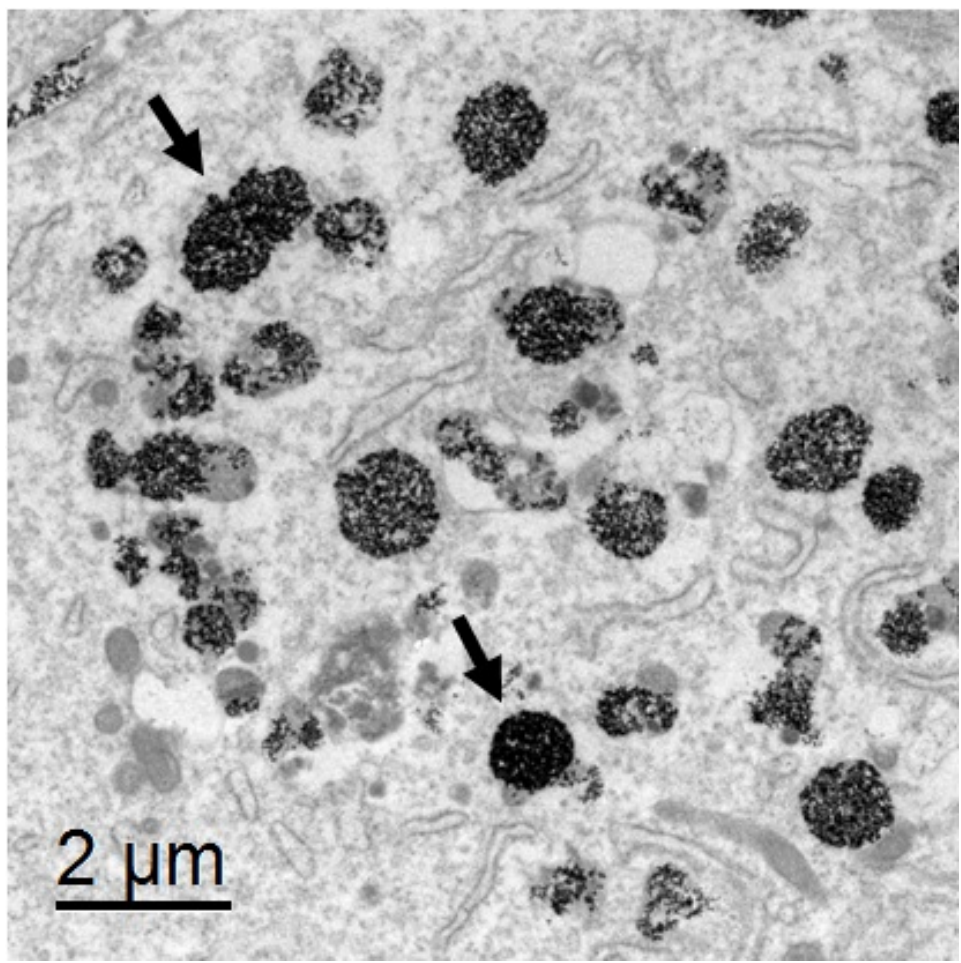

**Figure 14 –** Distribution of citric acid coated iron oxide nanoparticles inside endosomes in DX3 human melanoma cells. Courtesy of C. Blanco-Andujar.

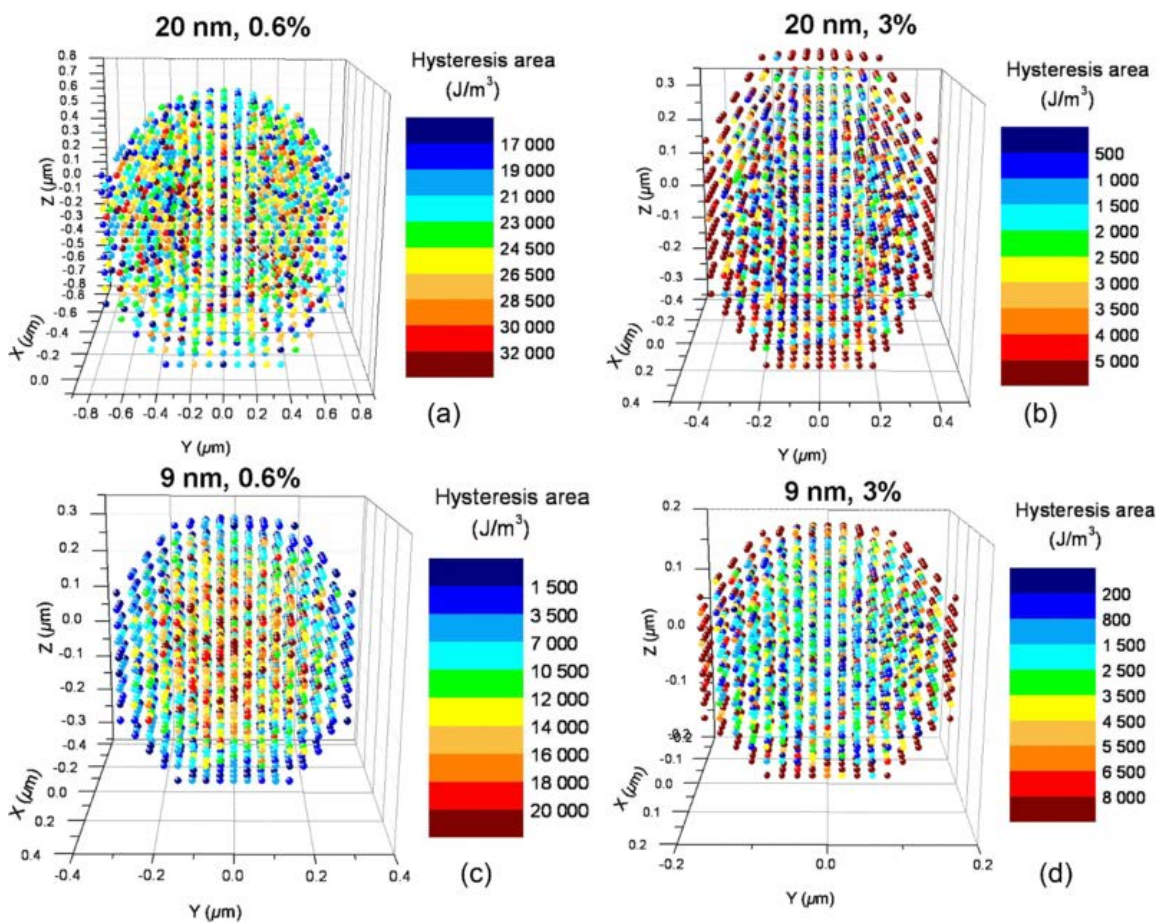

**Figure 15** – Heating power of MNPs inside a lysosome. The plotted heating power corresponds to an average over 50 hysteresis loops with a change in the anisotropy axis direction and the exact NP position between each cycle. The MNPs are shown positioned on a cubic lattice, which is thus their *average* position. The heating power displayed here is actually that of each NP and is not spatially averaged. The size of the MNPs in the figure has been chosen for clarity reasons and does not match their true size. Only half of the lysosome is shown so the reader faces the hemisphere. (a) $d$=20nm, $\varphi$=0.6%. (b) $d$=20nm, $\varphi$=3%. (c) $d$=9nm, $\varphi$=0.6%. (d) $d$=9nm, $\varphi$=3%. $d$ is the particle diameter, $\varphi$ and the volumetric concentration of nanoparticles inside the lysosome [153]. Reproduced with permission from Phys. Rev. B 90, 214421 (2014). Copyright 2014 American Physical Society.

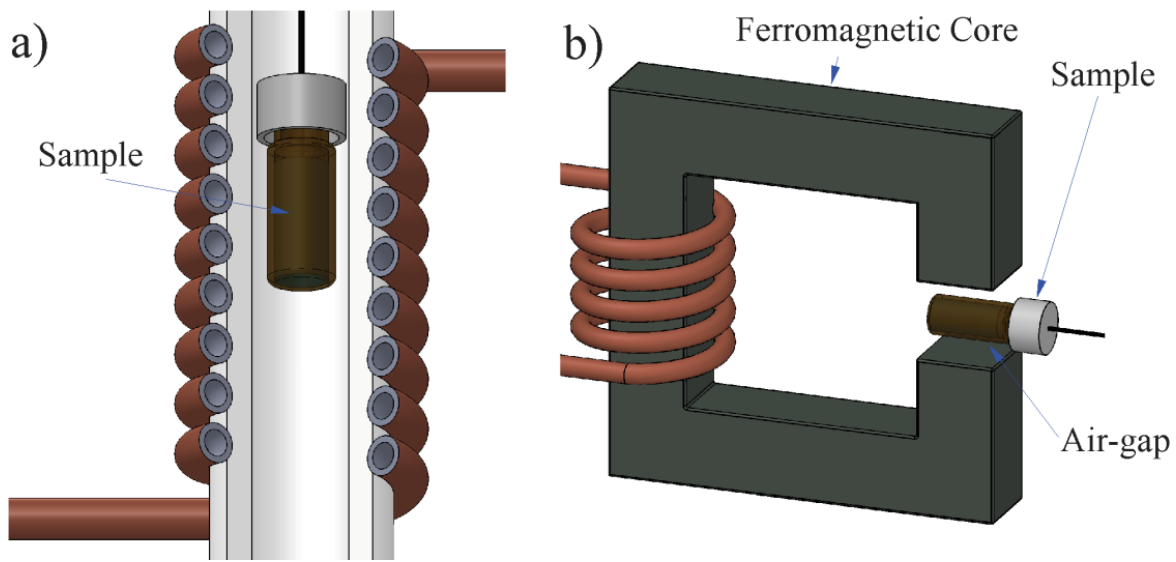

**Figure 16 –** Two types of inductors. a) Air coil to generate a AMF with the sample placed in the center of the inductor. b) A ferromagnetic core to concentrate the field into an air gap, where the sample is placed. In both cases, an alternating current $I_{AC}$ crosses the conductor.

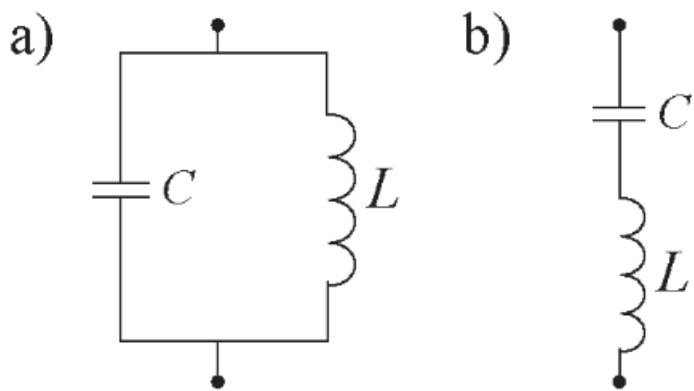

**Figure 17** – LC resonant circuits: a) series and b) parallel.

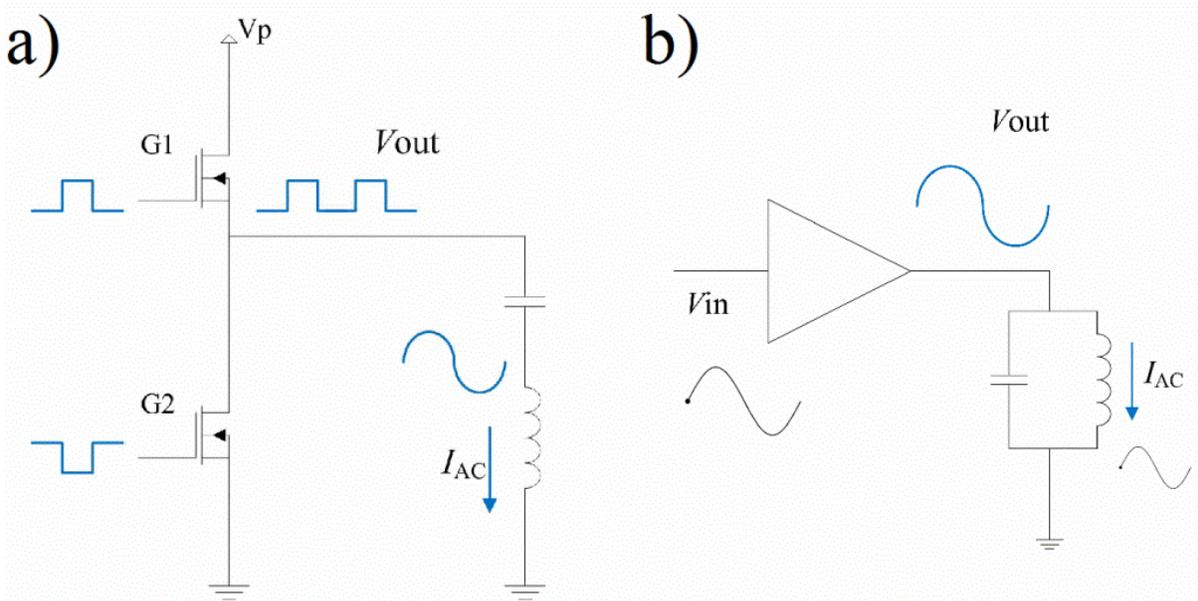

**Figure 18** – a) Example of a half h-bridge amplifier with two MOSFET transistors as switchers. The amplifier feeds a series resonant circuit. b) A linear power amplifier connected to a parallel resonant circuit. Note that the output signal ($V_{out}$) of the h-bridge amplifier is pulsed, whereas the output signal of the linear amplifier is nearly sinusoidal.

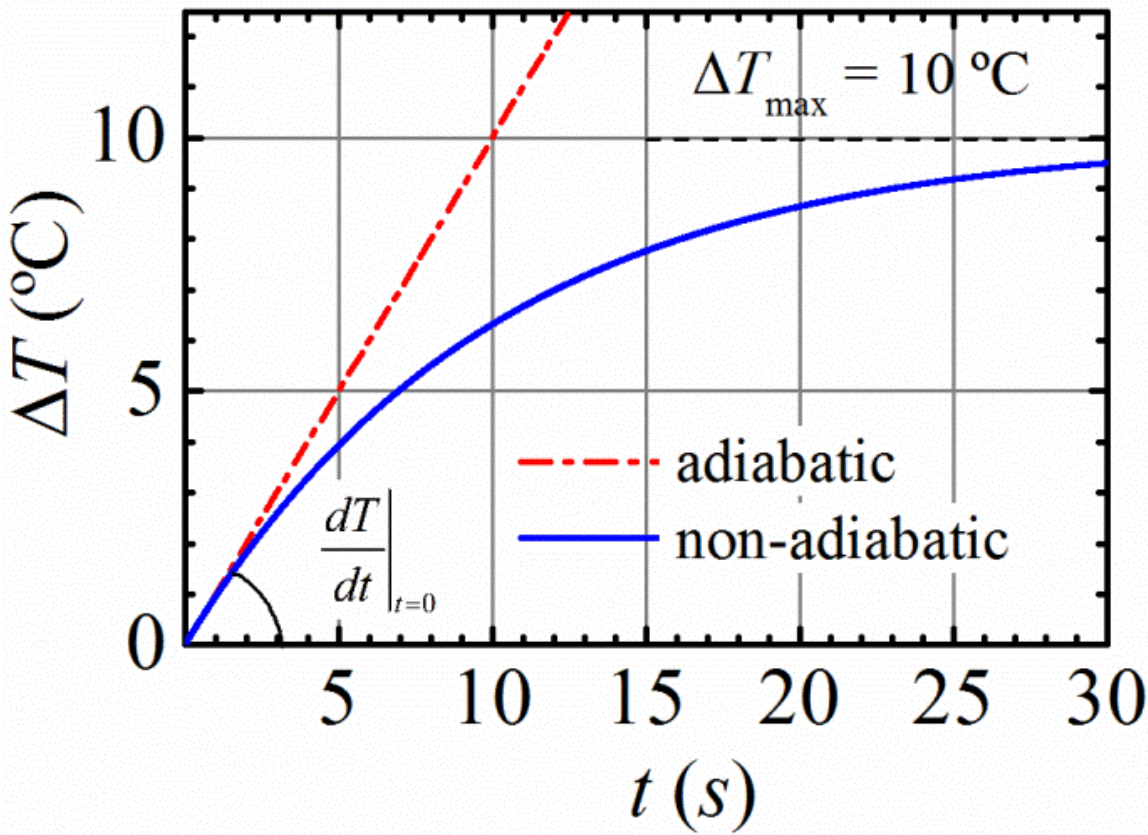

**Figure 19** – Temperature evolution for an adiabatic sample (Eq. 4.6) and for a non-adiabatic one (Eq. 4.7). The initial temperature derivative over time is 1 ºC·s⁻¹, whereas $\lambda_Q$ in Eq. 4.7 is 10 s. The *y*-axis represents the temperature difference ($\Delta T = T - T_0$). Note that in the non-adiabatic case, there is a plateau when temperature reaches the steady state at around 10ºC. Also, note that the initial temperature slope is the same for both situations: adiabatic and non-adiabatic.

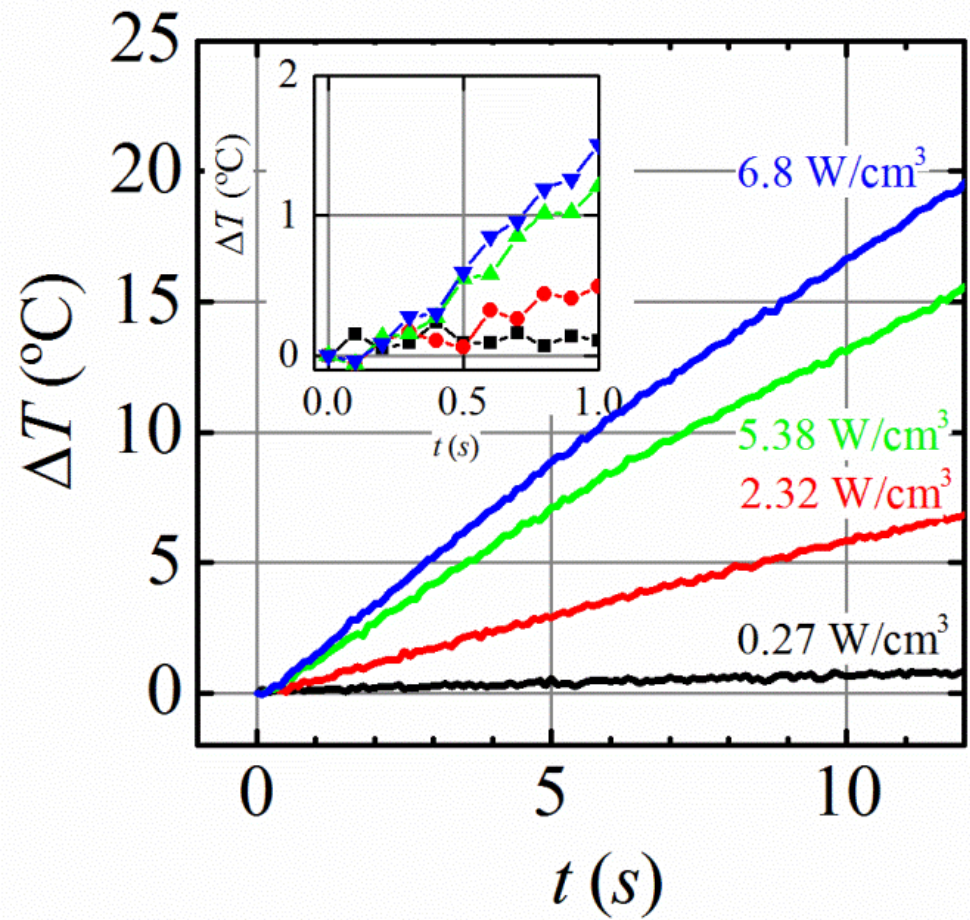

**Figure 20** – Time evolution of sample's temperature increment for different measured $P_{vd}$ in Wcm$^{-3}$. The inset displays the temperature evolution during the first second [178]. Reproduced with permission from J. Magn. Magn. Mater. 368, 432 (2014). Copyright 2014 Elsevier.

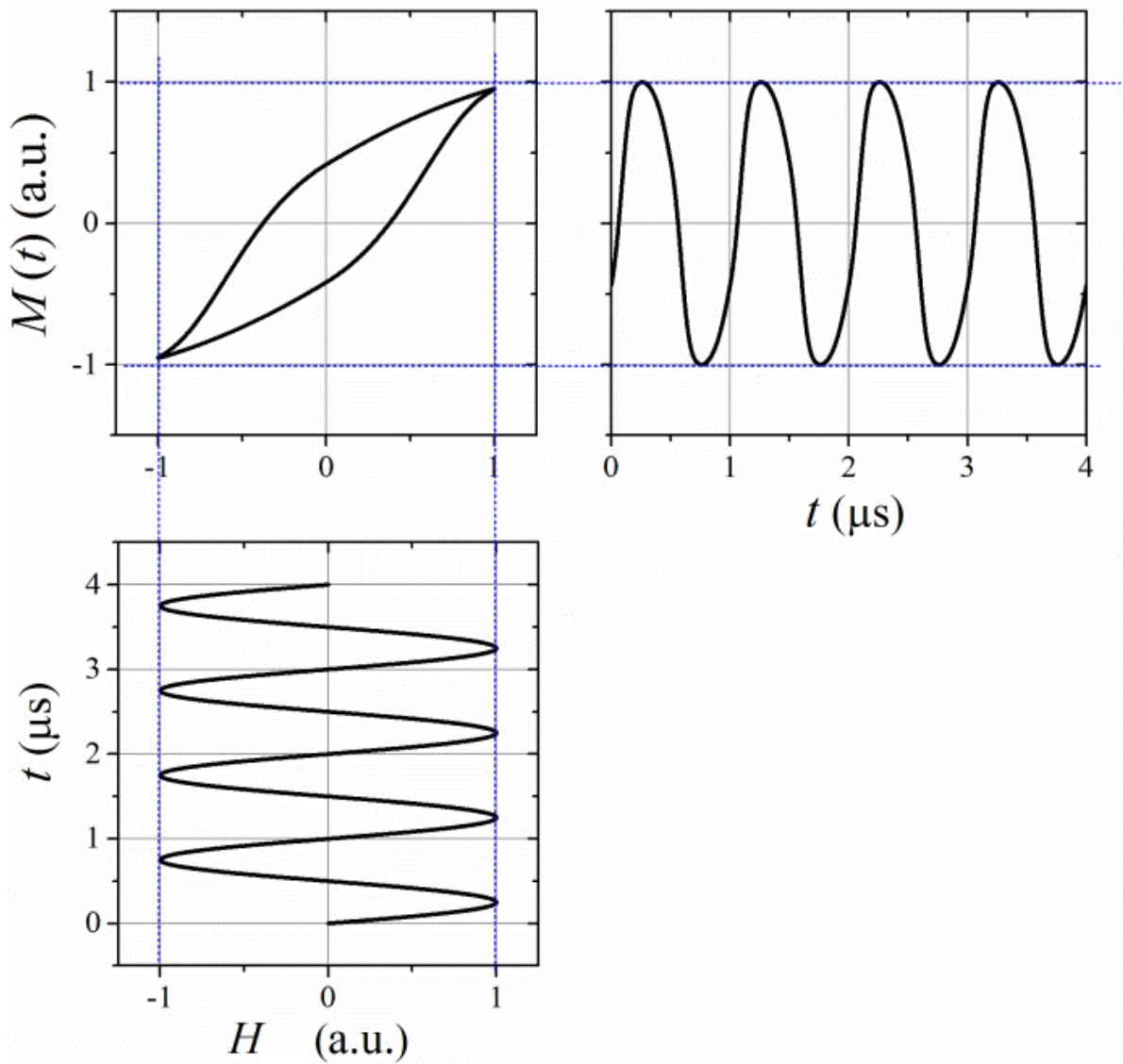

**Figure 21** – AC hysteresis loops obtained plotting the dynamic magnetization *M(t)* as function of the applied field. The absorbed power is proportional to the area of the loop.

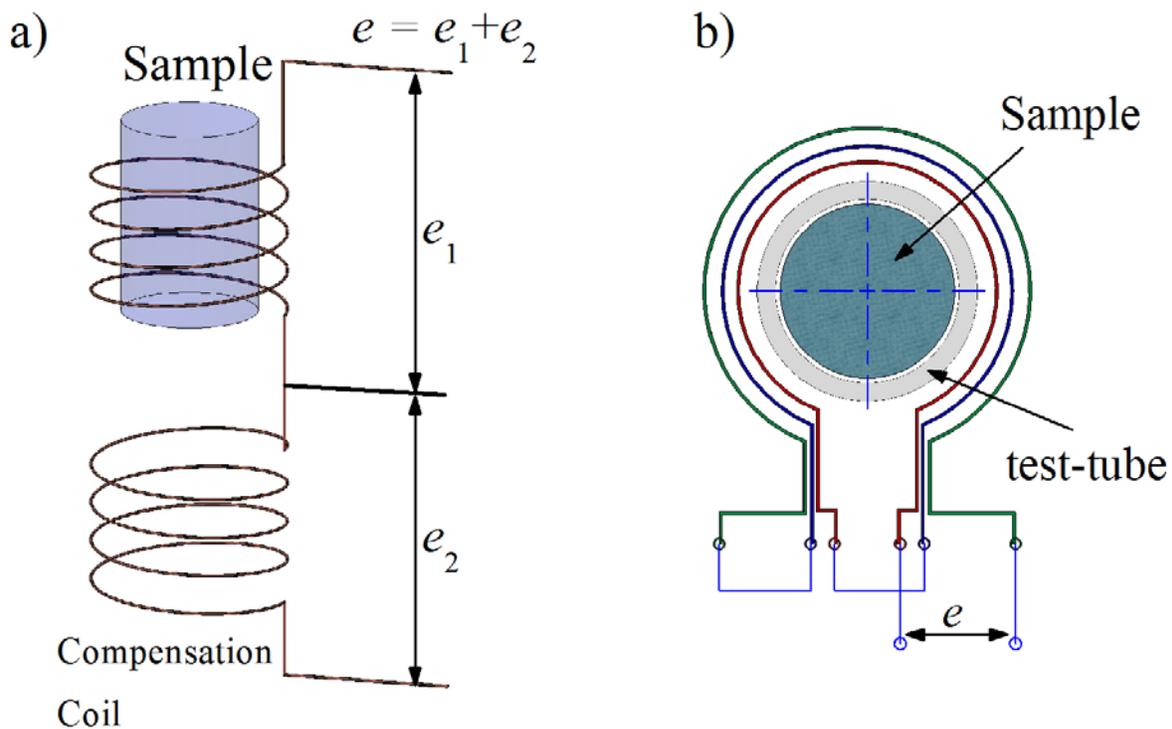

**Figure 22 –** Diagram of the pick–up coil together with the compensation coil. Because the two coils are symmetric, the total induced voltage is null ($e_1 + e_2 = 0$) when the sample is absent. The drawings in this figure are adapted from the cited references [163,166].

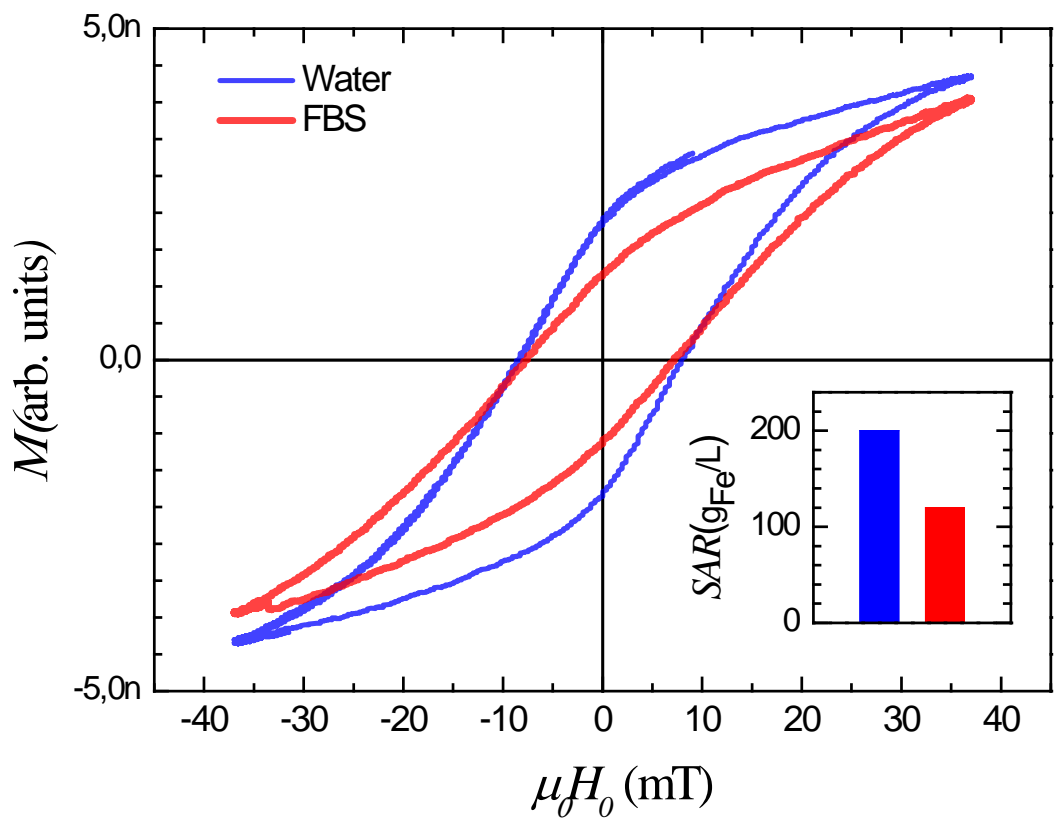

**Figure 23** – Magnetization cycles of 20 nm size γ−Fe$_2$O$_3$ NP dispersed in water (blue color) and FBS medium (red color) at $f$ = 55 kHz and 27.8 kAm$^{-1}$ (~350 Oe) and 1g$_{Fe}$/L. Inset: *SAR* values of 20 nm size γ−Fe$_2$O$_3$ NP dispersed in water (blue color) and FBS medium (red color) at $f$ = 185 kHz and 19.9 kAm$^{-1}$ (~250 Oe) and 1g$_{Fe}$/L. Data kindly provided by Antonio Aires and David Cabrera.

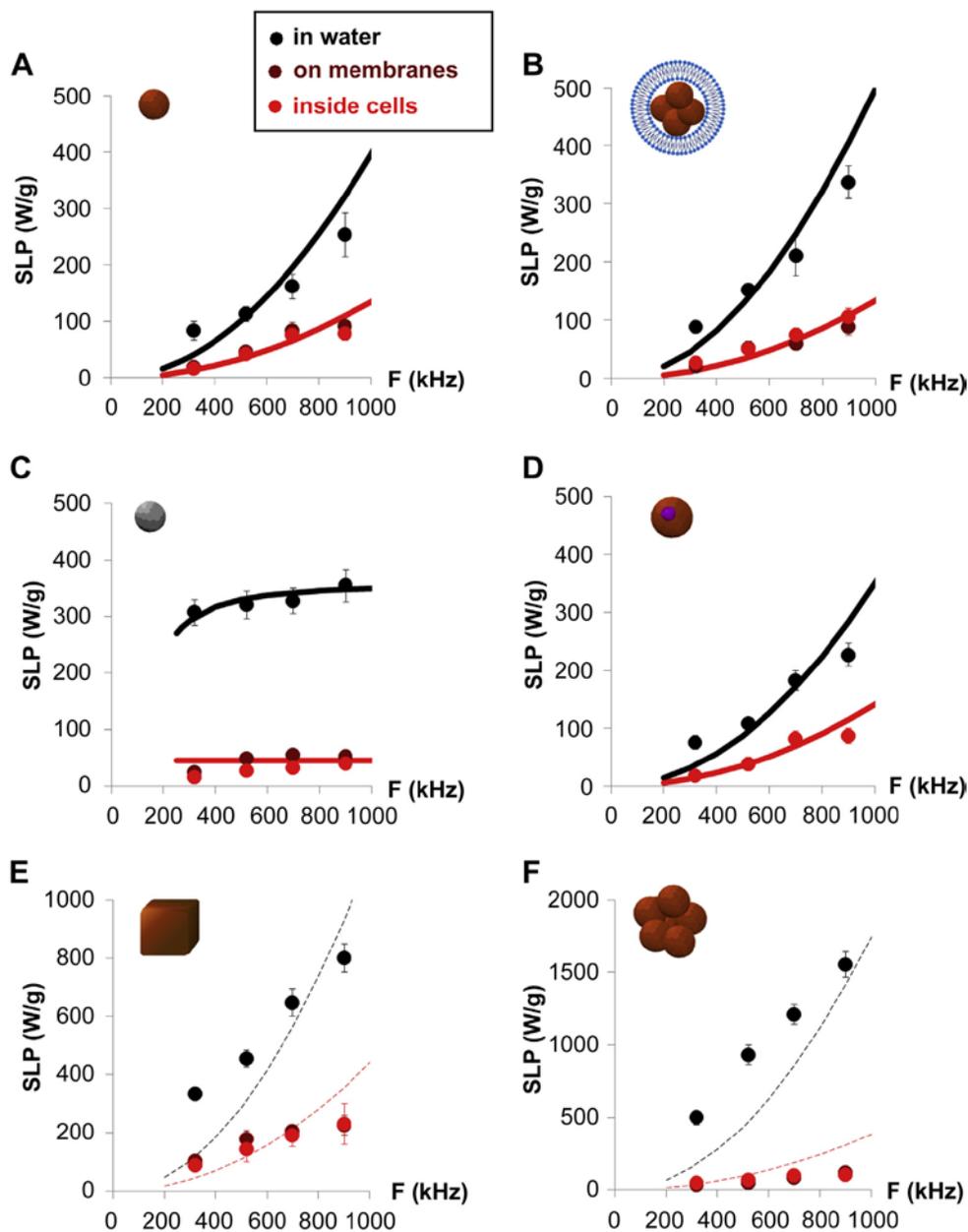

**Figure 24 –** Frequency dependence of *SLP* values obtained from MNPs colloids dispersed in water (black color), located onto the cell membrane (brown color) or internalized into cells (red color): A) maghemite nanoparticles, B) in liposomes, C) cobalt ferrite nanoparticles, D) iron oxide/gold dimers, E) iron oxide nanocubes, F) iron oxide nanoflowers. Lines correspond to fits according to the linear response theory [109]. Reproduced with permission from Biomat. 35, 6400 (2014). Copyright 2014 Elsevier.

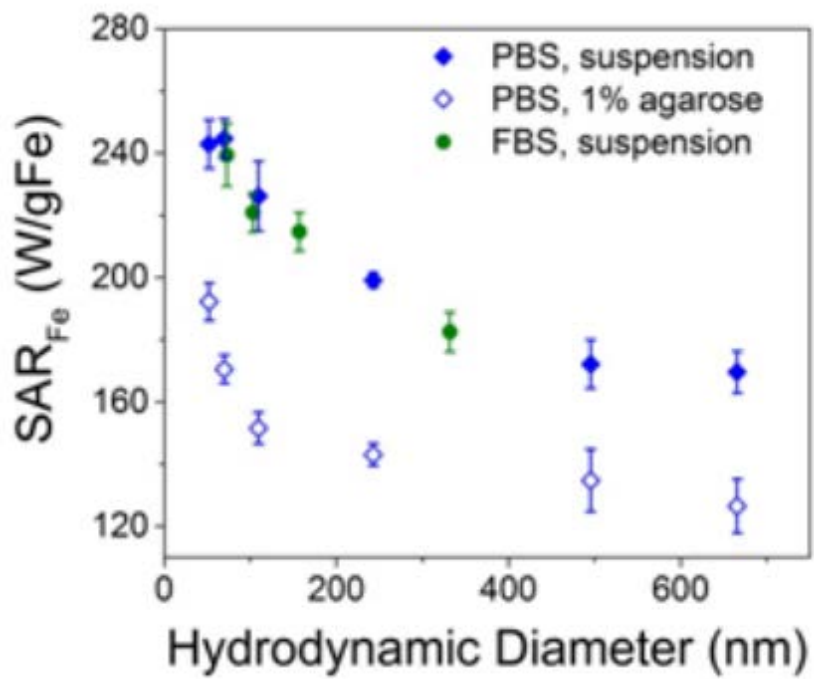

**Figure 25 –** Aggregate size dependence of *SAR* of MNPs dispersed in different media applying an AMF at 190 kHz and 20 kAm$^{-1}$ (~ 250 Oe) [139]. Reproduced with permission from Technology 2, 214 (2014). Copyright 2014 WSPC.

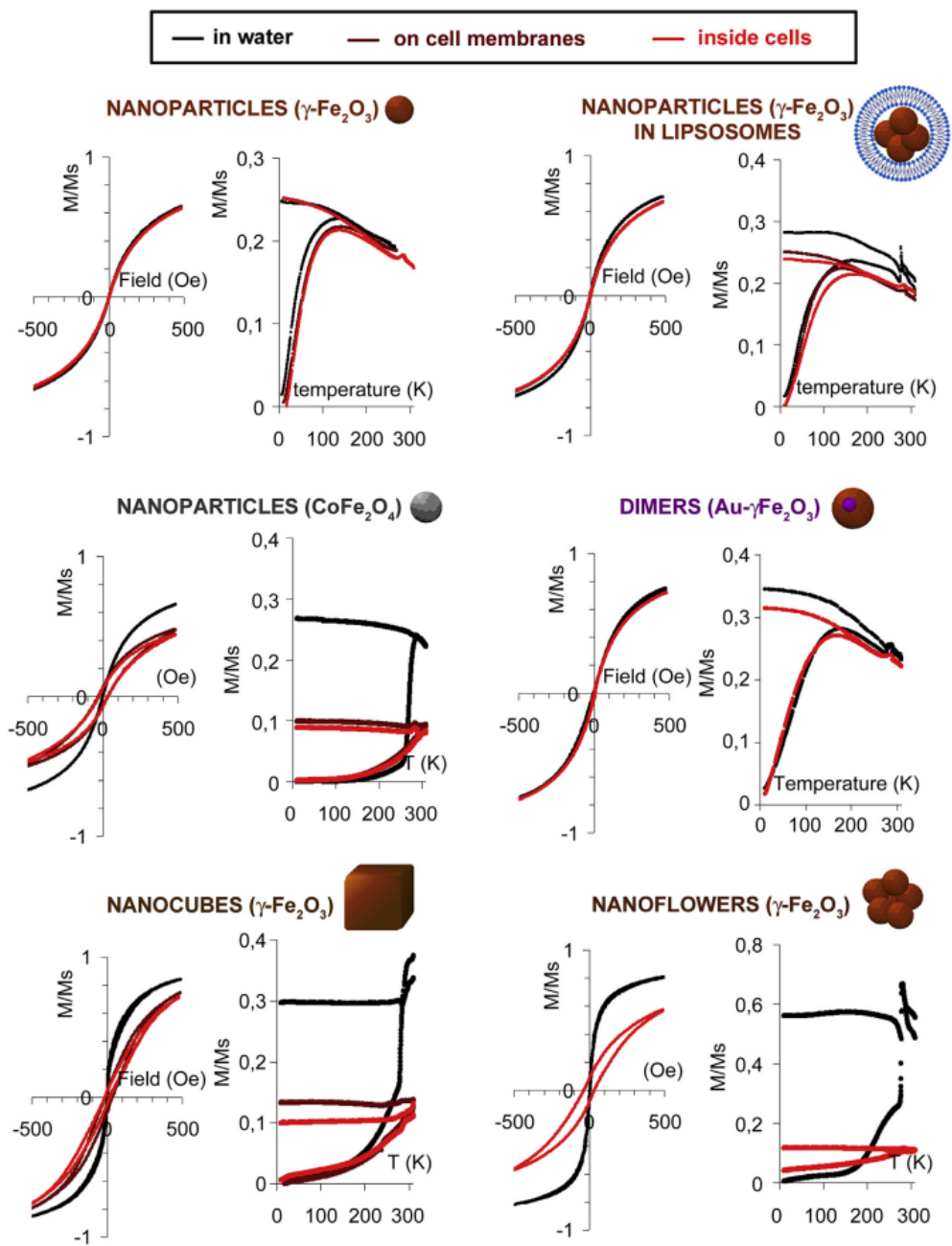

**Figure 26 –** Magnetic properties of different MNPs colloids dispersed in solution, located on the cell membrane, and internalized inside cells. Magnetization cycles measured at *T*= 300 K (left), ZFC-FC magnetic susceptibility, measured at 4 kAm$^{-1}$ (50 Oe) (right) [109]. Reproduced with permission from Biomat. 35, 6400 (2014). Copyright 2014 Elsevier.

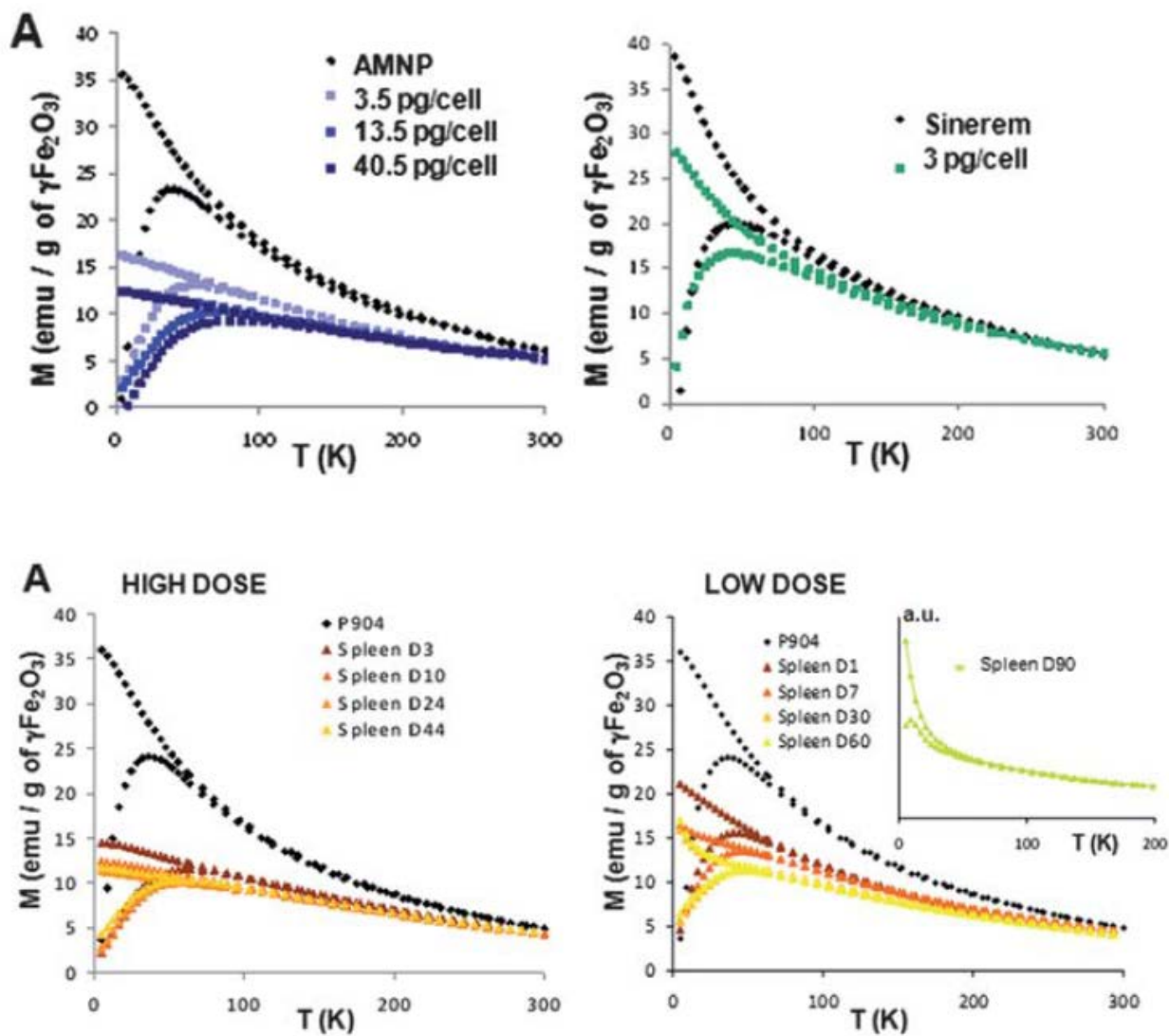

**Figure 27 –** Upper A) ZFC/FC magnetization curves from AMNP dispersed in water (black dots) and internalized into macrophages at different iron loads per cell (color dots) or with Sinerem_ MNP (green dots, right). Lower A) ZFC/FC magnetization curves of spleen at different times after high dose or low MNP dose injection of P904 [188]. Reproduced with permission from Nanoscale 3, 4402 (2011). Copyright 2014 Royal Society of Chemistry.

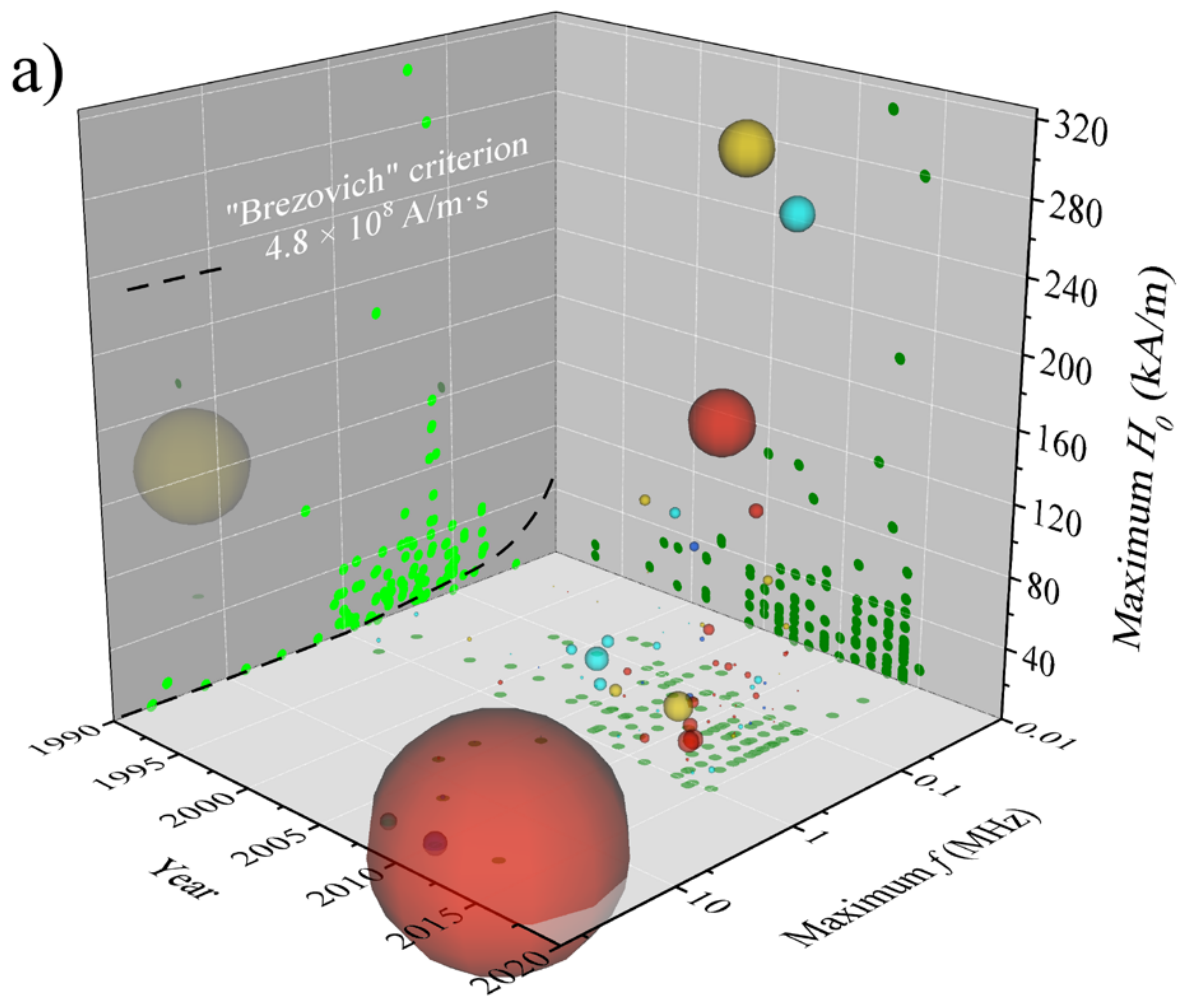

**Figure 28 –** (a) 3D scatter plot showing the main experimental conditions reported in a set of 120 publications on MH spanning over the last 25 years. Scatter size is proportional to the $H_0 \cdot f$ product. Green dots of different intensities are the 2D projection of the central scatter over the XY, XZ and YZ planes. The color code of the spheres is related to the type of assay described in the corresponding publication: red refers to *in vivo* tests, yellow to *in vitro* tests, blues to combined *vivo/in vitro* tests and cyan to *SAR* measurements. The dashed black curve superimposed to the field intensity vs maximum frequency plane indicates the "Brezovich" or $H_0 \cdot f$ criterion, along which the $H_0 \cdot f = 4.8 \times 10^8$ A(ms)$^{-1}$ condition is fulfilled [note that the experiments complying with this criterion are those below the curve].

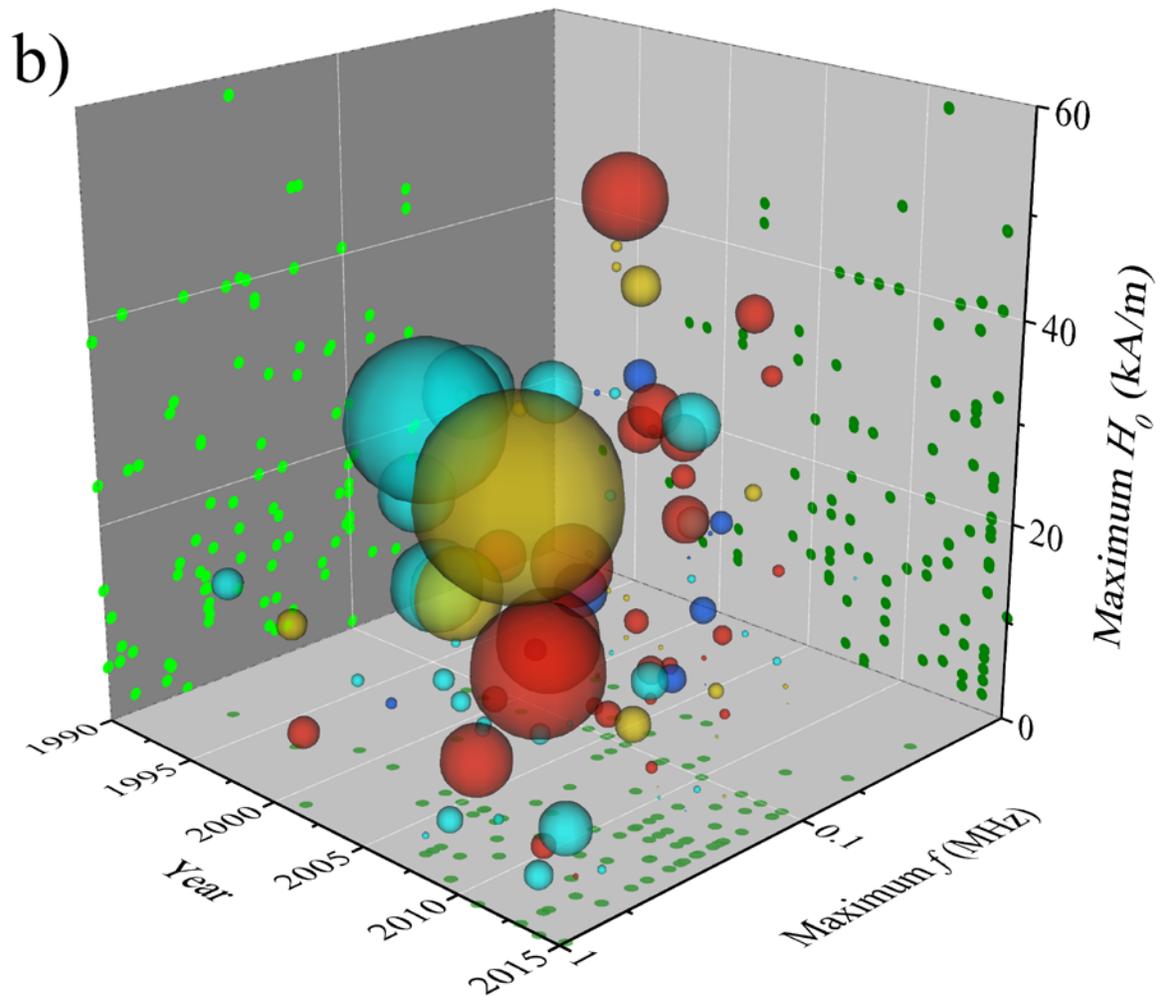

**Figure 28 –** (b) Zoom of the most populated region of (a), where the concentration of *SAR* measurements in the 2005-2009 period an *in vivo* tests in the 2010-2015 period is shown.

$$\left. \begin{array}{l} \text{Homogeneous}: \phi(r) = \phi_H \\[2ex] \text{Linear}: \phi(r) = 4\phi_H \left(1 - \dfrac{r+R}{r}\right) \\[2ex] \text{Parabolic}: \phi(r) = \dfrac{5}{2}\phi_H \left(1 - \left(\dfrac{r+R}{r}\right)^2\right) \end{array} \right\} \text{where} \quad \phi_H = \dfrac{\dfrac{m_{NP}}{\rho_{NP}}}{\dfrac{m_t}{\rho_t}}$$

**Table 1** – Nanoparticle volume distributions across a generic tumor considered by Lahonian and Golneshan [133]. Notation: $\phi_H$ is the homogeneous volume fraction, $m_{NP}$ is the nanoparticle mass, $\rho_{NP}$ the nanoparticle density, $m_t$ the tumor mass and $\rho_t$ the tumor density.

| | Description | Inductor | Field parameters |
|---|---|---|---|
| Jordan et al. [159] | Human-sized prototype | Ferromagnetic core | 100 kHz<br>up to 18 kAm$^{-1}$ |
| Connord et al. [160] | Coupled with confocal microscopy | Ferromagnetic core | - |
| Garaio et al. [161] | Parallel LCC resonator | Air-coil | 149 – 1030 kHz<br>up to 35 kAm$^{-1}$ |
| Lacroix el al. [162] | Series LC resonator | Ferromagnetic core. Litz-wire | 100 – 500 kHz<br>up to 3.8 kAm$^{-1}$ |
| Chieng-Chi Tai et al. [164] | Half-bridge inverter configuration | Ferromagnetic core | - |
| Cano et al. [165] | Full-bridge inverter configuration | Air-coil | 206 kHz.<br>up to 12 kAm$^{-1}$ |
| Bekovic et al. [166] | Rotational magnetic field | Air-coil | 20 – 160 kHz<br>up to 4.1 kAm$^{-1}$ |
| Dürr et al. [167] | Experiments with small animals | 2 flat pancake coils | 200 kHz<br>up to 6.76 kAm$^{-1}$ |

**Table 2** – Cited electromagnetic applicators.

**Table 1.** Evolution of nanoparticle characteristics over time in the lysosome-like medium.

| Time (Days) | $M_S$ (A.m².kg⁻¹) | $d_{mag}$ (nm) ($\sigma$) | $d_H$ (nm) (sd) | SAR (W/g) | $d_{NMR}$ (nm) |
|---|---|---|---|---|---|
| 0* | 82 | 14.0 (0.25) | 35.0 (0.80) | 1604 (56) | 22.4 |
| 0 | 71 | 14.0 (0.30) | 91.5 (0.60) | 1761 (35) | 21.2 |
| 1 | 64 | 13.0 (0.30) | 101.4 (0.44) | – | 20.4 |
| 2 | 51 | 13.2 (0.30) | 100.3 (0.46) | – | – |
| 3 | – | – | – | – | 20.8 |
| 6 | 35 | 13.1 (0.20) | 20.8 (0.90) | 532 (58) | 20.4 |
| 14 | 19 | 13.1 (0.19) | 19.6 (0.90) | 192 (24) | 13.2 |
| 23 | 14 | 12.2 (0.17) | 18.4 (0.90) | 19 (5) | – |

*Measurement realized in water.

**Table 3 –** Evolution of MNPs characteristics over time in the lysosome-like medium. Table extracted from Ref. [217]. Reproduced with permission from Small 10, 3325 (2014). Copyright 2014 John Wiley and Sons.